%% file: main.tex
\documentclass[opre]{informs3_copy}  

\OneAndAHalfSpacedXI 

\usepackage{endnotes}
\let\footnote=\endnote
\let\enotesize=\normalsize
\def\notesname{Endnotes}%
\def\makeenmark{$^{\theenmark}$}
\def\enoteformat{\rightskip0pt\leftskip0pt\parindent=1.75em
  \leavevmode\llap{\theenmark.\enskip}}

\usepackage{natbib}
 \bibpunct[, ]{(}{)}{,}{a}{}{,}%
 \def\bibfont{\small}%
 \def\newblock{\ }%
\newcommand{\vect}[1]{\boldsymbol{#1}}
\usepackage{booktabs}
\usepackage{amsmath,amssymb}
\usepackage{changes}
\usepackage{diagbox}
\usepackage{graphicx}
\usepackage{bm}
\usepackage[shortlabels]{enumitem}
\usepackage{float}
\usepackage{listings}
\usepackage{amsfonts}
\usepackage{mathtools}
\usepackage{mathrsfs}
\usepackage{makecell}
\usepackage{url}
\usepackage{ulem}
\usepackage{bbm}
\usepackage{lipsum}
\usepackage[framemethod=TikZ]{mdframed}
\usepackage{theoremref}
\usepackage[margin=1in]{geometry}
\usepackage{pgfplots}
\usepackage{breqn}
\usepackage{tikz}
\usepackage{tkz-euclide}
\usepackage{cases}
\usepackage{setspace}
\usepackage{multirow} 
\usepackage{hyperref}
\usepackage{cleveref}
\pgfplotsset{compat=1.15}

\DeclarePairedDelimiter\floor{\lfloor}{\rfloor}
\usetikzlibrary{decorations,calligraphy}

\makeatletter
\newcommand*{\rom}[1]{\expandafter\@slowromancap\romannumeral #1@}
\makeatother

\newcommand{\yding}{\textcolor{black}}
\newcommand{\granot}{\textcolor{black}}
\newcommand{\yifeng}{\textcolor{black}}
\newcommand{\discuss}{\textcolor{black}}
\newcommand{\td}{\textcolor{black}}

\newcommand{\bpr}{{\bf Proof.} \hspace{1 em}}
\newcommand{\epr}{\hfill $\blacksquare$ \medskip}   

\TheoremsNumberedThrough     

\ECRepeatTheorems

\EquationsNumberedThrough    

\begin{document}


\RUNTITLE{Tight Bounds for The Price of Fairness}

\TITLE{Tight Bounds for The Price of Fairness}

\ARTICLEAUTHORS{%
\AUTHOR{Yifeng Cao}
\AFF{Sauder School of Business, University of British Columbia, Vancouver, BC V6T 1Z2, \EMAIL{yifeng.cao@sauder.ubc.ca}} 

\AUTHOR{Yichuan Ding}
\AFF{Desautels Faculty of Management, McGill University,  Montreal, QC H3A 1G5, \EMAIL{daniel.ding@mcgill.ca}} 

\AUTHOR{Daniel Granot}
\AFF{Sauder School of Business, University of British Columbia, Vancouver, BC V6T 1Z2, \EMAIL{daniel.granot@sauder.ubc.ca}}
} 

\ABSTRACT{%
A central decision maker (CDM), who seeks an efficient allocation of scarce resources among a finite number of players, often has to incorporate fairness criteria to avoid unfair outcomes. Indeed, the Price of Fairness (POF), a term coined in the seminal work by \citet{priceoffairness}, refers to the efficiency loss due to the incorporation of fairness criteria into the allocation method. Quantifying the POF would help the CDM strike an appropriate balance between efficiency and fairness. In this paper we improve upon existing results in the literature, by providing tight bounds for the POF for the proportional fairness criterion for any $n$, when the maximum achievable utilities of the players are equal or are not equal. Further, while \citet{priceoffairness} have already derived a tight bound for the max-min fairness criterion for the case that all players have equal maximum achievable utilities, we also provide a tight bound in scenarios where these utilities are not equal. \yifeng{For both criteria, we characterize the conditions where the POF reaches its peak and provide the supremum bounds of our bounds over all maximum achievable utility vectors, which are shown to be asymptotically strictly smaller than the supremum of the \citet{priceoffairness} bounds. Finally, we investigate the sensitivity of our bounds and the bounds in \citet{priceoffairness} for the POF to the variability of the maximum achievable utilities.}
}%


\maketitle

\section{Introduction}

In this paper, we consider a problem facing a central decision maker (CDM) who needs to allocate scarce resources among finitely many players. There are many real-life applications wherein the CDM cannot use the most efficient allocation due to a fairness constraint. 
Instead, the CDM often has to allocate resources while recognizing the tradeoffs between efficiency and fairness \citep{zenios2000,alphabertsimas}.



An illustrative example of the efficiency-fairness trade-off is the allocation of deceased-donor kidneys. Kidney transplantation is a desired treatment for individuals with end-stage renal disease (ESRD). However, the reality is that the demand for kidneys far exceeds the available supply. 
To illustrate, in 2021, 43,617 new patients registered to the kidney waiting list, while only 25,490 transplants were performed that year\footnote{See https://optn.transplant.hrsa.gov/data/}. This significant discrepancy between the number of patients in need and the available transplants highlights the significance of allocating deceased-donor kidneys in an efficient and fair manner. Efficiency, in this context, refers to maximizing the overall utility, usually measured by the total quality-adjusted life-years (QALY). The problem of maximizing the sum of the utilities among all candidates on the waitlist has been extensively studied in the operations research literature; 
see, e.g., \citet{derman1972}, \citet{righter1989} and \citet{davidYechiali1995}. However, a policy solely focused on maximizing total QALYs may inadvertently disadvantage patient groups with lower expected QALYs gained from transplant. If the policy maker seeks to decrease such inequities, they must contend with the trade-off of potentially reducing the overall number of life-years saved, e.g., \citet{yichuan2021,su2006,bertsimas2013fairness}.





Similar trade-offs between efficiency and fairness naturally emerge in many other applications. For example, in air traffic flow management, the CDM needs to fairly allocate limited air space to different airlines \citep[\text{e.g.,}][]{barnhart2012,vossen2003,jiang2022, scarce2021}. In congested airports, the CDM has to fairly allocate runways and aprons to different airlines \citep{Fairbrother2020}. Other equitable resource allocation problems arise in finance, where
simultaneous trading with multiple accounts is carried out and the CDM's goal is to treat all clients fairly while maximizing collective interests \citep{oc2006}, water allocation, where a CDM attempts to balance unsatisfied demand of all consumers \citep{udias2012cost}, medical funds allocation, where equity across different regions and populations is a major concern \citep{earnshaw2007linear}, spectrum allocation in a communication system where a common frequency band is shared by multiple users \citep{ye2014competitive}, and kanban allocation in production systems, where both minimizing the rate of lost sales and balancing lost sales across product types are under consideration \citep{ryan2005allocating}. Finally, efficiency-fairness tradeoffs can arise in classical optimization problems such as location, vehicle routing, transportation and scheduling, as elaborated in the excellent survey papers by \citet{KARSU2015343} and \citet{luss1999}.

\yding{Naturally, quantifying the efficiency loss resulting from the incorporation of fairness concerns is essential for the CDM who strives to select the appropriate balance between efficiency and fairness when facing resource allocation challenges. \citet*{priceoffairness} \discuss{(BFT 2011)} introduced the notion of the price of fairness (POF), which is defined as the relative efficiency loss, compared to the utility-maximization solution (i.e., utilitarian solution), of a fair solution. They considered two fairness notions that have been widely used in the literature: Proportional Fairness (PF) and Max-Min Fairness (MMF), and quantified the POF for both. That is, they provided upper bounds of the POF for both the PF and the MMF cases, which we refer to as the BFT bounds.}




\discuss{When all players have equal maximum achievable utilities, the BFT bound for the PF criterion, when the number of players, $n$, is a square of an integer, and the BFT bound for the MMF criterion for an arbitrary number of players were verified to be tight for a specific resource allocation problem. By contrast, we theoretically derive tight upper bounds for the POF both for PF and MMF. Indeed, our approach is shown to reveal the source for the gap between our bound and the BFT bound for the POF for PF when the number of players, $n$, is not a square of an integer.}

For PF, we prove that the improvement of our bound over the BFT bound achieves a local maximum when $n$ is a product of two consecutive integers, e.g., $n = 2, 6, 12$, etc. Both bounds increase as a function of $n$, and as noted by \citet*{priceoffairness}, for a small number of players, the price of proportional fairness is small. Indeed, for example, for $n=2$, which corresponds to Nash original two-player bargaining problem, the price of fairness according to our bound is at most $6.7\%$. By comparison, we note that according to the BFT bound, the price of proportional fairness for $n=2$ is at most $8.6\%$. We further show that for a fixed number of players, both for PF and MMF, our bounds and the BFT bounds increase when the variability of the maximum achievable utilities increases. However, the BFT bounds increase at a faster rate.

\yding{The assumption that players have equal maximum achievable utilities is appropriate in settings in which the utility functions merely represent the players’ preferences among the various resource allocation options.  These preferences are invariant to positive affine transformations of the corresponding utility functions, and thus, permit the assumption that in such settings, the maximum achievable utilities of all players are equal. However, there are many applications in which the utility functions have some intrinsic values, which facilitates interpersonal comparisons of utilities. For instance, in the context of organ transplantation, as previously discussed, the utility of each patient is commonly quantified by QALYs \citep{su2005,zenios2000}. In air traffic flow management, the disutility of an airline is measured by the total time delays \citep{barnhart2012}, and in most business applications, a firm’s utility can be measured by a monetary value. In all such examples, the utilities of players cannot be normalized and the assumption of equal maximum achievable utilities cannot be made without loss of generality.}

\yding{When the maximum achievable utilities are unequal, the BFT bounds for the price of PF and MMF are, in general, not tight. We note, though, that for the price of MMF, the BFT bound is attained in an example of bandwidth allocation in a communication network, where the maximum achievable utilities follow a special pattern. By contrast, our bounds for both the price of PF and MMF are tight in the sense that for any collection of maximum achievable utilities by the players, we prove the existence of a utility set $U$ for which our bound is attained.} \granot{The improvement of our bounds over the BFT bounds for the unequal case are demonstrated in our study of the sensitivity of the bounds to the variability of the maximum achievable utilities. Both bounds are shown to increase with such variability, but for a wide range of the maximum achievable utility vectors, our bounds are shown to be significantly smaller than the BFT bounds both for the PF and MMF criteria. Additionally, we have studied the supremum of our bounds and the BFT bounds over all possible maximum achievable utilities, and have proven that our bounds are asymptotically strictly smaller than the BFT bounds.} \yding{Specifically, for a large $n$, we prove that the BFT bound for the POF is of the order of $1-O(1/n^{1.25})$ for the PF criterion and $1-O(1/n^{2})$ for the MMF criterion, while our derived tight bounds are of the order of $1-O(1/n)$ for both fairness criteria. See Table \ref{tab:compare} for a summary of the comparison between our bounds and the BFT bounds.}

\begin{table}[ht]
\centering
\parbox{0.8\textwidth}{\centering BFT Bounds}
\begin{minipage}{1\textwidth}
\centering
\vspace{0.3cm}
\begin{tabular}{c c cc}
    \toprule
    \multirow{2}{*}{BFT Bounds} & \makecell{Equal Maximum \\ Achievable Utilities} & \multicolumn{2}{c}{\makecell{Unequal Maximum \\ Achievable Utilities}} \\ 
    \cmidrule(lr){2-4}
                                & Tight?                                & \multicolumn{1}{c}{Tight?}     & Asymptotic Order      \\ 
    \midrule
    Proportional Fairness       & Yes only when $\sqrt{n}$ is integral     & \multicolumn{1}{c}{No}           & $1-O(1/n^{1.25})$     \\ 
    Max-Min Fairness            & Yes                                     & \multicolumn{1}{c}{\hspace{0.15cm}No$^*$}           & $1-O(1/n^2)$          \\ 
    \bottomrule
\end{tabular}
\vspace{0.5cm}
    \end{minipage}

    \parbox{0.8\textwidth}{\centering Our Bounds}
    \vspace{0.3cm} 

    \begin{minipage}{0.8\textwidth}
        \centering
        \begin{tabular}{c c cc}
    \toprule
    \multirow{2}{*}{Our Bounds} & \makecell{Equal Maximum \\ Achievable Utilities} & \multicolumn{2}{c}{\makecell{Unequal Maximum \\ Achievable Utilities}} \\ 
    \cmidrule(lr){2-4}
                                & Tight?                                & \multicolumn{1}{c}{Tight?}     & Asymptotic Order      \\ 
    \midrule
    Proportional Fairness       & Yes      & \multicolumn{1}{c}{Yes}           & $1-O(1/n)$     \\ 
    Max-Min Fairness            & Yes$^{**}$                                       & \multicolumn{1}{c}{\hspace{0.15cm}Yes$^*$}           & $1-O(1/n)$          \\ 
    \bottomrule
\end{tabular}
        \vspace{0.3cm}
    \end{minipage}
    \vspace{0.5cm}
    \caption{Summary of BFT bounds and our bounds \\
    \raggedright $^*$ ``Yes" or ``No" refers to whether the bound is tight for all maximum achievable utility vectors.\\
    $^{**}$ Our tight bound for the max-min unequal case reduces to the BFT tight bound for the equal case.}\label{tab:compare}
\end{table}

Our paper makes the following contributions:
\begin{enumerate}

\item Our paper introduces a fractional programming whose optimal solution delivers tight bounds for the POF, while a relaxation of this programming yields the BFT bounds, effectively delineating the source of the gap in the BFT's estimation. By characterizing the optimal solution to this fractional programming, we establish the first tight bounds for the POF under both PF and MMF criteria in general scenarios involving an arbitrary number of players with potentially unequal maximum achievable utilities. Our results complement the work of BFT (2011), who demonstrated the tightness of their bounds when players' maximum utilities are all equal or follow special structures. Under generalized conditions of maximum utilities, our bounds asymptotically improve over the BFT bounds for both PF and MMF criteria.

\item \yding{The tight bounds for the POF we derived provide qualitative insights into how the POF changes with problem's parameters, which allows a CDM to evaluate and compare the POF across different application scenarios. We show that when players have different maximum achievable utilities, the POF is relatively large when the distribution of the maximum achievable utilities exhibit a large variance. In fact, we show that the POF attains its peak as when one player has a large maximum achievable utility while all others have a very small maximum achievable utilities. Our results strengthen BFT (2011) conclusion that the POF is likely to be small for a small number of players. Indeed, for the two-player case we prove that the POF for PF with equal maximum achievable utilities is at most 6.7\%.}

\item A growing body of research, exemplified by studies such as \citep{hasankhani2022proportionally,ma2023fairness,Agnetis2019}, has been concerned with calculating the POF for special problem instances. Our tight bounds for the POF in general settings could serve not only as a cap but also as a benchmark for the POF estimates in those papers. This underscores the broader applicability and relevance of our findings within the literature on the POF across various applications.

\end{enumerate}


\newblock

\section{Fairness}


\subsection{Related Literature}


There are two stream of literature closely related to our study of the POF. The first stream is concerned with the various fairness criteria. It is generally understood that different equity/fairness criteria may be required in different contexts/applications \citep[see, e.g.,][]{sen1997economic,young1994}, and that no universal criterion of fairness can be applied in all settings. Nevertheless, there are several widely accepted criteria for fairness.
  


A simple and common fairness criterion, based on the Rawlsian principle \citep{rawls}, is the MMF criterion. According to this criterion we seek a solution which is the lexicographically largest vector, whose elements are either the allocations to the different players or the performance function values corresponding to the different activities, which are arranged in a non-increasing order.

Some studies handle fairness using an inequality index, which is a function that maps a resource allocation instance to a scalar value representing the level of inequality. For example, \citet{kozanidis2009solving} uses the difference between the upper bound and the lower bound of outcomes as the inequality index, and \citet{TURKCAN2011780} uses variance to measure fairness.

Another approach to achieve fairness optimizes an objective function which is some aggregation of the allocations to the different players. Indeed, proportional fairness (PF), which we study in this paper, is achieved by maximizing the sum of the logarithms of the utility outcomes corresponding to the allocations to the players. It is a generalization of the Nash bargaining solution, see, e.g., \citet{nashbargaining1950}, which has an axiomatic basis as we further elaborate in the sequel. Finally for a classification of the extensive literature concerned with fairness in terms of the fairness criteria they employ, such the Rawlsian principle, its lexicographic extension or an aggregation function, see the survey by \citet{KARSU2015343}.

The second stream of literature related to our study of the POF focuses on the assessment of  the POF in particular contexts. Recent papers in this stream include, for example, \citet{hasankhani2022proportionally}, who studied the US heart transplant system and, using a fluid model, have  quantified the POF for both the PF and MMF fairness criteria. They have shown, for example, that consistent with known theoretical results, the price of PF is smaller than that of MMF. \citet{liu2022fairness} calculated the POF for a facility location problem, considering disutility as an aggregate of transit time and queueing delays. \citet{Agnetis2019} were the first to investigate the POF in scheduling problems, providing tight bounds within their defined context. \citet{zhang2020price} extended this exploration to a two-agent scheduling game variant, specifically examining scenarios where one of the two players has exactly two jobs. \discuss{\citet{donahue2020fairness} provided provable upper bounds on the gap between the maximum possible utilization of the resource and the maximum utilization achievable under a natural fairness condition in resource allocation problems with uncertain demand, showing that for certain distributions, this gap exists but is bounded by a constant factor.} \citet{elzayn2019fair} studied the POF using the Philadelphia Crime Incidents dataset and fully characterized a worst-cast variant of the POF. In the context of indivisible goods, \citet{barman2020optimal} studied the POF for two well-established fairness notions, envy-freeness up to one good (EF1) and approximate maximin share (MMS), and \citet{feldman2024optimal} studied the tradeoffs between fairness and efficiency, where envy-freeness up to any item (EFX) is used as a fairness criterion and the Nash welfare is used as an efficiency criterion. \citet{dickerson2014price} showed that the POF in the standard kidney exchange model is small and empirically explored the POF under two natural definitions of fairness using both real and simulated data. \citet{nicosia2017price} characterized the POF for a specific discrete allocation problem with two agents for multiple fairness notions.

As noted above, our tight bounds for the POF for the PF and MMF criteria, and our related investigation about the sensitivity of the POF to the distribution of the maximum achievable utilities could be helpful in future investigations into the POF in different contexts and application areas.

\subsection{Fairness Notions}

We quantify the price of fairness for two fairness criteria - PF and MMF.

\subsubsection{Proportional Fairness}

Under PF, the preferred allocation, $u^{PF}(U)=(u^{PF}_1(U),u^{PF}_2(U), $ $\dots,u^{PF}_n(U))$, from a utility set $U$, is such that for any other feasible allocation of utilities $u$, the aggregate of proportional changes is zero or negative, i.e.,

\begin{equation}
    \sum_{i=1}^n \frac{u_i-u^{PF}_i(U)}{u^{PF}_i(U)} \leq 0,\qquad \text{for any }\vect{u} \in U.
\label{2.1}
\end{equation}

For a convex utility set $U$, $u^{PF}(U)$ is a generalization on the Nash bargaining solution for $n$ players, and can be obtained by maximizing the product of the players' utilities, i.e., $$u^{PF}(U):=\argmax_{(u_1,u_2,\dots,u_n)\in U} \ \prod_{i=1}^n u_i.$$

Equivalently, $u^{PF}(U)$ can be derived by maximizing the logarithmic transformation of the above objective function, 
i.e.,
\begin{equation}
    u^{PF}(U):=\argmax_{(u_1,u_2,\dots,u_n)\in U} \ \sum_{i=1}^n \log u_i.
\label{2.2}
\end{equation}

Note that \eqref{2.1} is the necessary and sufficient condition for the optimality of $u^{PF}(U)$ for the optimization problem associated with \eqref{2.2}. As mentioned above, such a solution is a generalization of the Nash bargaining solution. It satisfies four axioms: Pareto optimality, symmetry, affine invariance and independence of irrelevant alternatives. Pareto optimality avoids a waste of resources. Symmetry guarantees that the solution does not distinguish between the players if the model does not distinguish between them. By the third axiom, the allocation is invariant to scaling of utilities, and thus would not be affected if different measurement units of utility are used by players. The fourth axiom implies that the preferred allocation remains unchanged when inferior allocations are removed, see, e.g., \citet{nashbargaining1950} and \citet{Roth1979}.

The first application of PF was in the telecommunications field \citep{mazu1991}, and the term PF was first coined by \citet{kelly1998rate}.


\subsubsection{Max-min Fairness}

MMF is motivated by the Kalai-Smorodinsky (KS) solution for two-person bargaining problems, axiomatically characterized by \citet{kalai1975other}. The KS solution is the unique solution satisfying the axioms of Pareto optimality, symmetry, invariance with respect to affine transformations of utility, and monotonicity. According to the KS solution, the players obtain the largest possible equal fraction of their respective maximum achievable utilities. However, \citet{roth1979impossibility} has shown that in $n$-person bargaining problems, such a solution may not satisfy Pareto optimality and that there does not exist a solution satisfying all the above five axioms. \citet{imai1983individual} has modified the KS solution to $n$-person bargaining problems by proposing a weaker set of axioms. Namely, he proposes a solution which satisfies Pareto optimality, symmetry and invariance under linear utility transformation. But, instead of the monotonicity axiom, originally proposed by \citet{kalai1975other}, he requires the axioms of independence of irrelevant alternatives other than ideal point and individual monotonicity.


The axiom of independence of irrelevant alternatives other than the ideal point is a less stringent requirement compared to the axiom of independence of irrelevant alternatives, which is satisfied by the Nash bargaining solution. It stipulates that the solution remains unchanged if the alternatives
that have been removed do not alter the maximum achievable utilities by all players. The other axiom, individual monotonicity, requires that if the utility set expands in a manner which keeps the projection of the utility set onto the $N\backslash \{i\}$-dimensional space unchanged, then the utility of the $i$-th player must increase.

\citet{imai1983individual} has proven that when all players have equal maximum achievable utility, his solution coincides with the MMF solution (referred to as lex-max-min in his paper), which, according to BFT (2011), maximizes the ratios of players’ utilities to their maximum achievable utilities in a lexicographical manner. That is, the CDM first maximizes the minimum ratio of the players’ utilities to their respective maximum achievable utilities. Subsequently, the CDM maximizes the second smallest ratio, the third smallest ratio, and so on. Thus, for two-person problems, the MMF solution coincides with the KS solution.


\subsection{The Price of Fairness}


The utilitarian solution, which maximizes the sum of the utilities to all the players, is viewed as a measure for system efficiency. Naturally, implementing a fair solution, instead of a utilitarian solution, will reduce system efficiency. To quantify the relative loss incurred by adopting a fair solution over the utilitarian solution, we employ the concept of the POF as introduced by BFT (2011).


Formally, consider a resource allocation problem with $n$ players, and let $U:\subseteq \mathbf{R}^n$ denote the set of all possible utility vectors of the players. Let $u_i$ denote the utility of player $i$. We assume $U$ to be convex and compact, which is a common assumption in the literature (see BFT 2011). Since the PF, MMF, and utilitarian solutions are all Pareto optimal, we can assume, without loss of generality, that $U$ is monotone. That is: 

\begin{definition}
A set $U$ is \textbf{monotone} if for any $\vect{u} \in U$ and $0 \leq \vect{v} \leq \vect{u}$, $\vect{v} \in U$.
\end{definition}

\subsubsection{Utilitarian Solution}
A utilitarian solution is an allocation based on classical utilitarianism, which maximizes the sum of utilities of all players. Thus, the utilitarian solution, denoted by $\vect{u}^*(U):=(u_1^*(U),\dots,u_n^*(U))$, is an optimal solution to the following problem:
\begin{equation}
\begin{aligned}
    &\max& \ &\sum_{i=1}^n u_i,\\
    &\text{ s.t.} & &(u_1,u_2,\dots,u_n)\in U.
    \label{utaplm}
\end{aligned}
\end{equation}

The sum of utilities to the players is often used as a measure of system efficiency. Thus, if the social planner aims to maximize efficiency without fairness considerations, the utilitarian solution should be used.

\subsubsection{Fairness Solution}

As previously mentioned, in many applications it would be inappropriate to implement solutions based exclusively on system efficiency. In these circumstances, the central decision maker will choose an allocation scheme based on some fairness criteria, trying to balance both efficiency and fairness.

We consider two notions in this paper, PF and MMF, and denote the allocation scheme with a set function $\vect{u}^{fair}:2^{\mathbb{R}^n_+} \rightarrow \mathbb{R}^n_+$, where the superscript, fair, could be $PF$ or $MMF$ to indicate whether we use PF or MMF.


\subsubsection{The Price of Fairness}

We use the notion of the price of fairness, introduced by BFT (2011) and denoted $POF(U;fair)$, to measure the relative loss resulting from an implementation of a fair solution rather than the utilitarian solution, i.e.,
\begin{equation}\label{eq:POFdef}
POF(U;fair)=\frac{\sum_{i=1}^n u^*_i(U)-\sum_{i=1}^n u^{fair}_i(U)}{\sum_{i=1}^n u^*_i(U)}=1-\frac{\sum_{i=1}^n u^{fair}_i(U)}{\sum_{i=1}^n u^*_i(U)},
\end{equation}

\noindent where $u^*_i$, resp., $u^{fair}_i, i = 1, \dots, n$, denotes the utilitarian, respectively, the fair solution which is being used.

BFT (2011) have derived upper bounds for the $POF(U;fair)$ for the cases where the fair solution is either the PF solution or the MMF solution, and for the cases when the maximum achievable utilities of the players are either equal or not equal. As will be clarified in the sequel, we improve upon the bounds derived by BFT (2011), by providing tight bounds for the price of proportional fairness when the maximum achievable utilities by the players are either equal or not necessarily equal, and for the price of MMF when the maximum achievable utilities by the players are not necessarily equal. 



\newblock

\section{Upper Bounds for The Price of Proportional Fairness}

\subsection{Equal Maximum Achievable Utilities}


We begin by considering the case where all players have equal maximum achievable utilities in the utility set $U$. For simplicity and without loss of generality, we set this maximum achievable utility to one, and denote by $N:=\{1,2,\dots,n\}$, the set of players. Consequently, $\max\{u_i|\vect{u} \in U\}=1$ for $i\in N$.

To derive an upper bound for the price of fairness (POF) in $U$, we employ a utility set $U'$, $U' \supseteq U$, such that the PF solutions in $U$ and $U'$ coincide. Consequently, since $U \subseteq U'$, the POF with respect to $U$ is bounded by the POF with respect to $U'$. Formally,  

\begin{proposition}\label{prop:1}
Suppose $U \subseteq \{\vect{u}:=(u_1,u_2,\dots,u_n)\, |\, 0 \leq u_i \leq 1, i \in N \}$ is a convex and compact utility set, and $\max\{u_i\, |\,\vect{u} \in U\}=1$ for $i\in N$. Then there exist $c_i \in [1/n,1],\ i\in N$, such that $U \subseteq U'=\{\vect{u}\, |\,\sum_{i=1}^n c_i u_i \leq 1, 0 \leq u_i \leq 1, i\in N\}$, and $\max \{ \sum_{i=1}^n \log{u_i} \, |\, \vect{u}\in U \} = \max \{ \sum_{i=1}^n \log{u_i} \, |\, \vect{u}\in U' \}$. Consequently, $POF(U;PF) \le POF(U';PF)$.
\end{proposition}

The proof of Proposotion \ref{prop:1} is in Appendix \ref{sec:proofprop14}. By Proposition \ref{prop:1}, we can derive an upper bound for the price of proportional fairness (PF) with respect to $U$, by finding this bound with respect to $U'$, $U'=\{(u_1,u_2,\dots,u_n) \, |\, \sum_{i=1}^n c_i u_i \leq 1,~0 \le u_i \le 1,~i\in N\}$. To that end, we need to find the utilitarian solution $u^*(U')$, which
is an optimal solution to the following Problem \eqref{eq:utilitarianeq}:
\begin{subequations}\label{eq:utilitarianeq}
\begin{alignat}{2}
&\max \quad  &&\sum_{i=1}^n u_i,\\
&\ \text{s.t.} \qquad  &&0 \leq u_i \leq 1, \qquad \forall i\in N, \\
& &&\sum_{i=1}^n c_i u_i \leq 1.
\end{alignat}
\end{subequations}

Without loss of generality, we assume that
\begin{equation}
    c_1 \leq c_2 \leq \cdots \leq c_n.
    \label{incasp}
\end{equation}

Following BFT (2011), we define
\begin{equation}
    l(c):=\max \{j \ |\  \sum_{i=1}^j c_i \leq 1\}, \qquad \delta(c):=\frac{1-\sum_{i=1}^{l(c)} c_i}{ c_{l(c)+1}} \in [0,1).
    \label{lddef}
\end{equation}
Then, since Problem \eqref{eq:utilitarianeq} is a linear relaxation of the 0-1 knapsack problem, the optimal solution, $u^*$, to Problem \eqref{eq:utilitarianeq} is $u_1^*=u_2^*=\dots=u_{l(c)}^*=1, u_{l(c)+1}^*=\delta(c), u_{l(c)+2}^*=u_{l(c)+3}^*=\dots=u_n^*=0$, and the optimal value of Problem \eqref{eq:utilitarianeq} is $l(c)+\delta(c)$.

On the other hand, the PF solution can be obtained by solving the following Problem \eqref{5.3}:
\begin{subequations}
\begin{alignat}{2}
&\max \quad  &&\sum_{i=1}^n log(u_i),\\
&\text{ s.t.} \qquad  &&0 \leq u_i \leq 1, \qquad \forall i \in N, \\
& &&\sum_{i=1}^n c_i u_i \leq 1.
\end{alignat}
\label{5.3}
\end{subequations}

Problem \eqref{5.3} has a unique optimal solution, $\vect{u}^{PF}$, which has the following explicit form:

\newblock

\begin{proposition}\label{prop2}
$\vect{u}^{PF}(U'):=(1/nc_1,1/nc_2,\dots,1/nc_n)$ is the unique optimal solution to Problem \eqref{5.3}.
\end{proposition}

\discuss{The proof of Proposition \ref{prop2} is in Appendix \ref{sec:proofprop14}.}



The utilitarian solution, $\vect{u}^*$, and the PF solution, $\vect{u}^{PF}$, we have derived with respect to $U'$, can be used to find an upper bound for the price of proportional fairness by minimizing $f_0(\vect{c},l):=1-POF(U';PF)=\sum_{i=1}^n u^{PF}_i(U') / \sum_{i=1}^n u_i^*(U')$, which is equivalent to Problem \eqref{op44}:
\begin{subequations}
\begin{alignat}{2}
&\min_{\vect{c},l}   && f_0(\vect{c},l):= \frac{\sum_{i=1}^n \frac{1}{nc_i}}{l+\frac{1-\sum_{i=1}^l c_i}{c_{l+1}}},\label{op44obj}\\
&\text{s.t.} \qquad  &&\frac{1}{n} \leq c_1 \leq c_2 \leq \cdots \leq c_n \leq 1, \label{d1} \\
& &&\sum_{i=1}^{l} c_i \leq 1, \label{d2}\\
& &&\sum_{i=1}^{l+1} c_i > 1, \label{d3}
\end{alignat}
\label{op44}
\end{subequations}
  \yding{where the numerator of the objective function can be interpreted as system efficiency, i.e., the sum of players' utilities, under the proportional fairness solution}, the denominator of the objective function stands for system efficiency under the utilitarian solution, and \eqref{d2}, \eqref{d3} restrict $l$ to satisfy the definition of $l(c)$, as given in the Expression \eqref{lddef}.

\yding{The optimal value of Problem \eqref{op44} thus gives a tight bound for the price of proportional fairness. Later, we will derive an explicit expression for the optimal value and show that it strictly improves upon the upper bound, $1-(2\sqrt{n} - 1) / n$, derived by BFT (2011). Before proceeding, we investigate the source of the gap between our bound and the BFT bound. To that end, we introduce a fractional programming formulation \eqref{newop} \yifeng{below}, which is an alternative formulation for Problem \eqref{op44}, and whose optimal value yields the tight bound for the price of proportional fairness. We then prove, in Proposition \ref{propbft}, that a relaxation of Problem \eqref{newop}, derived after the removal of Constraints \eqref{11c} therefrom, yields the BFT bound. Thus, the difference between the two bounds can be entirely attributed to the removal of Constraints \eqref{11c} from Problem \eqref{newop}.}


\yding{In Problem \eqref{newop}, we introduce new variables, $y_i$'s, $y_i \in [0,1]$, to replace the superscript variable $l$. Each such variable $y_i$ represents the portion of good $i$ that is included in the knapsack. We also introduce new variables, $d_{2i-1},d_{2i}$ such that $d_{2i-1}=d_{2i}=c_i$. Problem $\eqref{newop}$ is defined as follows:}

\begin{subequations}
\begin{alignat}{2}
&\min_{\vect{d},\vect{y}}   &&f_1(\vect{d},\vect{y}):=\frac{\frac{1}{n}\sum_{i=1}^n (\frac{y_i}{d_{2i-1}}+\frac{1-y_i}{d_{2i}})}{\sum_{i=1}^n y_i},\label{newopobj}\\
&\text{s.t.} \qquad  &&\frac{1}{n} \leq d_1,d_2 \leq d_3,d_4 \leq \cdots \leq d_{2n-1},d_{2n} \leq 1, \label{11b} \\
& && d_{2i-1}=d_{2i}, \qquad i=1,2,\dots,n, \label{11c}\\
& &&\sum_{i=1}^{n} y_i d_{2i-1} = 1, \label{11d}\\
& &&y_i \geq y_{i+1}, \qquad i=1,2,\dots,n-1,\label{11e}\\
& &&0\leq y_i \leq 1,\label{11f}\\
& && \sum_{i=1}^n (\sqrt{y_i}-y_i)=\prod_{i=1}^n (\sqrt{y_i}-y_i+1)-1 \label{11g}.
\end{alignat}
\label{newop}
\end{subequations}

\yifeng{Note that Constraints \eqref{11b} correspond to Constraints \eqref{d1}, Constraints \eqref{d2} and \eqref{d3} correspond to Constraints \eqref{11d}, and Constraints \eqref{11e}-\eqref{11g} ensure that $y_i$ is decreasing in $i$, and that there is at most one variable $y_i$ which is strictly between 0 and 1. Thus, the $y_i$ variables act as a substitute for $l$. Formally, the following proposition, \discuss{whose proof is in Appendix \ref{sec:proofprop14}}, states the equivalence of the two formulations.}

\begin{proposition}\label{prop:3}
    Problems \eqref{op44} and \eqref{newop} are equivalent.   
\end{proposition}

\yifeng{In Proposition \ref{propbft} below, we prove that the optimal value of Problem \eqref{newop} \discuss{from which Constraints \eqref{11c} were removed} yields the BFT bound for the price of proportional fairness. For brevity, let $k$ denote the largest integer that is not larger than the square root of $n$, i.e., $k=\lfloor\sqrt{n}\rfloor$.}

\begin{proposition}\label{propbft}
    The optimal solution to Problem \eqref{newop} \discuss{from which Constraints \eqref{11c} have been removed} is given by $$\overline{y}_1=\dots=\overline{y}_k=1, \ \overline{y}_{k+1}=\sqrt{n}-k, \ \overline{y}_{k+2}=\dots=\overline{y}_n=0,$$
$$\overline{d}_{2i-1}=\frac{1}{\sqrt{n}} \text{ for }i\in \{i:y_i \neq 0\},\ \overline{d}_{2i}=1  \text{ for }i\in \{i:y_i \neq 1\},$$
\begin{center}
    all other variables $\overline{d}_i$ can assume any feasible value.
\end{center}
  with a corresponding optimal value, $(2\sqrt{n}-1)/n$, which yields the BFT bound for the price of proportional fairness, $1-(2\sqrt{n}-1)/n$.
\end{proposition}

\bpr
By Cauchy–Schwarz inequality, we have
\begin{equation}\label{caus}
    (\sum_{i=1}^n \frac{y_i}{d_{2i-1}})(\sum_{i=1}^{n} y_i d_{2i-1})\geq (\sum_{i=1}^n y_i)^2.
\end{equation}

By \eqref{11d}, $\sum_{i=1}^{n} y_i d_{2i-1}=1$, and thus, \eqref{caus} reduces to
\begin{equation}\label{p41}
    \sum_{i=1}^n \frac{y_i}{d_{2i-1}} \geq (\sum_{i=1}^n y_i)^2,
\end{equation}
 \yifeng{where equality holds when all $d_{2i-1}$ variables are equal for $i\in \{i:y_i \neq 0\}$.}

In addition,
\begin{equation}\label{p42}
    \sum_{i=1}^n \frac{1-y_i}{d_{2i}}\geq \sum_{i=1}^n (1-y_i) = n - \sum_{i=1}^n y_i,
\end{equation}\yifeng{where equality holds when $d_{2i}=1$ for $i\in \{i:y_i \neq 1\}$.}

Thus by \eqref{p41} and \eqref{p42},
\begin{equation*}
\begin{aligned}
    f_1(\vect{d},\vect{y})=\frac{\frac{1}{n}(\sum_{i=1}^n \frac{y_i}{d_{2i-1}}+\sum_{i=1}^n \frac{1-y_i}{d_{2i}})}{\sum_{i=1}^n y_i} &\geq \frac{(\sum_{i=1}^n y_i)^2+n - \sum_{i=1}^n y_i}{n\sum_{i=1}^n y_i} \\&= \frac{1}{n}( \sum_{i=1}^n y_i+\frac{n}{\sum_{i=1}^n y_i}-1)\geq \frac{2\sqrt{n}-1}{n},
\end{aligned}
\end{equation*}
\yifeng{where the first inequality \discuss{is tight} when the equality conditions in \eqref{p41} and \eqref{p42} are satisfied, and the second inequality \discuss{is tight} when $\sum_{i=1}^n y_i=\sqrt{n}$.}

For convenience, let $k=\lfloor \sqrt{n}\rfloor$ and define $S^E_1=\{i\,|\,y_i=1\}, S^E_2=\{i\,|\,0<y_i<1\}, S^E_3=\{i\,|\,y_i=0\}$, and $d_{2i-1}=x$ for $i\in S^E_1\cup S^E_2$. In addition, \discuss{we have $d_{2i}=1$ for $i\in S^E_2\cup S^E_3$ by the equality condition of \eqref{p42}}.

From \eqref{11d},
$$\sum_{i=1}^{n} y_i d_{2i-1} = \sum_{i \in S^E_1\cup S^E_2} y_i d_{2i-1} = x \cdot \sum_{i \in S^E_1\cup S^E_2} y_i=x \cdot \sum_{i=1}^n y_i=x\sqrt{n}=1 \Rightarrow x=\frac{1}{\sqrt{n}}.$$

\td{Constraint \eqref{11g} ensures that $S^E_2$ is either a singleton set or an empty set. Given that $y_i$ is non-increasing with $i$ and $\sum_{i=1}^n y_i =\sqrt{n}$, the only possible scenario for the optimal solution is
$$\overline{y}_1=\dots=\overline{y}_k=1, \ \overline{y}_{k+1}=\sqrt{n}-k, \ \overline{y}_{k+2}=\dots=\overline{y}_n=0,$$
$$\overline{d}_{2i-1}=\frac{1}{\sqrt{n}} \text{ for }i\in \{i:y_i \neq 0\},\ \overline{d}_{2i}=1  \text{ for }i\in \{i:y_i \neq 1\},$$
\begin{center}
    all other variables $\overline{d}_i$ can assume any feasible value.
\end{center}
}


\yifeng{Then the optimal value is 
$$f_1(\overline{\vect{d}},\overline{\vect{y}})=\frac{2\sqrt{n}-1}{n}.$$}
\epr  





In the absence of Constraints \eqref{11c}, both the first term and the second term in the numerator of the objective function \eqref{newopobj} can reach their respective minima simultaneously even when $y_i$ is neither 0 or 1, as shown in \eqref{p41} and \eqref{p42}. \yifeng{Specifically, for $i$ such that $y_i$ is strictly between 0 and 1, $d_{2i-1}$ should be equal to $1/\sqrt{n}$ to allow the first term in the numerator of \eqref{newopobj} reach its minimum with fixed $\sum_{i=1}^n y_i$, as shown in \eqref{p41}, while $d_{2i}$ should be equal to 1 to allow the second term in \eqref{newopobj} reach its minimum with fixed $\sum_{i=1}^n y_i$, as shown in \eqref{p42}. However, in the presence of Constraints \eqref{11c}, $d_{2i-1}$ is constrained to be equal to $d_{2i}$, for each $i$, preventing the first and second terms in the numerator of $\eqref{newopobj}$ from being minimized simultaneously. This discrepancy, stemming from the possible existence of Constraints \eqref{11c}, creates the difference between our bound and the BFT bound for the price of PF.} \td{However, if $\sqrt{n}$ is an integer, then $\overline{y}_{k+1}=0$, and thus, one optimal solution could be $\overline{d}_{2i-1}=\overline{d}_{2i}=1/\sqrt{n}$ for $i\in [1,k]$, and $\overline{d}_{2i-1}=\overline{d}_{2i}=1$ for $i\in [k+1,n]$, which already satisfies Constraints \eqref{11c}. Therefore, in this case, Problem \eqref{newop} and Problem \eqref{newop} without Constraints \eqref{11c} have the same optimal value.}

\yifeng{Next, we proceed to solve Problem \eqref{newop}. Recall that $k=\lfloor \sqrt{n} \rfloor$. Claims \ref{psc1} and \ref{psc2} below assert that the optimal values of the $y_i$ variables, $i \neq k+1$, coincide with the optimal values of the $\overline{y}_i$ variables in the relaxation of Problem \eqref{newop}. Note that for $i\neq k+1$, \discuss{the $d_{2i-1}=d_{2i}$} constraints are easily satisfied since either $d_{2i-1}$ or $d_{2i}$ can assume any feasible value. However, $d_{2k+1}$ and $d_{2k+2}$ do not satisfy \eqref{11c}. We will show that the difference between the optimal solutions of Problem \eqref{newop} and its relaxation stems from the values of $y_{k+1}$ and $d_i,i\leq 2k+2$. The proofs of Claims \ref{psc1} and \ref{psc2} are provided in Appendix \ref{sec:proofclaims16}}.

Let $(\vect{d}^*,\vect{y}^*)$ denote an optimal solution to Problem \eqref{newop}.
\begin{claim}
    $\sum_{i=1}^n y^*_i \in [k,k+1]$. \label{psc1}
\end{claim}


\begin{claim}
    $y^*_1=y^*_2=\dots=y^*_k=1, y^*_{k+2}=\dots=y^*_n=0.$  \label{psc2}
\end{claim}






\yifeng{Next we will derive an explicit expression for the optimal value of Problem \eqref{newop}. To that end, we will keep the constraints $d_{2i-1}=d_{2i}$ for $i=1,\dots,n$, and for simplicity, let $c_i=d_{2i-1}=d_{2i}$. By Claims \ref{psc1} and \ref{psc2}, we can replace $y_{k+1}$ with $(1-\sum_{i=1}^k c_i)/c_{k+1}$, leading to Problem \eqref{newop2}:}

\begin{subequations}
\begin{alignat}{2}
&\min_{\vect{c}}   &&g_1(\vect{c}):=\frac{\frac{1}{n}\sum_{i=1}^n \frac{1}{c_i}}{k +\frac{1-\sum_{i=1}^k c_i}{c_{k+1}}},\label{newop2obj}\\
&\text{s.t.} \qquad  &&\frac{1}{n} \leq c_1 \leq c_2 \leq \cdots \leq c_{n} \leq 1, \label{12a} \\
& &&\sum_{i=1}^{k} c_{i} \leq 1, \label{12c}\\
& &&\sum_{i=1}^{k+1} c_{i} \geq 1. \label{12d}
\end{alignat}
\label{newop2}
\end{subequations}

A key step towards the derivation of the optimal value is the characterization of the optimal solution, $c_i^*, i=1,\dots,n$, to Problem \eqref{newop2}, as stated in Proposition \ref{prop:cstar}.


\begin{proposition}\label{prop:cstar}
(1) When $n=2$, $c_1^*=\sqrt{3}-1, c_2^*=1$.\\ 
(2) When $n \geq 3$ and $ n \leq k(k+1), c_1^*=\dots=c_{k}^*=1/k, c_{k+1}^*=\dots=c_n^*=1$.\\
(3) When $n \geq 3 $ and $ n > k(k+1), c_1^*=\dots=c_{k+1}^*=1/(k+1), c_{k+2}^*=\dots=c_n^*=1$.
\end{proposition}

\bpr

Claims \ref{c1} - \ref{c6}, whose proofs are provided in Appendix \ref{sec:proofclaims16}, are used in the proof of Proposition \ref{prop:cstar}.

\begin{claim}
$c^*_i=1$ for $i=k+2,\dots,n$.\label{c1}
\end{claim}

\begin{claim}
$c_1^*=c_2^*=\dots=c_k^*$. \label{c2}
\end{claim}

Now, for convenience, \discuss{let $x:=c_1(=c_2=\dots=c_k)$, 
and let $h_1(x,c_{k+1})$ denote the objective function \eqref{newop2obj} in terms of variables, $x$ and $c_{k+1}$, after removing the constant $1/n$.} Then, 
\begin{equation}\label{def:h1}
    h_1(x,c_{k+1}):=\frac{\frac{k}{x}+\frac{1}{c_{k+1}}+n-k-1}{k+\frac{1-kx}{c_{k+1}}}.
\end{equation}

\discuss{We next characterize an optimal solution $(x^*,c_{k+1}^*)$ in the next three claims. }

\begin{claim}
$c_{k+1}^*$ is either $\max\{x,1-kx\}$ or $1$.\label{c3}
\end{claim}

\begin{claim}
When $n\geq 3$, $h_1(x,c_{k+1})$ reaches its minimum either at $x=1/(k+1), c_{k+1}=1/(k+1)$ or at $x=1/k, c_{k+1}=1$.
\label{c4}
\end{claim}

Comparing the objective function values at the two solutions, we have 
$$h_1(\frac{1}{k},1)\leq h_1(\frac{1}{k+1},\frac{1}{k+1}) \Rightarrow n\leq k(k+1).$$

\begin{claim}
    When $n=2$, $c_1^*=\sqrt{3}-1,c_2^*=1$.
    \label{c6}
\end{claim}   

When $n=2$, Proposition \ref{prop:cstar} coincides with Claim \ref{c6}. When $n \geq 3$, we have $x^*=1/k,c_{k+1}^*=1$ if $ n \leq k(k+1) $, and otherwise, $x^*=1/(k+1),c_{k+1}^*=1/(k+1)$, by Claim \ref{c4}. 

\epr


Proposition \ref{prop:cstar} provides a complete characterization of the optimal solution to Problem \eqref{op44}, which leads to the first main result on the upper found of the POF for the proportional fairness criterion.

\begin{theorem}\label{t1}
Consider a resource allocation problem with $n$ players, $n \geq 2$, for which the utility set $U \subseteq [0,1]^n$ is compact and convex, and the maximum achievable utilities of all players are equal.

Let $\sqrt{n}=k+\epsilon$, where $k\in\mathbb{N}$ is the integral part of $\sqrt{n}$ and $\epsilon \in [0,1)$ is the fractional part. Then, \discuss{the} tight bound of the price of proportional fairness solution is:

\begin{enumerate}[(a)]
    \item For $n=2$,
    $$POF(U;PF) \leq \frac{2-\sqrt{3}}{4}.$$
    \item For $n\geq 3:$
    \begin{enumerate}[(1)]
    \item For  $n < k(k+1)$,
    $$POF(U;PF) \leq 1-\frac{2\sqrt{n}-1+\frac{\epsilon^2}{k}}{n}.$$
    \item For  $n \geq k(k+1)$,
    $$POF(U;PF) \leq 1-\frac{2\sqrt{n}-1+\frac{(1-\epsilon)^2}{k+1}}{n}.$$
    \end{enumerate}
\end{enumerate}
\end{theorem}

\bpr
(a) For $n=2$, we have $g_1(\vect{c}^*)=(2+\sqrt{3})/4$ by Proposition \ref{prop:cstar}, and $$POF(U;PF)\leq 1-g_1(\vect{c}^*)=1-\frac{2+\sqrt{3}}{4}=\frac{2-\sqrt{3}}{4}.$$
(b) For $n\geq 3$, by Proposition \ref{prop:cstar}, we have

\begin{equation*}
   g_1(\vect{c}^*)=  \begin{cases}
     \cfrac{k-1}{n}+\cfrac{1}{k},  &n\leq k(k+1),\\
        &\\
      \cfrac{k}{n}+\cfrac{1}{k+1}, &n > k(k+1).
 \end{cases}
\end{equation*}



Thus, if $n < k(k+1)$, then $g_1(\vect{c}^*)=(n+k(k-1))/nk=(2\sqrt{n}-1+\epsilon^2/k)/n$, and 
$$POF(U;PF) \leq 1-\frac{2\sqrt{n}-1+\frac{\epsilon^2}{k}}{n}.$$

If $n \geq k(k+1)$, then $g_1(c^*)=(n+k(k+1))/(n(k+1))=(2\sqrt{n}-1+(1-\epsilon)^2/(k+1))/n$, and 
$$POF(U;PF) \leq 1-\frac{2\sqrt{n}-1+\frac{(1-\epsilon)^2}{k+1}}{n}.$$
The above bounds are tight, as we have explicitly derived in Proposition \ref{prop:cstar} the optimal solution to Problem \eqref{newop2}.

\epr


\begin{figure}
        \centering
        \includegraphics[width=0.7\linewidth]{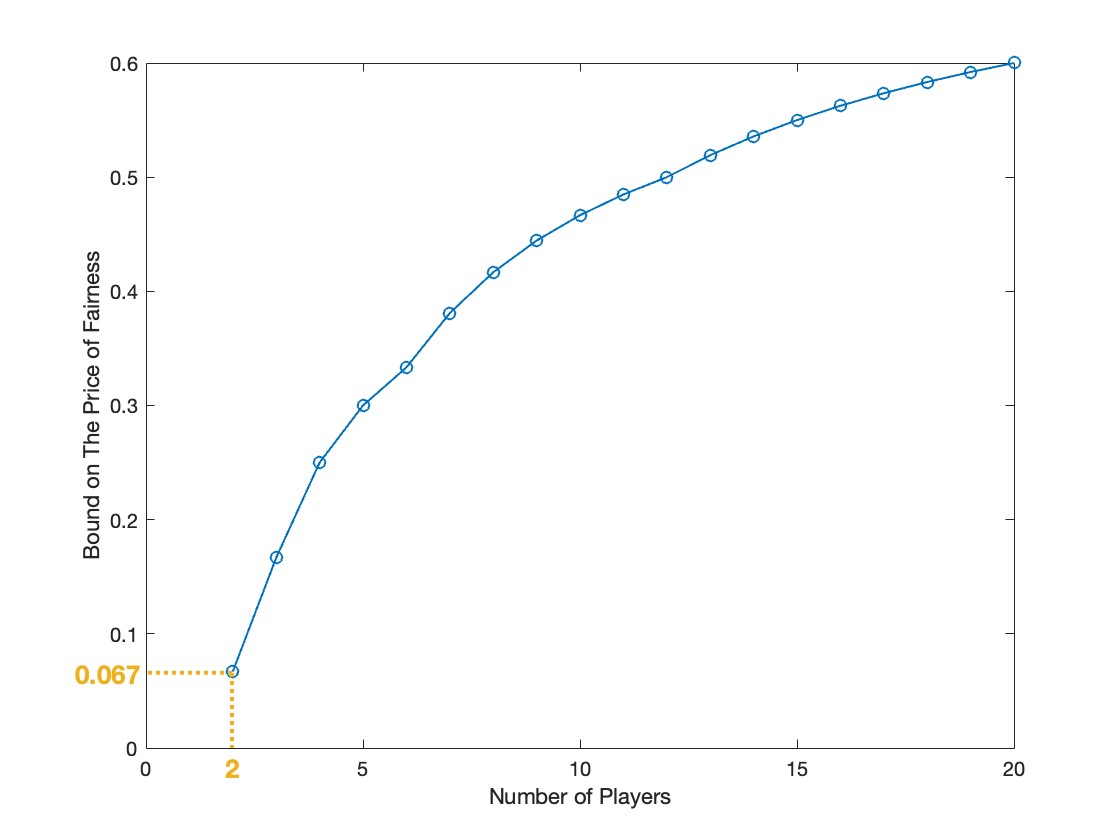}
        \caption{Upper bound of the Proportional POF as a function of the number of players.}
        \label{fig:pfequal}
\end{figure}

\begin{figure}
    \centering
    \includegraphics[width=0.7\linewidth]{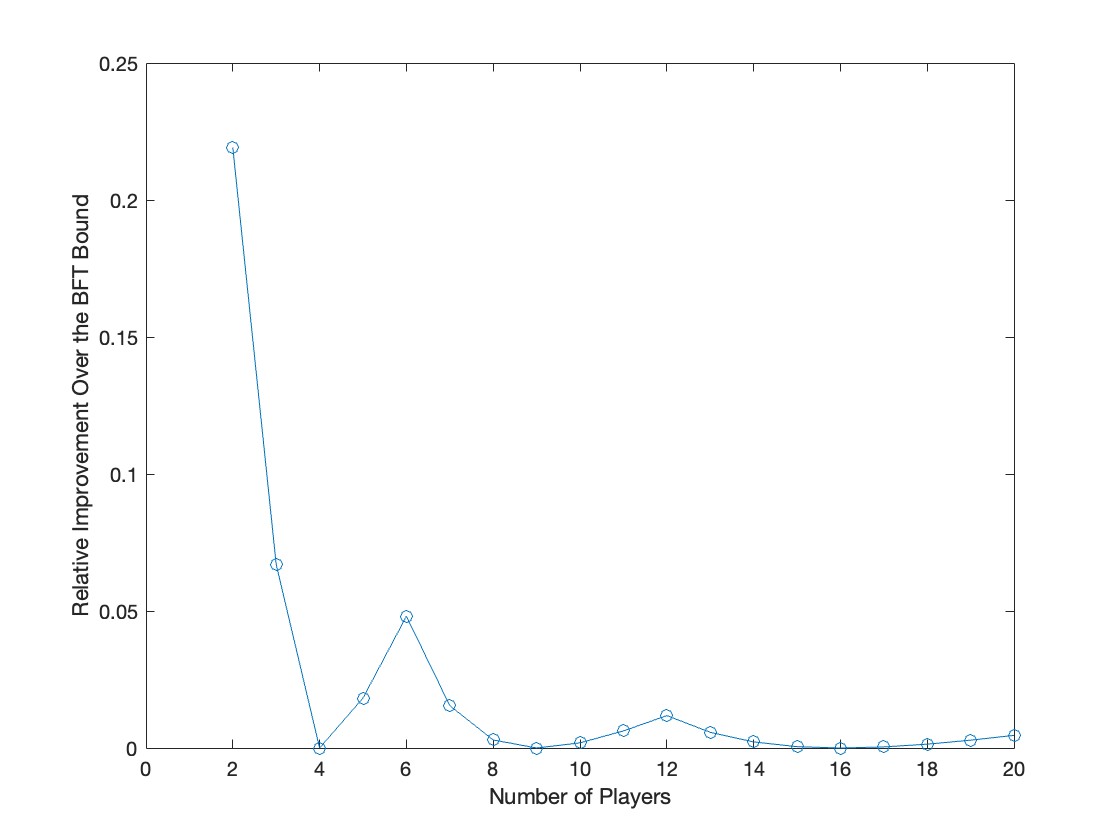}
    \caption{The relative improvement of our upper bound of the proportional POF over the BFT bound.}
    \label{fig:diff}
\end{figure}


Figure \labelcref{fig:pfequal} plots our bound for the price of proportional fairness, $POF(U;PF)$, as a function of the number of players, $n$. Evidently, the bound increases in $n$.

Let us consider the relative improvement, $\Delta(n)$, of our bound over the BFT bound, and note that for $\sqrt{n}\in \mathbb{N}$, i.e., $\epsilon=0$, the BFT bound coincides with our bound. \td{Indeed, when $\sqrt{n}$ is an integer, there exists an optimal solution for Problem \eqref{newop} without Constraints \eqref{11c} which already satisfies Constraints \eqref{11c}. Thus, the optimal values of Problem \eqref{newop} with or without Constraints \eqref{11c} coincide, both yielding the BFT bound. 
}
\begin{equation}\label{delta}
    \Delta(n):=\frac{\text{BFT bound}-\text{our bound}}{\text{BFT bound}}=\begin{cases}
    \frac{\epsilon^2}{k(n-2\sqrt{n}+1)}, & n < k(k+1);\\
    \frac{(1-\epsilon)^2}{(k+1)(n-2\sqrt{n}+1)}, & n \geq k(k+1). \end{cases}
\end{equation}
Figure \labelcref{fig:diff} plots $\Delta(n)$ as a function of $n$, and we note that it reaches a local maximum when the number of players, $n$, is a product of two adjacent integers, that is, when $n = a(a+1)$ for some $a \in \mathbb{N}$, as in the cases when $n= 6, 12, \dots$, etc. In Lemma \ref{lem:Deltan} below, whose proof is in Appendix \ref{sec:Deltan}, we formally prove this result.


\begin{lemma} \label{lem:Deltan}
    For any $a\in \mathbb{N}^+$, $\Delta(n)$ reaches a local maximum when $n=a(a+1)$.
\end{lemma}

Finally, as BFT (2011) have noted, the price of proportional fairness is relatively small when $n$ is small. In particular, for $n=2$, which corresponds to the Nash Bargaining two-player game setting, the BFT bound is 8.6\%. Our results strengthen BFT (2011) observation. Indeed, as shown in Figures \labelcref{fig:pfequal} and \labelcref{fig:diff}, for $n=2$ the POF for our bound is 6.7\%, an improvement of 22\% over the BFT bound.


\subsection{Unequal Maximum Achievable Utilities}

We next consider the case where players have unequal maximum achievable utilities. Denote the maximum achievable utility for player $i$ by $L_i$. Thus, $0 \leq u_i \leq L_i$ for all $\vect{u} \in U$. 

\discuss{To derive an upper bound for the POF in this case, we again start by employing a utility set $U' \supseteq U$, as stated in the following proposition. The proof can be found in Appendix \ref{sec:proofprop689}.}

\begin{proposition}
Suppose $U \subseteq \{\vect{u}\, |\, 0 \leq u_i \leq L_i, i \in N\}$ is a convex and compact utility set, and $\max_{\vect{u} \in U} u_i=L_i$ for $i \in N$. Then there exist $c_i \in [1/nL_i,1/L_i],~i \in N$, such that $U \subseteq U'=\{\vect{u}\, |\, \sum_{i=1}^n c_i u_i \leq 1, \,0 \leq u_i \leq L_i, \,i \in N \}$, and $\max \{ \sum_{i=1}^n \log{u_i} \,| \,\vect{u}\in U \} = \max \{ \sum_{i=1}^n \log{u_i}\, |\, \vect{u}\in U' \}$. Consequently, $POF(U;PF) \le POF(U';PF)$.
\label{prop:4}
\end{proposition}

\subsubsection{Proportional Fairness}
To find the PF solution in the unequal maximum achievable utilities case we need to solve the following optimization Problem \eqref{eq:unequalPF1}: 

\begin{subequations}
\begin{alignat}{2}
&\max_{(u_1,u_2,\dots,u_n)} \quad  &&\sum_{i=1}^n log(u_i),\\
&\qquad \text{s.t.} \qquad  &&0 \leq u_i \leq L_i, \qquad \forall i \in N, \\
& &&\sum_{i=1}^n c_i u_i \leq 1.
\end{alignat}\label{eq:unequalPF1}
\end{subequations}

Let $k_i=u_i/L_i$, then Problem \eqref{eq:unequalPF1} can be written as Problem \eqref{eq:unequalPF2}:
\begin{subequations}
\begin{alignat}{2}
&\max_{(k_1,k_2,\dots,k_n)} \quad  &&\sum_{i=1}^n log(k_i)+\sum_{i=1}^n log(L_i),\\
&\qquad \text{s.t.} \qquad  &&0 \leq k_i \leq 1, \qquad \forall i \in N, \\
& &&\sum_{i=1}^n c_i L_i k_i \leq 1.
\end{alignat}\label{eq:unequalPF2}
\end{subequations}

The PF solution in the unequal maximum achievable utilities case is provided in the next proposition. Its proof is essentially identical to the proof of Proposition \ref{prop2} and is therefore omitted.

\begin{proposition}\label{prop:PFunequal}
The unique optimal solutions to Problems \eqref{eq:unequalPF1} and \eqref{eq:unequalPF2} are $\vect{k}^{PF}=(1/(n c_1 L_1),$ $ 1/(n c_2 L_2), \dots, 1/(n c_n L_n))$ and $\vect{u}^{PF}=(1/(n c_1), 1/(n c_2), \dots, 1/(n c_n))$, respectively.
\end{proposition}

\subsubsection{Utilitarian Solution}
By Proposition \ref{prop:4}, the utilitarian solution in the unequal maximum achievable utilities case is the optimal solution to
\begin{subequations}
\begin{alignat}{2}
&\max_{(u_1,u_2,\dots,u_n)} \quad  &&\sum_{i=1}^n u_i,\\
&\qquad \text{s.t.} \qquad  &&0 \leq u_i \leq L_i, \qquad \forall i \in N, \\
& &&\sum_{i=1}^n c_i u_i \leq 1.
\end{alignat}
\end{subequations}

Let $k_i=u_i/L_i$, then the above optimization problem can be written as Problem \eqref{b2}:
\begin{subequations}
\begin{alignat}{2}
&\max_{(k_1,k_2,\dots,k_n)} \quad  &&\sum_{i=1}^n L_i k_i,\\
&\qquad \text{s.t.} \qquad  &&0 \leq k_i \leq 1, \qquad \forall i \in N, \\
& &&\sum_{i=1}^n c_i L_i k_i \leq 1.
\end{alignat}
\label{b2}
\end{subequations}

Problem \eqref{b2} is a linear relaxation of the 0-1 knapsack problem with rewards $L_i$ and costs $c_i L_i$. Thus, its optimal solution is
\begin{equation*}
k_{\sigma(i)}=
    \begin{cases}
1&i=1,\dots,l(c)\\ 
\delta(c) &i=l(c)+1\\ 0 &i=l(c)+2,\dots,n,
\end{cases}
\end{equation*}
where
\begin{equation}\label{defldt2}
l(c):=\max\{j\ |\ \sum_{i=1}^j c_{\sigma(i)} L_{\sigma(i)}\leq 1 \},\qquad \delta(c):=\frac{1-\sum_{i=1}^{l(c)} c_{\sigma(i)} L_{\sigma(i)}}{c_{\sigma(l(c)+1)}L_{\sigma(l(c)+1)}},
\end{equation}
and $\sigma(i)$ is the index of the $i^{th}$ smallest element in vector $c$.

Without loss of generality, we assume that the $L_i$'s are decreasing, i.e. $L_1 \geq L_2 \geq \cdots \geq L_n$. Then let us consider the following Problem \eqref{b3},
\begin{subequations}
\begin{alignat}{2}
&\max_{(k_1,k_2,\dots,k_n)} \quad  &&\sum_{i=1}^n L_i k_i,\\
&\qquad \text{s.t.} \qquad  &&0 \leq k_i \leq 1, \qquad  i \in N, \\
& &&\sum_{i=1}^n c_{\sigma(i)} L_i k_i \leq 1.
\end{alignat}
\label{b3}
\end{subequations}

Problem \eqref{b3} is a knapsack problem with rewards $L_i$ and costs $c_{\sigma(i)} L_i$. Compared to Problem \eqref{b2}, in Problem \eqref{b3} items with larger rewards, $L_i$, have smaller costs, $c_{\sigma(i)}L_i$. 
Thus, the optimal value of Problem \eqref{b3} exceeds that of Problem \eqref{b2}. 
Note further that replacing all $c_i$ with $c_{\sigma(i)}$ does not affect the sum of the players' utilities in the PF solution. Thus, optimality would be achieved when $c_1 \leq c_2 \leq \cdots \leq c_n$, and we can therefore assume in the sequel that $c_1 \leq c_2 \leq \cdots \leq c_n$.

\subsubsection{The Price of Proportional Fairness}
Similar to our analysis of the equal maximum achievable utilities case, to calculate the upper bound of the price of proportional fairness, we solve the following optimization Problem \eqref{pofpfue}:
\begin{subequations}
\begin{alignat}{2}
&\min_{c,l}   &&(\frac{\sum_{i=1}^n u^{PF}_i}{\sum_{i=1}^n u_i^*}=)\frac{\sum_{i=1}^n \frac{1}{n c_i}}{\sum_{i=1}^{l} L_i + \frac{1-\sum_{i=1}^{l} c_i L_i}{c_{l+1}}},\label{pofpfueobj} \\
&\text{s.t.} \qquad  &&\frac{1}{n L_1} \leq c_1 \leq c_2 \leq \cdots \leq c_n \leq \frac{1}{L_n},\label{c1b} \\
& &&\frac{1}{n} \leq c_i L_i \leq 1, \qquad i \in N,\label{c1c}\\
& &&\sum_{i=1}^{l} c_i L_i \leq 1, \label{c1d}\\
& &&\sum_{i=1}^{l+1} c_i L_i > 1.\label{c1e}
\end{alignat}
\label{pofpfue}
\end{subequations}
Constraints \eqref{c1d} and \eqref{c1e} ensure that $l$ satisfies the definition of $l(c)$, given in Expression \eqref{defldt2}. 

Let $(\vect{c}^*:=(c_1^*,c_2^*,\dots,c_n^*),l^*)$ denote an optimal solution for \eqref{pofpfue}.
The derivation of $(\vect{c}^*,l^*)$ would yield a tight bound of the POF for the proportional fairness case when players have unequal maximum achievable utilities. 

Recall that, for a resource allocation problem with $n \ge 2$ players and a utility set $U$ with a corresponding positive vector of maximum achievable utilities $\vect{L}=(L_1,L_2,\dots, L_n)>0$, we denote by $POF(U;PF)$ the price of proportional fairness of a problem instance with a utility set $U$. The next theorem derives an explicit expression for \td{the upper bound of $POF(U;PF)$, $UB(n,\vect{L};PF)$, with given parameters $n$ and $L$, }and proves its tightness.

\yding{\begin{theorem}\label{t2}
For any $n \ge 2$ and $\vect{L}=(L_1,L_2,\dots,L_n)$ \td{with $L_1\geq L_2\geq \cdots \geq L_n>0$}, we have 
\begin{equation}\label{eq:tightPF}
\begin{aligned}
    \max\bigg\{POF(U;PF)\,|\, U \mbox{ is convex and compact},\, \dim(U)=n,& \\ \max(U)=(\max_{\vect{u} \in U}u_1, \dots,\max_{\vect{u} \in U}u_n)=\vect{L}\bigg\} & = UB(n,\vect{L};PF),
\end{aligned}
\end{equation}
where
\begin{equation*}
UB(n,\vect{L};PF) =\left\{\begin{array}{ll} 
1- \frac{(\sum_{i=1}^{\ell} \sqrt{L_i})^2 +\sum_{i=\ell+1}^n L_i}{n\sum_{i=1}^\ell L_i},  & \mbox{ if } \sum_{i=1}^{n} L_i \cdot L_{2} \leq L_1^2, \\
\qquad \\
  1- \min  \bigg\{  \frac{(\sum_{i=1}^{\ell} \sqrt{L_i})^2 +\sum_{i=\ell+1}^n L_i}{n\sum_{i=1}^\ell L_i},   & \\ \qquad \qquad  \frac{(\sqrt{L_2^2+2 L_1 L_2 +(L_1+L_2) \sum_{i=3}^n L_i}+\sqrt{L_1 L_2})^2}{n (L_1+L_2)^2} \bigg\}, & \mbox{ if }\sum_{i=1}^{n} L_i \cdot L_{2} > L_1^2,
    \end{array}\right.
    \end{equation*}
\begin{equation*}
\ell =\begin{cases}
1,  & \mbox{ if } S^P_1 =\varnothing, \\
\max(S^P_1),   & \mbox{ otherwise },
    \end{cases}
    \end{equation*}
and $S^P_1:=\{l\in \mathbb{N}^+\cap [2,n-1]\,|\, \sqrt{L_{l}}[(\sum_{i=1}^l \sqrt{L_i})^2+\sum_{i=l+1}^n L_i] \geq 2 \sum_{i=1}^l L_i \cdot \sum_{i=1}^{l-1} \sqrt{L_i}\}$. 
\end{theorem}}

\yding{According to Equality \eqref{eq:tightPF}, our bound, $UB(n,\vect{L};PF)$, is tight in the sense that for any given $n\ge 2$ and $\vect{L} >0$, there exists a utility set $U$ for which $POF(U;PF)=UB(n,\vect{L};PF)$. Equality \eqref{eq:tightPF} does not hold for the BFT bound except when all entries of $\vect{L}$ are equal and $\sqrt{n}$ is an integer.  
} 

\bpr

\yding{To characterize an optimal solution, $(c^*,l^*)$, to Problem \eqref{pofpfue}, we start by fixing $l$ and characterizing the corresponding minimizers, $\overline{c}_i$ ($i=1,\dots,n$). The next two claims characterize $\overline{c}_i$ for $i \le l$ and $i \ge l+2$, respectively, except for $c_{l+1}$ whose value cannot be determined.} The proofs of the two claims are provided in Appendix \ref{sec:proofclaims712}. 



\begin{claim}
$\overline{c}_i=1/L_i$, for $i=l+2,\dots,n.$
\label{claim:cstaruneq}
\end{claim}

\begin{claim}
    ${\overline{c}_1} \sqrt{L_1}={\overline{c}_2} \sqrt{L_2}=\dots={\overline{c}_{l}} \sqrt{L_{l}}$.
    \label{clm:cstartol}
\end{claim}

Given Claim \ref{clm:cstartol}, we define $y:=\overline c_1\sqrt{L_1}$. We also define $x:=c_{l+1}$. Then, Problem \eqref{pofpfue} can be written as follow:
\begin{subequations}
\begin{alignat}{2}
&\min_{x,y,l}   &&\frac{\frac{1}{y} \sum_{i=1}^{l} \sqrt{L_i}+ \frac{1}{x}+\sum_{i=l+2}^n L_i}{\sum_{i=1}^{l} L_i + \frac{1-y \sum_{i=1}^{l} \sqrt{L_i}}{x}},\\
&\text{s.t.} \qquad  && \frac{y}{\sqrt{L_{l}}} \leq x \leq \frac{1}{L_{l+1}} \label{cons:remove},\\
& &&x \geq \frac{1}{n L_{l+1}},\\
& &&\frac{1}{n \sqrt{L_{l}}} \leq y \leq \frac{1}{\sum_{i=1}^{l} \sqrt{L_i}} \label{cons:remove2} ,\\
& &&y \sum_{i=1}^{l} \sqrt{L_i}+x L_{l+1} > 1 \label{cons:remove3}.
\end{alignat}
\label{pb2}
\end{subequations}

For convenience, we define $A(l):=\sum_{i=1}^{l} \sqrt{L_i}$, $B(l):=\sum_{i=l+2}^n L_i$, and $M(l):=\sum_{i=1}^{l} L_i$. Problem \eqref{minxy} below is a relaxation of Problem \eqref{pb2}, 
derived therefrom by removing the left-hand-side constraints of \eqref{cons:remove} and \eqref{cons:remove2}.
\begin{subequations}
\begin{alignat}{2}
&\min_{x,y,l}   &&f_2(x,y,l;\vect{L}):=\frac{\frac{1}{y} A(l)+ \frac{1}{x}+B(l)}{M(l) + \frac{1-y A(l)}{x}},\label{minxyobj}\\
&\text{s.t.} \qquad  &&\frac{1}{n L_{l+1}} \leq x \leq \frac{1}{L_{l+1}}\label{rangex},\\
& &&y \leq \frac{1}{A(l)},\label{15c}\\
& &&y A(l)+x L_{l+1} > 1 \label{yx1}.
\end{alignat}
\label{minxy}
\end{subequations}

We next characterize an optimal solution, $(x^*,y^*,l^*)$, to Problem \eqref{minxy}. \yding{Later, we will show that $(x^*,y^*,l^*)$ is also a feasible solution to the original Problem \eqref{pb2}, and thus, it is an optimal solution to Problem \eqref{pb2} as well. The characterization of $(x^*,y^*,l^*)$ leads to a complete characterization of an optimal solution, $(c^*,l^*)$, to the original Problem \eqref{pofpfue}.} We start by characterizing $x^*$ and $y^*$ for a given $l^*$ in the next two claims, whose proofs are in Appendix \ref{sec:proofclaims712}.

\begin{claim}
    $x^*=1/L_{l^*+1}$.
    \label{clmx}
\end{claim}

\begin{claim}
(1) $y^*=1/A(l^*)$ when $l^*= 1$ with $\sum_{i=1}^n L_i \cdot L_{2} \leq L_1^2$, or when $l^* \geq 2$. \\ (2) $y^*=\tilde{y}:=(-\sqrt{L_1}+\sqrt{B(0) + L_1+ B(0)L_1/L_2})/B(0)$ when $l^*=1$ with $ \sum_{i=1}^n L_i \cdot L_{2} > L_1^2$.
    \label{clmy}
\end{claim}

Finally, it remains to characterize $l^*$. If $\sum_{i=1}^n L_i \cdot L_{2} \leq L_1^2$, then, by Claims \ref{clmx} and \ref{clmy}, the objective function \eqref{minxyobj} attains the minimum at $x^*=1/L_{l^*+1}, y^*=1/A(l^*)$, regardless of whether $l^*=1$ or $l^*\geq 2$. 
\yding{Knowing that, we may rewrite the objective function \eqref{minxyobj} as}
$$h_2(l):=f_2(\frac{1}{L_{l+1}},\frac{1}{A(l)},l;\vect{L})=\frac{(A(l))^2+L_{l+1}+B(l)}{M(l)}=\frac{(A(l))^2+B(l-1)}{M(l)}.$$
In Claim \ref{clml} below, whose proof is provided in Appendix \ref{sec:proofclaims712}, we prove that \yding{the objective function} $h_2(l)$ \discuss{is unimodal, i.e., there exists $\ell \in [1,n-1]$ such that $h_2(l)$ is decreasing for $l \in [1,\ell]$ and increasing for $l \in [\ell,n-1]$}. Thus, we can \yding{search for a minimizer $\ell$ by comparing $h_2(l)$ for successive values of $l$}. If $\sum_{i=1}^n L_i \cdot L_{2} > L_1^2$, then $x=1/L_2$, $y=\tilde{y}$, $l=1$, as well as $x=1/L_{l+1}$, $y=1/A(l)$, $l=2,3,\dots,n-1$, are all candidates for an optimal solution. So we need to compare the values of $f_2(1/L_2,\tilde{y},1;\vect{L})$ and $h_2(l)$ for $l=2,\dots,n-1$. 

\begin{claim}\label{clml}
$h_2(l)$ is unimodal, i.e., there exists $\ell\in [1,n-1]$ such that $h_2(l)$ is decreasing for $l \in [1,\ell]$, increasing for $l \in [\ell,n-1]$, and 
\begin{equation*}
\ell =\begin{cases}
1,  & \mbox{ if } S^P_1 =\varnothing, \\
\max(S^P_1),   & \mbox{ otherwise },
    \end{cases}
    \end{equation*}
where $S^P_1:=\{l\in \mathbb{N}^+\cap [2,n-1]\,|\, \sqrt{L_{l}}[(\sum_{i=1}^l \sqrt{L_i})^2+\sum_{i=l+1}^n L_i] \geq 2 \sum_{i=1}^l L_i \cdot \sum_{i=1}^{l-1} \sqrt{L_i}\}$. 
\end{claim}

\discuss{Thus, if $\sum_{i=1}^n L_i \cdot L_{2} \leq L_1^2$, then $l^*=\ell$. If $\sum_{i=1}^n L_i \cdot L_{2} > L_1^2$, then $l^*$ is either $\ell$, if $h_2(\ell) \leq f_2(1/L_2,\tilde{y},1;\vect{L})$, or 1 otherwise.}

\yding{So far we have characterized an optimal solution $(x^*,y^*,l^*)$ to Problem \eqref{minxy}. It remains to show that $(x^*,y^*,l^*)$ is feasible to the original optimization Problem \eqref{pb2}, and thus, it is an optimal solution to Problem \eqref{pb2}, since Problem \eqref{minxy} is a relaxation of Problem \eqref{pb2}.} 

\begin{claim}\label{clmfeasible}
$(x^*,y^*,l^*)$, as characterized in Claims \ref{claim:cstaruneq}-\ref{clml}, is feasible and thus optimal to Problem \eqref{pb2}.
\end{claim}

To characterize, $c^*$, we set $c^*_i=y^*/\sqrt{L_i}$ for $i=1,2,\dots,l^*, c^*_{l^*+1}=x^*$ and $c^*_i=1/L_i$ for $i=l^*+2,\dots,n$. We thus obtain an explicit construction of an optimal solution, $(c^*,l^*)$, to Problem \eqref{pofpfue}.
\yding{
This also confirms the tightness of our upper bound for the price of proportional fairness, as for any given $(L_1,\dots,L_n)$, the upper bound can always be achieved by constructing a utility set $U:=\{u\, |\, \sum_{i=1}^n c_i^* u_i \leq 1, \,0 \leq u_i \leq L_i, \,i \in N \}$, which is non-empty since it was verified in the proof of Claim \ref{clmfeasible} that $c_i^*\in [1/nL_i,1/L_i]$ for all $i$.}

\epr



The relationship between Theorem 1 and Theorem 2 is formalized in the following corollary, whose proof is provided in Appendix \ref{sec:prooflemma2}.

\begin{corollary}\label{crl1}
When the players have equal maximum achievable utilities, the bound in Theorem \ref{t2} reduces to the bound in Theorem \ref{t1}. 
\end{corollary}

\subsubsection{Worst-Case Bound for the Price of Proportional Fairness}
\yding{Next, \discuss{in Proposition \ref{proppfworst}}, we derive the supremum of $UB(n,\vect{L};PF)$ across all possible values of $\vect{L}$, for a fixed $n$, under the condition that $1 \ge L_1 \ge L_2 \geq \cdots \geq L_n >0$. The proof can be found in the Appendix \ref{sec:proofprop689}. Investigating this supermum is crucial for understanding the worst-case price of proportional fairness and its dependence on the distribution of players' maximum achievable utilities.} 

\begin{proposition}\label{proppfworst}
    $$\sup_{\vect{L}:~1 \ge L_1  \geq ... \geq L_n>0} UB(n,\vect{L};PF) = 1- \frac{1}{n},$$
\yding{where the supremum is approached when $\vect{L}=(1-\epsilon, \epsilon/(n-1), \dots, \epsilon/(n-1))$ with $\epsilon \rightarrow 0$.}
\end{proposition}

\yding{According to Proposition \ref{proppfworst}, the worst case for our bound for the price of PF is when one player has a very large maximum achievable utility $L_1$ and all other players have very small \discuss{maximum achievable} utilities.} 


\yding{Next, we calculate the supremum of the BFT bound for proportional fairness among all $\vect{L}$ under the same condition $1 \ge L_1 \geq \cdots \geq L_n >0$, and compare it with ours.} Let $BFT(n,\vect{L};PF): \mathbb{R}^n \rightarrow \mathbb{R}$ denote the BFT bound of the price of PF for a given $n$ and $\vect{L}$. \yding{The supremum of $BFT(n,\vect{L};PF)$ over all $\vect{L}$ has a more complicated expression than the supremum of our bound, but it admits a simple asymptotic characterization when $n$ is large; see the next proposition, \discuss{whose proof is in the Appendix \ref{sec:proofprop689}.}}

 \begin{proposition}\label{rmk1}

    $$\sup_{\vect{L}:~1 \ge L_1  \geq ... \geq L_n>0} BFT(n,\vect{L};PF) = 1-\frac{1}{n}+\frac{\left(1-\sqrt{\frac{2\sqrt{n}-1}{n}}\right)^2}{n-1}=1-O(\frac{1}{n^{1.25}}),$$
where the supremum is reached when 
\begin{equation}\label{pfbftwc}
\vect{L}=\left(\sqrt{\delta}, \frac{1-\sqrt{\delta}}{n-1},\dots, \frac{1-\sqrt{\delta}}{n-1}\right) \text{ and } \delta=\frac{2\sqrt{n}-1}{n}.
\end{equation}
\end{proposition}

The comparison of the supremums of the upper bounds derived in Proposition \ref{proppfworst} and \ref{rmk1} shows that the supremum of our bound, which exhibits an asymptotic order $1-O(1/n)$, strictly improves the BFT bound characterized by an asymptotic order $1-O(1/n^{1.25})$. \discuss{This improvement is achieved by deriving a tighter bound for the second term in \eqref{eq:POFdef}, which represents the ratio of the total utility gained by a proportional fairness solution to the maximum possible total utility.}

\yding{We observe that the worst-case $\vect{L}$ values for our bound and the BFT bound follow the same pattern, with $L_1 \ge L_2=\dots=L_n$. However, our bound reaches its supremum when $L_1$ is much larger than $L_i$, whereas the BFT bound reaches its supremum when $L_1$ is moderately larger than the $L_i$'s.
In the following section, we will numerically investigate how these two bounds vary with the values of $L_1$ and $L_i$ $(i\geq 2)$ when $\vect{L}$ satisfies $L_1 \ge L_2=\dots=L_n$.}

\subsubsection{Numerical Illustration}

\discuss{We compare our bound with the BFT bound for the price of PF under varying structures of the maximum achievable utility vector $\vect{L} = (L_1, \dots, L_n)$.}

\begin{figure}
    \centering
    \includegraphics[width=0.69\linewidth]{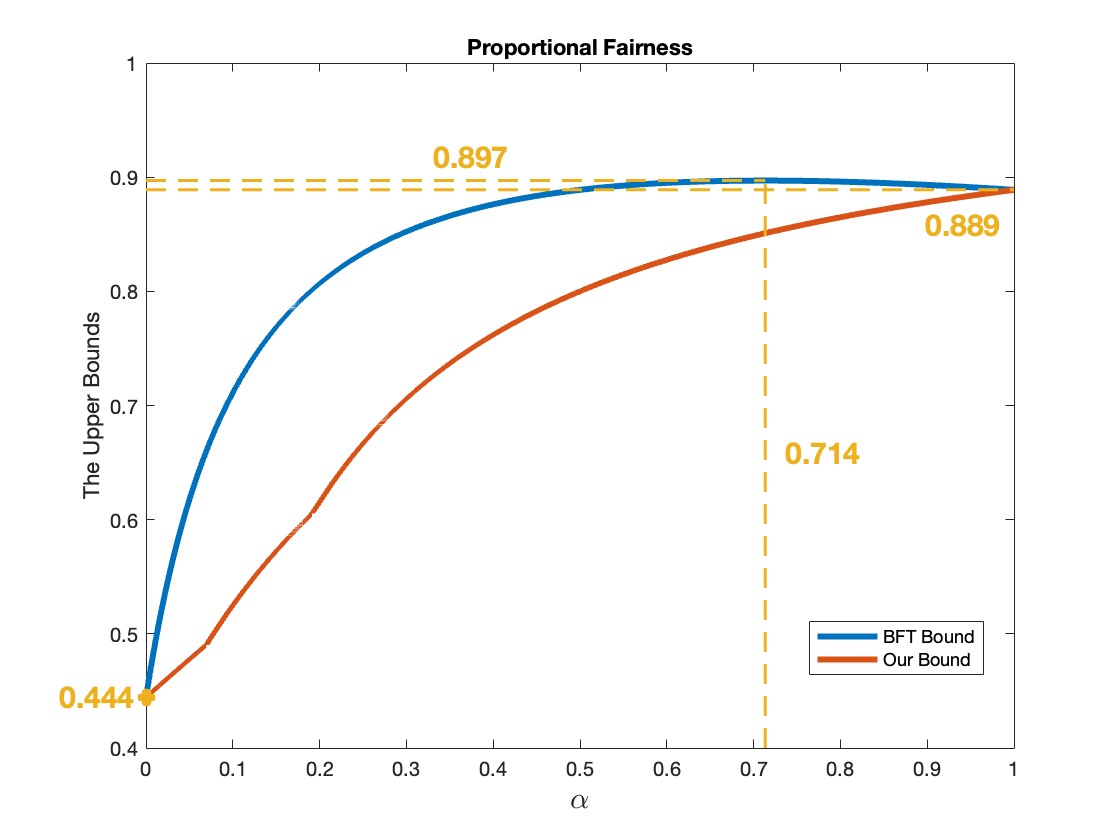}
    \caption{The two upper bounds of the price of proportional fairness as a function of the variance for $n=9$ and for a vector $\vect{L}$ for which $L_1 \geq L_2=\dots=L_n$.}
    \label{fig:pf9}
\end{figure}

\yding{We first examine the case where $\vect{L}$ is structured as $L_1 \ge L_2=\dots=L_n$, which was shown to derive the worst case for our bound and the BFT bound. In the numerical analysis, we consider a scenario with $n=9$ where $\vect{L}$ has the parametric form $(\alpha+(1-\alpha)/n, (1-\alpha)/n,\dots, (1-\alpha)/n)$ for $\alpha \in [ 0,1)$. By varying $\alpha$, we generate a family of vectors that maintain the structure $L_1 \ge L_2 \geq \cdots \geq L_n$ and satisfy $\sum_i L_i=1$. The corresponding values of our bound and the BFT bound are plotted in Figure \ref{fig:pf9}.}

\yding{From Figure \ref{fig:pf9}, we observe that  at $\alpha=0$, both bounds coincide at a value of $4/9$, approximately $0.444$. This coincidence occurs because both bounds are tight when the players' maximum achievable utilities are equal, i.e., $\vect{L}=(1/9,\dots,1/9)$, and $n$ is a square of an integer. As $\alpha$ increases, leading to greater variance across the $L_i$ values, both bounds increase, although the BFT bound increases more rapidly. Specifically, at $\alpha=(3\sqrt{5}-1)/8\approx 0.714$, 
when $\vect{L}$ coincides with the expression in \eqref{pfbftwc}, the BFT bound peaks at its maximum of 0.897, indicated by the upper yellow dashed line. As $\alpha$ continues to increase and approaches $1$, the gap between the two bounds narrows. At the limit, as $\alpha \rightarrow 1$ and $\vect{L}$ approaches $(1,0,\dots,0)$, our bound approaches its supremum, as delineated in Proposition \ref{proppfworst}, marked by the lower yellow line. At this point, the BFT bound converges to our bound, both equal to $1-1/9 \approx 0.889$.}



\begin{figure}
    \centering
    \includegraphics[width=0.69\linewidth]{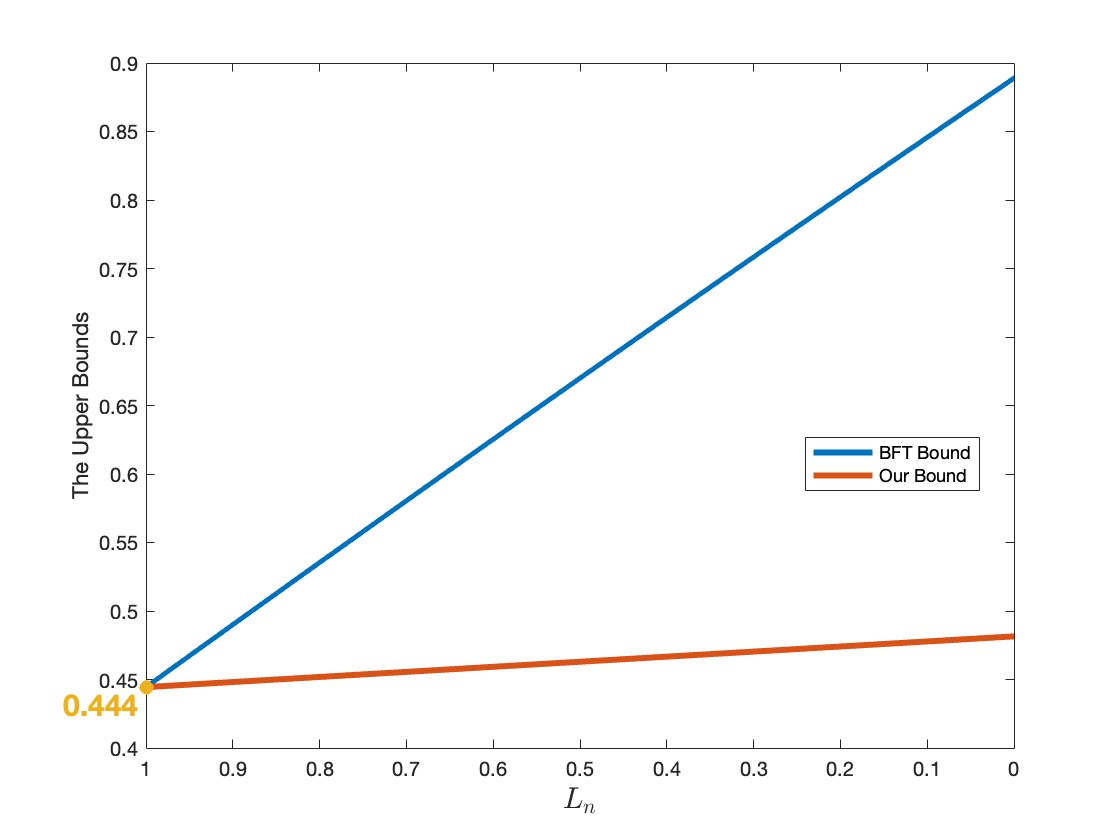}
    \caption{The two bounds for the price of PF when $n=9$ and $L_1=L_2=\dots=L_{n-1}=1, L_n \in (0,1]$.}
    \label{fig:pfln}
\end{figure}

\yding{When $\vect{L}$ adopts forms other than $L_1 \ge L_2=\dots=L_n$, the improvement of our bound over the BFT bound becomes even more pronounced. Figure \ref{fig:pfln} illustrates the comparison of the two bounds for $n=9$ when $\vect{L}$ assumes the structure of $(1,1,\dots,1,L_n)$, with $L_n$ varying from $1$ to $0$. As expected, in the case where $L_n=1$ and all players' maximum achievable utilities are equal, both bounds are tight, having the value of $4/9=0.444$. As $L_n$ decreases, the BFT bound increases faster than our bound. This reveals a greater sensitivity of the BFT bound to the variance of the \discuss{maximum} utility values than our bound for the case when $\vect{L}$ has this structure. The performance of the two bounds behaves similarly for utility vectors of the form $\vect{L}=(1,1,\dots,1,L_{n-k+1},\dots,L_n)$ when $k$ is a small integer and $L_{n-k+1},\dots,L_n$ varying between 1 and 0. 
\td{Thus, in scenarios where the maximum achievable utilities of most players are equal and is very high, but the maximum achievable utilities of a few players are much smaller and thus possibly difficult to accurately estimate, our bound offers a more robust and significantly lower estimate for the price of PF compared to the BFT bound.}}

\yding{Finally, we explore scenarios where $\vect{L}$ does not follow a specific structure. In this study, we fix $n=100$ and generate 100 random maximum achievable utility vectors $\vect{L}$. The entries of the $t^{th}$ vector for $t=1,\dots,100$ are independently drawn from a truncated normal distribution with mean $1$ and a standard deviation $\sigma_t=0.01(t-1)$. The distribution is truncated to ensure  that all elements of $\vect{L}$ remain positive.}

\begin{figure}
    \centering
    \includegraphics[width=0.69\linewidth]{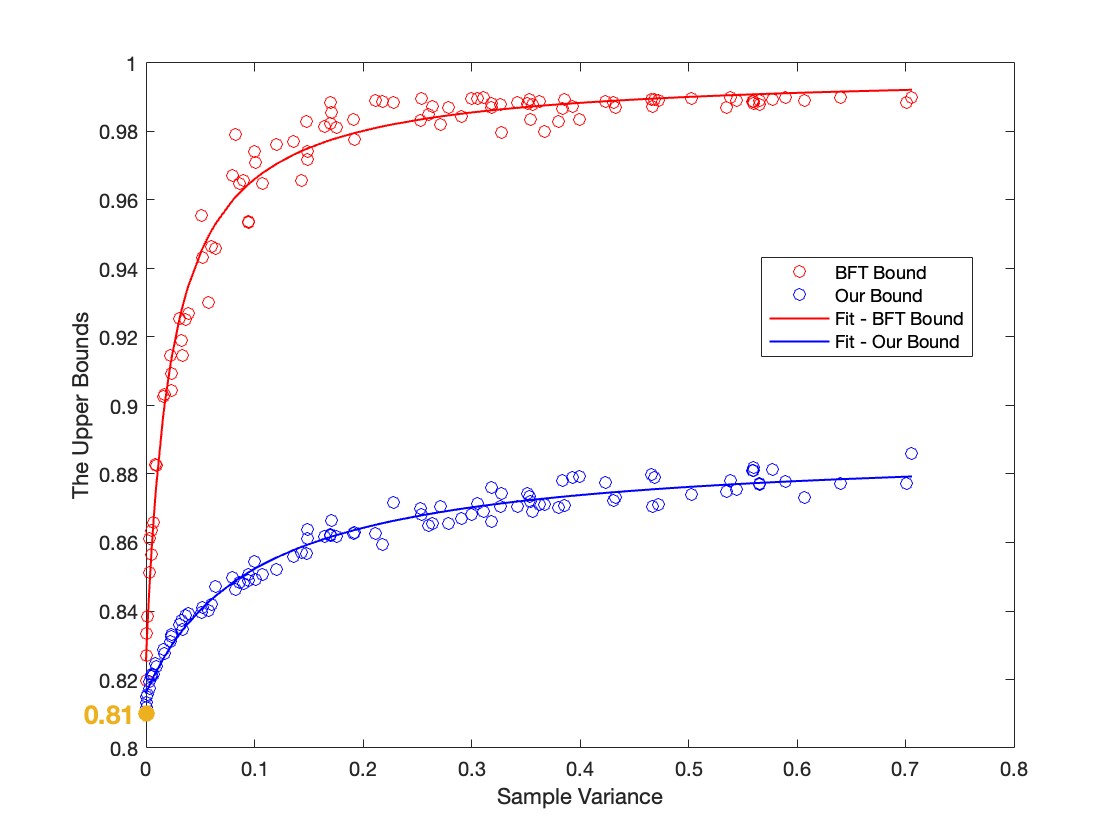}
    \caption{Comparison of two bounds for the price of PF for varying $\sigma_t$}
    \label{fig:uqbd}
\end{figure}

\yding{Figure \ref{fig:uqbd} displays the two bounds for the 100 $\vect{L}$ vectors with varying $\sigma_t$. At $\sigma_t=0$, the situation reduces to the case of equal maximum achievable utilities $\vect{L}=(1,1,\dots,1)$, where both bounds are tight. However, as $\sigma_t$ increases, the value of the BFT bound increases more rapidly than our bound, significantly widening the gap between them. The solid lines in the figure represent a \discuss{nonlinear regression} fit 
to the values of each bound for the 100 instances.}

\yding{From these numerical examples, it is evident that our bound consistently outperforms the BFT bound in situations involving unequal maximum achievable utilities\discuss{, whether for small or large values of $n$}. The advantage of our bound becomes more pronounced as the variation among players' maximum utilities increases, affirming its robustness and effectiveness across a broad spectrum of utility vector structures.}



\newblock

\section{Upper Bounds for The Price of Max-Min Fairness} \label{ubks}

\discuss{In this section we improve upon the BFT bound for the price of max-min fairness when the players maximum achievable utilities are not necessarily equal. Specifically, we prove that the bound we derive is tight and we investigate the gap between our bound and the BFT bound for the unequal case. Further, we prove that when all players have equal maximum achievable utilities, our tight bound for the unequal case reduces to the bound derived by BFT (2011) for the equal case. Thus, our result theoretically confirms the tightness of the BFT bound for the case of equal maximum achievable utilities, which was verified by a specific resource allocation problem.}



\subsection{The Max-Min Fairness Solution}
Similar to the PF case, we first utilize a super set $U'$ of the utility set $U$, to derive an upper bound for the price of MMF in a tractable form, \discuss{as stated in the following proposition. The proof can be found in the Appendix \ref{sec:proofprop1012}.}  

\begin{proposition}
Suppose $U \subseteq \{u\, |\, 0 \leq u_i \leq L_i, i\in N\}$ is a convex and compact utility set, and $\max\{u_i\,|\, \vect{u} \in U\}=L_i$ for $i \in N$. Then there exist $c_i \in \mathbb{R}$ for $i \in N$, such that $U \subseteq U':=\{u\, |\, \sum_{i=1}^n c_i u_i \leq 1, 0 \leq u_i \leq L_i,~ i \in N\}$, and $\max \{ \min_{i\in N} (u_i/L_i) \,|\, \vect{u} \in U \} = \max \{ \min_{i\in N} (u_i/L_i) \,|\, \vect{u} \in U' \}$. Then,  
\begin{equation}\label{mmfobj}
    POF(U;MMF) \leq 1-\frac{\sum_{i=1}^n L_i}{\sum_{i=1}^n c_i L_i \sum_{i=1}^n u^*(U')_i}.
\end{equation}
\label{prop6}
\end{proposition}

\subsection{The Utilitarian Solution}
The approach to find the utilitarian solution for $U'$ is similar to that used to derive the price of proportional fairness, which involves solving the following optimization Problem \eqref{29}:
\begin{subequations}\label{29}
\begin{alignat}{2}
&\max_{(u_1,u_2,\dots,u_n)} \quad  &&\sum_{i=1}^n u_i,\\
&\qquad \text{s.t.} \qquad  &&0 \leq u_i \leq L_i, \qquad \forall i \in N, \\
& &&\sum_{i=1}^n c_i u_i \leq 1.
\end{alignat}
\end{subequations}

Let $k_i=u_i/L_i$ for all $i$. Then Problem \eqref{29} can be formulated as Problem \eqref{ksb2}:
\begin{subequations}
\begin{alignat}{2}
&\max_{(k_1,k_2,\dots,k_n)} \quad  &&\sum_{i=1}^n L_i k_i,\\
&\qquad \text{s.t.} \qquad  &&0 \leq k_i \leq 1, \qquad \forall i \in N, \\
& &&\sum_{i=1}^n c_i L_i k_i \leq 1.
\end{alignat}
\label{ksb2}
\end{subequations}

Problem \eqref{ksb2} is a linear relaxation of the knapsack problem with rewards $L_i$ and costs $c_i L_i$. Without loss of generality, we assume that $c_i$ is increasing in $i$, i.e., 
\begin{equation}
     c_1 \leq c_2 \leq \cdots \leq c_n.
     \label{cincreasing}
 \end{equation}


As before, following BFT (2011), we define 
\begin{equation}
    l(c):=\max\{j\ |\ \sum_{i=1}^j c_i L_i \leq 1\}, \qquad \delta(c):=\frac{1-\sum_{i=1}^{l(c)} c_i L_i}{ c_{l(c)+1} L_{l(c)+1}} \in [0,1).
    \label{defldt}
\end{equation}

Then the optimal solution to Problem \eqref{ksb2} is given by
\begin{equation*}
k_i=
    \begin{cases}
1&i=1,\dots,l(c)\\ \delta(c) &i=l(c)+1\\ 0 &i=l(c)+2,\dots,n,
\end{cases}
\end{equation*}
and the optimal value is $\sum_{i=1}^{l(c)} L_i + \delta(c) L_{l(c)+1}$.


\subsection{The Price of MMF}
In order to calculate the upper bound of the price of max-min fairness, it suffices to minimize the negative term \discuss{on the right-hand side of \eqref{mmfobj}}, by incorporating the optimal solution to Problem \eqref{ksb2} derived above. Let $l$ denote $l(c)$, and consider the optimization Problem \eqref{originalpb}:

\begin{subequations}
\begin{alignat}{2}
&\min_{c,l}   && f_3(c,l;L):= \frac{c_{l+1} \sum_{i=1}^n L_i}{\left(\sum_{i=1}^n c_i L_i\right) \cdot \left( c_{l+1}\sum_{i=1}^{l} L_i + 1-\sum_{i=1}^{l} c_i L_i\right)},\label{66a} \\
&\text{s.t.} \qquad  &&c_1 \leq c_2 \leq \cdots \leq c_n, \label{66b} \\
& &&c_i L_i \leq 1, \qquad i \in N,\label{66c}\\
& &&\sum_{i=1}^{l} c_i L_i \leq 1,\label{66d}\\
& &&\sum_{i=1}^{l+1} c_i L_i > 1 \label{66e}.
\end{alignat}
\label{originalpb}
\end{subequations}

The first Constraint, \eqref{66b}, is from \eqref{cincreasing}. Since $(0,\dots,L_i,\dots,0) \in U$ for all $i$ and $U \subseteq U'$, we have $(0,\dots,L_i,\dots,0) \in U'$, which implies Constraint \eqref{66c}. Constraints \eqref{66d} and \eqref{66e} ensure that $l$ satisfies the definition of $l(c)$, given in Expression \eqref{defldt}.

We next derive a tight bound for the prices of max-min fairness for a resource allocation problem with $n \ge 2$ players, where the utility set $U$ is assumed to be compact and convex, \discuss{and the maximum achievable utilities are assumed, without loss of generality, to satisfy $L_1 \ge L_2 \geq \cdots \geq L_n > 0$. We will show below that we can simultaneously assume both $c_1 \leq c_2 \leq \cdots \leq c_n$ and $L_1 \ge L_2 \geq \cdots \geq L_n$.} Recall that $POF(U;MMF)$ denotes the price of max-min fairness for a problem instance with a utility set $U$, and let $UB(n,\vect{L}; MMF)$ denote the upper bound that we will specify in the next theorem.




\begin{theorem}
For any $n \ge 2$ and $\vect{L}=(L_1,L_2,\dots,L_n)$ \td{with $L_1\geq L_2\geq \cdots \geq L_n>0$}, we have 
\begin{equation}\label{eq:tightMMF}
\begin{aligned}
    \max \bigg\{ POF(U;MMF)\,|\, U \mbox{ is convex and compact},\, \dim(U)=n, & \\   \max(U)=(\max_{\vect{u} \in U}u_1, \dots,\max_{\vect{u} \in U}u_n)=\vect{L}\bigg\}  &= UB(n,\vect{L};MMF),
\end{aligned}
\end{equation}
where
\begin{equation*}
UB(n,\vect{L};MMF) =\begin{cases}
1- \frac{4L_{l^*+1}\sum_{i=1}^n L_i}{(\sum_{i=1}^{l^*} L_i+(n-l^*+1)L_{l^*+1})^2},  & \mbox{ if } S^M_1 \neq \varnothing, \\
\qquad\\
  1- \frac{\sum_{i=1}^n L_i}{\sum_{i=1}^{l^*} L_i(n-l^*+1)},   & \mbox{ if } S^M_1 = \varnothing,
    \end{cases}
    \end{equation*}
\begin{equation*}
l^* =\begin{cases}
\max S^M_1,  & \mbox{ if } S^M_1 \neq \varnothing, \\
  \min S^M_2,   & \mbox{ if } S^M_1 = \varnothing,
    \end{cases}
    \end{equation*}
and $S^M_1:=\{l\in \mathbb{N}^+\,|\,(n-l-1)L_{l+1}< \sum_{i=1}^l L_i \leq (n-l+1)L_{l+1}\},$ $S^M_2:=\{l\in \mathbb{N}^+\,|\,\sum_{i=1}^l L_i>(n-l+1)L_{l+1}\}$. 
\label{thme}
\end{theorem}

\bpr

The proofs of all claims used to prove Theorem \ref{thme} are in Appendix \ref{sec:proofclaims1317}. 

Let $(\vect{c}^*:=(c^*_i)_{i=1,2,\dots,n},l^*)$ denote an optimal solution of Problem \eqref{originalpb}. As before, we first characterize the minimizer $\overline{c}_i$ for $i \ge l+2$ corresponding to a given $l$. 

\begin{claim}
$\overline{c}_i=1/L_i$ for $i=l+2, \dots, n$.
\label{clm1}
\end{claim}


To simplify the notation, we define $$x:=c_{l+1}, y:=\sum_{i=1}^l c_i L_i,\text{ and } A:=\sum_{i=1}^l L_i.$$ According to Claim \ref{clm1}, Problem \eqref{originalpb} can be reformulated as Problem \eqref{op6}:
\begin{subequations}
\begin{alignat}{2}
&\min_{x,y,l}  &&\frac{x \sum_{i=1}^n L_i}{(y+n-l-1+x L_{l+1})(Ax+1-y)},\label{yxca} \\
&\text{s.t.} \qquad  &&y \leq A x,\label{yxcb} \\
& &&0 \leq x L_{l+1} \leq 1,\label{yxcc}\\
& &&0 \leq y \leq 1,\label{yxcd}\\
& &&y+x L_{l+1} > 1 \label{yxce}.
\end{alignat}
\label{op6}
\end{subequations}

To solve Problem \eqref{op6}, we first fix the value of $l, l\in [1,n-1]$ and characterize the corresponding optimal values of $x,y$. Note that if $l=n$, then $\sum_{i=1}^n c_i L_i \leq 1$ and thus $\vect{L}:=(L_1,L_2,\dots,L_n) \in U$. In this case, both the utilitarian solution and the MMF solution are $\vect{L}$, with a corresponding POF \discuss{of 0}, which is not an interesting case to explore. 

Let $g_3(x,y;l)$ denote the objective function without the constant $\sum_{i=1}^n L_i$ in \eqref{yxca}, i.e.,
$$g_3(x,y;l):=\frac{x}{(y+n-l-1+x L_{l+1})(A x+1-y)}.$$ 
For a given $l$, we have the following characterization of $x^*$, $y^*$, and $g_3(x^*,y^*;l)$.

\begin{claim}
    If $A \leq (n-l-1)L_{l+1} $, $$x^*\in [\frac{1}{A+L_{l+1}}, \frac{1}{L_{l+1}}],\quad y^*=1-x^*L_{l+1},\quad g_3(x^*,y^*;l)=\frac{1}{(n-l)(A+L_{l+1})};$$

    If $(n-l-1)L_{l+1} < A \leq (n-l+1)L_{l+1}$, $$x^*=\frac{1}{L_{l+1}},\quad y^*=\frac{1}{2} (\frac{A}{L_{l+1}}-n+l+1),\quad g_3(x^*,y^*;l)=\frac{4L_{l+1}}{(A+(n-l+1)L_{l+1})^2};$$

    If $A > (n-l+1)L_{l+1} $, $$x^*=\frac{1}{L_{l+1}},\quad y^*=1,\quad g_3(x^*,y^*;l)=\frac{1}{A(n-l+1)}.$$
    \label{clm2}
\end{claim}

We next characterize the order of the $L_i's$ at optimality. For convenience, we introduce $p(A,L_{l+1}; l)$, defined as follows:
\begin{equation}
p(A,L_{l+1};l)(:=g_3(x^*,y^*;l))=
\begin{cases}
    \frac{1}{(n-l)(A+L_{l+1})}, & A\leq (n-l-1)L_{l+1},\\
    \qquad \\
    \frac{4L_{l+1}}{(A+(n-l+1)L_{l+1})^2}, & (n-l-1)L_{l+1} < A \leq (n-l+1)L_{l+1},\\
    \qquad \\
    \frac{1}{A(n-l+1)}, & A> (n-l+1)L_{l+1}.
\end{cases} \label{funcp}
\end{equation}

\newblock

\begin{claim}
    $p(A,L_{l+1};l)$ is decreasing both in $A$ and $L_{l+1}$.
    \label{clm3}
\end{claim}


Since $A=\sum_{i=1}^l L_i$, the internal order of $L_1$ to $L_l$ does not affect the value of $A$ and thus, does not affect the objective function value. Therefore, without loss of generality, we assume that $L_1$ to $L_l$ are arranged in a descending order. Then, according to Claim \ref{clm3}, the permutation of the $L_i's$ that would minimize $p(A,L_{l+1};l)$ should satisfy:
$L_1 \geq L_2 \geq \cdots \geq L_l \geq L_{l+2}$,
$L_{l+1} \geq L_{l+2}$ and $L_{l+2} \geq L_{l+3} \geq \cdots \geq L_n$. In fact, in Claim \ref{clm9} below we prove that the $L_i's$ should be in descending order overall, i.e., $L_1\geq L_2\geq \cdots \geq L_n$.

\begin{claim}
The value of $p(A,L_{l+1};l)$ is minimized with respect to the $L_i, i=1,\dots,n$, if they are arranged in a descending order.
    \label{clm9}
\end{claim}

Next we discuss the value of $l$ at optimality. For convenience, we define

\begin{equation*}
    \begin{aligned}
        &S^M_0:=\{l\in \mathbb{N}^+\,|\,\sum_{i=1}^l L_i \leq (n-l-1)L_{l+1}\},\\
        &S^M_1:=\{l\in \mathbb{N}^+\,|\,(n-l-1)L_{l+1}< \sum_{i=1}^l L_i \leq (n-l+1)L_{l+1}\},\\
        &S^M_2:=\{l\in \mathbb{N}^+\,|\,\sum_{i=1}^l L_i>(n-l+1)L_{l+1}\}.
    \end{aligned}
\end{equation*}

\begin{claim}
    Let $l^*$ denote an optimal value of $l$. If $S^M_1\neq \varnothing$, $l^*=\max S^M_1$, otherwise $l^*=\min S^M_2$. \label{clm10}
\end{claim}

Now, $l^*=\max S^M_1$, corresponds to the second case in Claim \ref{clm2}. Thus, $g(x^*,y^*;l^*)=4L_{l^*+1}/(A+(n-l^*+1)L_{l^*+1})^2$ and $1-POF(U;MMF)$, which is the objective function of Problem \eqref{op6}, is given by $4 L_{l^*+1}\sum_{i=1}^nL_i/(A+(n-l^*+1)L_{l^*+1})^2$. Similarly, $l^* = \min S^M_2$ corresponds to the third case of Claim \ref{clm2}. Thus, $g(x^*,y^*;l^*)=1/A(n-l^*+1)$ and $1-POF(U;MMF)=\sum_{i=1}^nL_i/A(n-l^*+1)$, and we have derived the expressions for the upper bound for the price of MMF given in Theorem \ref{thme}. 

Finally, for any vector of maximum achievable utilities, $\vect{L}$, we construct below an instance of $U$ to illustrate that the above values of $l^*$, $x^*$, and $y^*$ (or $\vect{c}^*$) can be achieved, which implies that the upper bound in Theorem \ref{thme} is tight. Specifically, for a given $\vect{L}=(L_1,L_2,\dots,L_n)$, we can derive the corresponding $S^M_1,S^M_2$, and $l^*$, and define:
\begin{equation*}
    Y:=\begin{cases}
        \frac{1}{2}(\frac{\sum_{i=1}^{l^*} L_i}{L_{l^*+1}}-n+l^*+1), &\text{if }l^*=\max S^M_1,\\
        1 &\text{if } l^*=\min S^M_2.
    \end{cases}
\end{equation*}

If $l^*=\max S^M_1$, $(n-l^*-1)L_{l^*+1}< \sum_{i=1}^{l^*} L_i $, and thus $Y$ is always positive. We then construct a utility set $U=\{u\, |\, 0\leq u_i \leq L_i,~\sum_{i=1}^n c_i u_i \leq 1\}$, where
\begin{equation*}
    c_i =\begin{cases}
        \frac{Y}{l^*L_i} &i=1,\dots,l^*,\\
        \frac{1}{L_i} &i=l^*+1,\dots,n.
    \end{cases}
\end{equation*}

Now, recall that to obtain the MMF solution, we initially maximize the minimum ratio of players' utilities to their corresponding maximum achievable utilities, i.e., $\max_{\vect{u} \in U} \min_{i\in N } u_i/L_i$. 
In the above specific example, the solution to this max-min optimization problem can be shown to be $\bar{\vect{u}} :=(L_1/(Y+n-l^*),L_2/(Y+n-l^*),\dots,L_n/(Y+n-l^*))$, at which, obviously, all the ratios of the players utilities to their maximum achievable utilities are equal. Since at $\bar{\vect{u}}$, the constraint $\sum_{i=1}^l c_iu_i \leq 1$ in $U$ is satisfied as equality, and since all $c_i's$ are positive, any increase in the utility of any of the players will violate this constraint. Therefore, $\bar{\vect{u}}$ must be the MMF solution.


Furthermore, since the utilitarian solution, $\vect{u}^*$, for this example is $\vect{u}^*=(L_1,\dots,L_{l^*},(1-Y)L_{l^*+1},0,\dots,0)$, the price of MMF for this example is
\begin{equation*}
\begin{aligned}
    POF(U;MMF)&=1-\frac{\sum_{i=1}^n u_i^{MMF}}{\sum_{i=1}^n u_i^*}\\&=1-\frac{\frac{\sum_{i=1}^n L_i}{Y+n-l^*}}{\sum_{i=1}^{l^*} L_i +(1-Y)L_{l^*+1}}=
    \begin{cases} 1-\frac{4L_{l^*+1} \sum_{i=1}^n L_i}{(\sum_{i=1}^{l^*} L_i+(n-l^*+1)L_{l^*+1})^2} &l^*\in S^M_1, \\
    \qquad \\
        1- \frac{\sum_{i=1}^n L_i}{\sum_{i=1}^{l^*}L_i (n-l^*+1)} &l^*\in S^M_2.
        \end{cases}
\end{aligned}
\end{equation*}

\epr

\yding{We conclude that \eqref{eq:tightMMF} is indeed satisfied for the above example, which establishes the tightness of our bound for the price of max-min fairness. In contrast, for the BFT bound, in general, \eqref{eq:tightMMF} is not satisfied. }

We note that in the special case when all players have equal maximum achievable utilities, that is, when $L_1=L_2=\dots=L_n$, our upper bound, $UB(n,\vect{L};MMF)$, reduces to the BFT bound. Therefore, in some sense, our result theoretically confirms the tightness of the BFT bound in this case, which was shown by BFT (2011) to be tight for a specific resource allocation problem. 

\discuss{BFT (2011) provide an example where the maximum achievable utilities vector $\vect{L}$ has a special structure, $L_1=L_2=\dots=L_{(n+1)/2}\geq 1, L_{(n+3)/2}=L_{(n+5)/2}=\dots=L_n=1$, for which their bound is achieved. As we show below in Corollary \ref{crl2}, whose proof is in Appendix \ref{sec:crl2}, also in this case, for a more general maximum achievable utility vector, $\vect{L}$, $L_1=L_2=\dots=L_{\lfloor n/2 \rfloor +1}$, equality \eqref{eq:tightMMF} holds and our bounds reduces to the BFT bound.}

\begin{corollary}\label{crl2}
    \begin{enumerate}
        \item [(a)] For $L_1=L_2=\ldots=L_n$, our upper bound reduces to the BFT bound for the case of equal maximum achievable utilities, i.e.,
        $$UB(n,\vect{L};MMF)=1-\frac{4n}{(n+1)^2}.$$
    \item [(b)] For $L_1=L_2=\dots=L_{\lfloor n/2 \rfloor +1} \td{\geq L_{\lfloor n/2 \rfloor +2} \geq \cdots \geq L_n}$, our upper bound reduces to the BFT bound for the case of unequal maximum achievable utilities, i.e.,
        $$UB(n,\vect{L};MMF)=1-\frac{4\sum_{i=1}^n L_i}{(n+1)^2 L_1}.$$
    \end{enumerate}
\end{corollary}

\subsection{\yding{Worst-Case Bound for the Price of Max-Min Fairness}}

\yding{Similar to the proportional fairness scenario, we investigate the worst-case upper bound for the price of max-min fairness, which is the supremum of $UB(n,\vect{L};MMF)$ across all possible vectors $\vect{L}$ under the condition that $1 \ge L_1 \geq \cdots \geq L_n> 0$. The forthcoming proposition demonstrates that the supremum of $UB(n,\vect{L};MMF)$ is also $1-1/n$, which coincides with the upper bound supremum for the price of proportional fairness. \discuss{The proof can be found in the Appendix \ref{sec:proofprop1012}.}}

\yding{\begin{proposition}\label{mmfwc}
    $$\sup_{\vect{L}:L_i \in (0,1], L_1 \geq \cdots \geq L_n} UB(n,\vect{L};MMF) = 1- \frac{1}{n}.$$
where the supremum is approached when $\vect{L}=(1-\epsilon, \epsilon/(n-1), \dots, \epsilon/(n-1))$ with $\epsilon \rightarrow 0$.
\end{proposition}}

\yding{Next, let $BFT(n,\vect{L};MMF)$ denote the BFT bound of the price of max-min fairness with input parameters $n$ and $\vect{L}$. The next proposition characterizes the supremum of $BFT(n,\vect{L};MMF)$ over all $\vect{L}$ such that $L_1\geq L_2 \geq \cdots \geq L_n>0$.} \discuss{ Its proof can be found in the Appendix \ref{sec:proofprop1012}.}

\yding{\begin{proposition}\label{mmfbftwc}
 $$\sup_{\vect{L}:~1 \ge L_1 \geq \cdots \geq L_n>0} BFT(n,\vect{L};MMF) = 1- \frac{4}{(n+1)^2}=1-O(\frac{1}{n^2}).$$
where the supremum is approached when $\vect{L}=(1-\epsilon, \epsilon/(n-1), \dots, \epsilon/(n-1))$ with $\epsilon \rightarrow 0$.
\end{proposition}}

\yding{For all $n\geq 2$, $1/n>4/(n+1)^2$. Thus, the supremum of the BFT bound is strictly larger than that of our bound for all $n\geq 2$. When $n$ is large, our bound, $1-1/n$, asymptotically improves the BFT bound, which is $1-O(1/n^2)$.}

\yding{Finally, it is important to note that for the price of max-min fairness, both our bounds and the BFT bounds approach their supremum values as $\vect{L}$ tends towards $(1,0,\dots,0)$. This contrasts with the proportional fairness scenario, where our bounds and the BFT bounds reach their supremum values under different worst-case $\vect{L}$ configurations.}

\subsection{Numerical Illustration}



\begin{figure}[H]
    \centering
    \includegraphics[width=0.69\linewidth]{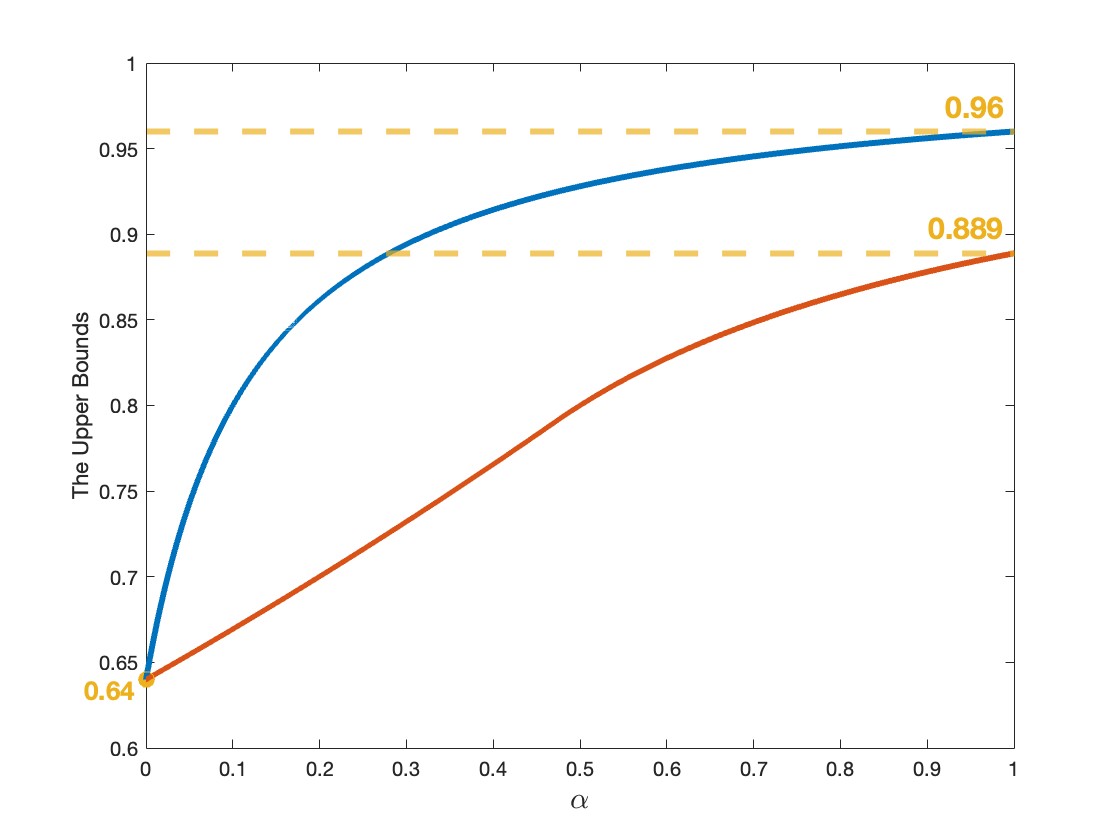}
    \caption{The two upper bounds of the price of MMF as a function of sample variance for $n=9$, and a maximum utility vector $\vect{L}$ for which $L_2=\dots=L_n$.}
    \label{fig:mmfwc}
\end{figure}

\yding{We compare the two upper bounds for the price of max-min fairness through numerical analysis. Similar to the PF scanario, we begin with $n=9$ and where $\vect{L}$ assumes the parametric form $(\alpha+(1-\alpha)/n, (1-\alpha)/n,\dots, (1-\alpha)/n)$ for $\alpha$ within the range $[0,1)$. Figure \ref{fig:mmfwc} illustrates the behaviour of our bound and the BFT bound as $\alpha$ varies from $0$ to $1$. At $\alpha=0$, $\vect{L}=(1/9,\dots,1/9)$, where both bounds are tight and yield a value of $0.64$. As $\alpha$ increases, so do both bounds. As $\alpha \rightarrow 1$, each bound approaches its respective supremum, indicated by the yellow dashed lines. The values of these supremums are detailed in Proposition \ref{mmfwc} and \ref{mmfbftwc}. }

\begin{figure}[H]
    \centering
    \includegraphics[width=0.7\linewidth]{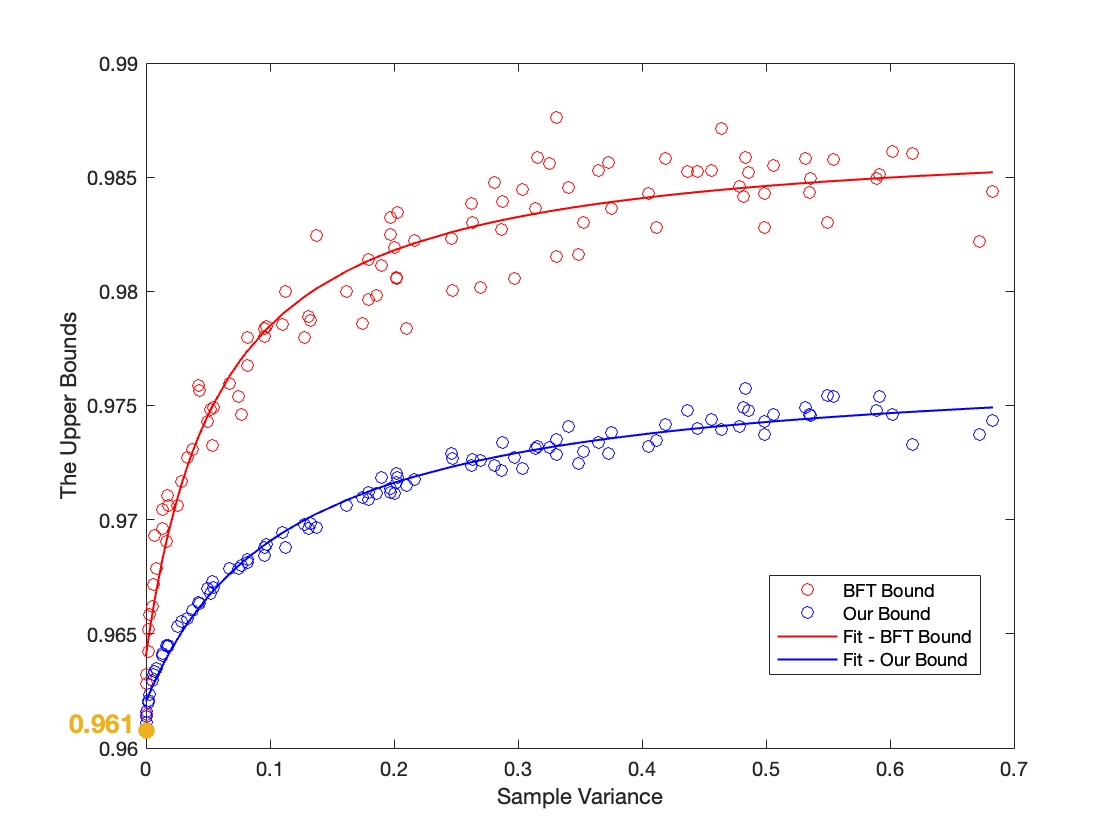}
    \caption{The two upper bounds of the price of MMF.}
    \label{fig:ksbd}
\end{figure}

\yding{We further evaluate the performance of the two bounds when $\vect{L}$ lacks a specific structure. Setting $n=100$, we generate 100 random vectors of maximum achievable utilities, $\vect{L}$. Every entry of the $t^{th}$ vector ($t=1,\dots,100$) is independently drawn from a truncated normal distribution with a mean of 1 and standard deviation of $\sigma_t=0.01(t-1)$. We compute the BFT bound and our bound for these 100 $\vect{L}$ vectors, with the results displayed in Figure \ref{fig:ksbd}. At $\sigma_t=0$, both bounds are tight and yield a value of 0.961. However, as $\sigma_t$ increases, a notable widening of the gap between the two bounds is observed in Figure \ref{fig:ksbd}. We also observe that the improvement of our bound over BFT bound for the price of max-min fairness is more significant compared to the PF scenario, which aligns with the theoretical result that the improvement is $1/n-O(1/n^{1.25})$ in the PF scenario and $1/n-O(1/n^{2})$ in the MMF scenario.}


\section{Conclusion}

We have derived tight bounds for the price of proportional fairness for all positive integers $n$ and for both the equal and the unequal maximum achievable utilities cases. As such, we improved on the bounds derived for this case by BFT (2011). Specifically, their bound for the price of proportional fairness is tight only when $\sqrt{n}$ is an integer and when all players have equal maximum achievable utilities.

We have also studied the price of the Max-Min fairness (MMF) criterion, which is a fairness in the spirit of Rawlsian justice and is consistent with the Kalai-Smorodinsky solution for the two-player case (see BFT 2011). BFT (2011) have derived a tight bound for the price of MMF only when the maximum achievable utilities for all players are equal. By contrast, we have derived a tight bound for the price of MMF for the unequal maximum achievable utilities case, which reduces to the tight bound derived by BFT (2011) for the case of equal maximum achievable utilities. 

For both fairness notions, we have also studied the sensitivity of our bounds and the BFT bounds to the variability of the maximum achievable utilities. \yding{Both bounds increase with the variability of the maximum achievable utilities, but the BFT bounds increase at a faster rate. The numerical experiments reveal that for maximum utility vectors with or without a special structure, our bounds significantly improve upon the BFT bounds for both fairness criteria.}

Finally, for the unequal maximum achievable utilities case, we have introduced, for both fairness criteria, the notion of the worst-case price of fairness in terms of the maximum achievable utilities. We have proven that for our bounds, both for the PF and MMF criteria, the worst-case of the POF is $1-1/n$. In contrast, we have proven that the BFT supremum bound for the price of PF is \yding{$1-O(n^{1.25})$, and that of MMF is $1-O(n^2)$. Therefore, our bounds asymptotically improve upon the BFT bound for both fairness criteria when there is a large number of players. The characterization of the worst-case maximum achievable utility vectors should also help the decision maker understand in which scenarios it is most costly to implement a fairness solution.
}
 





\newpage
\theendnotes

\bibliographystyle{plainnat}
\bibliography{ref}


\input{Ecompanion.tex}

\end{document}

%% file: Ecompanion.tex
\renewcommand{\theHchapter}{A\arabic{chapter}}
\ECSwitch


\section{Proof of Propositions \ref{prop:1} - \ref{prop:3}}\label{sec:proofprop14}

\begin{repeatproposition}[Proposition \ref{prop:1}]
Suppose $U \subseteq \{\vect{u}:=(u_1,u_2,\dots,u_n)\, |\, 0 \leq u_i \leq 1, i \in N \}$ is a convex and compact utility set, and $\max\{u_i\, |\,\vect{u} \in U\}=1$ for $i\in N$. Then there exist $c_i \in [1/n,1],\ i\in N$, such that $U \subseteq U'=\{\vect{u}\, |\,\sum_{i=1}^n c_i u_i \leq 1, 0 \leq u_i \leq 1, i\in N\}$, and $\max \{ \sum_{i=1}^n \log{u_i} \, |\, \vect{u} \in U \} = \max \{ \sum_{i=1}^n \log{u_i} \, |\, \vect{u} \in U' \}$. Consequently, $POF(U;PF) \le POF(U';PF)$.
\end{repeatproposition}

\bpr
Let $\vect{v}$ denote the optimal solution to $\max \{ \sum_{i=1}^n \log{u_i} \, |\, \vect{u} \in U \}$. Define $\nu(\vect{u}):=\sum_{i=1}^n \log{u_i}$ and denote by $\nabla(\nu(\vect{v}))$ the gradient of $\nu(\cdot)$ at $\vect{v}$. Since $U$ is convex and compact, and the objective function $\nu(\cdot)$ is concave, the necessary and sufficient optimality condition is 
\begin{equation} \label{3.1}
(\vect{u}-\vect{v})^T \nabla(\nu(\vect{v})) = \sum_{i=1}^n (\frac{u_i}{v_i} -1)  \le 0    \Leftrightarrow \sum_{i=1}^n c_i u_i  \le 1, ~~~\forall~\vect{u} \in U,
\end{equation}
where $c_i:=1/(nv_i)$. Therefore, the optimality condition implies that $U \subseteq U':=\{\vect{u}\,|\,\sum_{i=1}^n c_i u_i \leq 1, 0 \leq u_i \leq 1, i\in N\}$, for $c_i=1/(nv_i)$. 

We claim that $\vect{v}$ is also the optimal solution to the optimization problem to derive the PF solution with respect to $U'$, i.e., $\max \{ \sum_{i=1}^n \log{u_i} \, |\, \vect{u} \in U' \}$. Indeed, since $\vect{v} \in U \subseteq U'$, $\vect{v}$ is feasible to this optimization problem. Further, for all $\vect{u} \in U'$, we have $\sum_{i=1}^n c_i u_i = \sum_{i=1}^n u_i/(nv_i)  \leq 1$. Thus, $\vect{v}$ satisfies the necessary and sufficient optimality condition \eqref{3.1}. Therefore, $\vect{v}$ is also the PF solution with respect to $U'$. We thus have $\vect{u}^{PF}(U)=\vect{u}^{PF}(U')$. 

 Given that $U \subseteq U'$, we have $\sum_i u_i^*(U) \le \sum_i u_i^*(U')$, where $u_i^*(U)$ and $u_i^*(U')$ denote player $i$'s utility at the utilitarian solution with respect to $U$ and $U'$, respectively. Therefore, by the definition of the POF, i.e., \eqref{eq:POFdef}, we have $POF(U;PF) \le POF(U';PF)$.

Finally, since $0 \leq v_i \leq 1$, we have $c_i =1/(nv_i) \geq 1/n$. Let $e_i$ denote a vector whose $i$-th element is one and all other elements as zero. By assumptions of $U$, $e_i \in U$ and thus $e_i \in U'$ for $i\in N$. Then $\sum_{i=1}^n c_i u_i \leq 1 \Rightarrow c_i \leq 1$.
\epr

\begin{repeatproposition}[Proposition \ref{prop2}]
$\vect{u}^{PF}(U'):=(1/nc_1,1/nc_2,\dots,1/nc_n)$ is the unique optimal solution to Problem \eqref{5.3}.
\end{repeatproposition}

\bpr
Since $c_i \ge 1/n$, we have $u_i^{PF}(U') \le 1$. So $\vect{u}^{PF}(U')$ is a feasible solution to \eqref{5.3}. For any feasible $\vect{u} \in U'$, $(\vect{u}-\vect{u}^{PF}(U'))^T \nabla (\nu(\vect{u}^{PF}(U'))) =\sum_{i=1}^n (u_i-u_i^{PF}(U'))/u_i^{PF}(U')=\sum_{i=1}^n nc_i u_i-n \leq 0$, which implies the optimality of $\vect{u}^{PF}(U')$. Finally, since in \eqref{5.3} we maximize a strictly concave function over a convex compact domain, the maximizer must be unique.
\epr

\begin{repeatproposition}[Proposition \ref{prop:3}]
    Problems \eqref{op44} and \eqref{newop} are equivalent.   
\end{repeatproposition}

\bpr 
(\uppercase\expandafter{\romannumeral1}) For any feasible solution, $(\vect{c},l)$, to Problem \eqref{op44}, we can associate a feasible solution, $(\vect{d},\vect{y})$, to Problem \eqref{newop} as follows:
\begin{equation}\label{p3fm}
    \text{For each }i, i=1,\dots,n, \qquad
    d_{2i-1}=d_{2i}=c_i, \text{ and } y_i=\begin{cases}
    1, &\text{if }i\leq l, \\
    \frac{1-\sum_{i=1}^{l} c_i}{c_{l+1}}, &\text{if }i= l+1,\\
    0, &\text{if }i \geq l+2.
\end{cases}
\end{equation}
We will show that $(\vect{d},\vect{y})$ is feasible for \eqref{newop}. By \eqref{p3fm}, \eqref{11c} hold, and by \eqref{d1}, \eqref{11b} are also satisfied at $(\vect{d},\vect{y})$. Further, \eqref{11d} holds since
$$\sum_{i=1}^{n} y_i d_{2i-1}=\sum_{i=1}^{l} c_i + \frac{1-\sum_{i=1}^{l} c_i }{c_{l+1}} \cdot c_{l+1}=1.$$
By \eqref{d2} and \eqref{d3}, we have 
$$y_{l+1}=\frac{1-\sum_{i=1}^{l} c_i}{c_{l+1}} \in [0,1).$$
Therefore, \eqref{11e} and \eqref{11f} are satisfied at $(\vect{d},\vect{y})$.

In addition, examining \eqref{11g}, we have for any $i \neq l+1$, $\sqrt{y_i}-y_i=0$, so 
$$\sum_{i=1}^n (\sqrt{y_i}-y_i)=\sqrt{y_{l+1}}-y_{l+1}=\prod_{i=1}^n (\sqrt{y_i}-y_i+1)-1.$$

Finally, we show that $f_1(\vect{d},\vect{y})=f_0(\vect{c},l)$ for any feasible solution $(\vect{c},l)$ to Problem \eqref{op44}, and the corresponding solution $(d,y)$ defined in \eqref{p3fm}. For the numerator, $$\frac{1}{n}\sum_{i=1}^n (\frac{y_i}{d_{2i-1}}+\frac{1-y_i}{d_{2i}})=\frac{1}{n}\sum_{i=1}^n (\frac{y_i}{c_i}+\frac{1-y_i}{c_i})=\sum_{i=1}^n \frac{1}{nc_i},$$ and for the denominator, $$\sum_{i=1}^n y_i=l+y_{l+1}+0=l+\frac{1-\sum_{i=1}^l c_i}{c_{l+1}}.$$

(\uppercase\expandafter{\romannumeral2}) Next we show that for any feasible solution, $(\vect{d},\vect{y})$, to Problem \eqref{newop} we can find a corresponding feasible solution, $(\vect{c},l)$, to Problem \eqref{op44}, \yding{such that they achieve the same objective function value}.
By Constraints \eqref{11e}-\eqref{11g}, $y_i$ is decreasing in $i$, and there is at most one $y_i$ which is neither 0 nor 1. Let $y_k$ be the last variable equal to 1, i.e., $y_k=1, y_{k+1}<1$. Then $y_i = 1$ for $i\leq k$, $0\leq y_{k+1} <1$ and $y_i=0 $ for $i \geq k+2 $. \yding{Then, a feasible solution $(\vect{c},l)$ of Problem \eqref{op44} that corresponds to $(\vect{d},\vect{y})$ has the following expression:}
$$c_i=d_{2i-1} \text{ for }i=1,2,\dots,n,\ l=k.  $$
By \eqref{11d}, we have
$$\sum_{i=1}^k d_{2i-1}=\sum_{i=1}^k y_i d_{2i-1} \leq \sum_{i=1}^n y_i d_{2i-1} = 1, \text{ and } \sum_{i=1}^{k+1} d_{2i-1}=\sum_{i=1}^k y_i d_{2i-1}+d_{2k+1}> \sum_{i=1}^n y_i d_{2i-1} = 1.$$
Therefore, 
$$\sum_{i=1}^l c_i=\sum_{i=1}^k d_{2i-1} \leq 1, \text{ and } \sum_{i=1}^{l+1} c_i=\sum_{i=1}^{k+1} d_{2i-1} >1,$$
and thus \eqref{d2},\eqref{d3} are satisfied.

Finally, we show that $f_0(\vect{c},l)=f_1(\vect{d},\vect{y})$. For the numerator, 
$$\sum_{i=1}^n \frac{1}{n c_i} = \sum_{i=1}^n \frac{1}{nd_{2i-1}}=\frac{1}{n}\sum_{i=1}^n (\frac{y_i}{d_{2i-1}}+\frac{1-y_i}{d_{2i}}),$$ and for the denominator, 
$$l+\frac{1-\sum_{i=1}^l c_i}{c_{l+1}}=k+\frac{1-\sum_{i=1}^k y_i d_{2i-1}}{d_{2k+1}}=k+y_{k+1}=\sum_{i=1}^n y_i.$$
\epr

\section{Proof of Claims \ref{psc1} - \ref{c6}}\label{sec:proofclaims16}

\newblock

\begin{repeatclaim}[Claim \ref{psc1}]
$\sum_{i=1}^n y^*_i \in [k,k+1]$. 
\end{repeatclaim}

\bpr
First, the following two solutions are feasible to the original Problem \eqref{newop}:

$(\vect{d}',\vect{y}'): d'_1=d'_2=\dots=d'_{2k}=\frac{1}{k}, d'_{2k+1}=\dots=d'_{2n}=1, y'_1=\dots=y'_k=1, y'_{k+1}=y'_{k+2}=\dots=y'_n=0$.

$(\vect{d}'',\vect{y}''): d''_1=d''_2=\dots=d''_{2k+2}=\frac{1}{k+1}, d''_{2k+3}=\dots=d''_{2n}=1, y''_1=\dots=y''_{k+1}=1, y''_{k+2}=y''_{k+3}=\dots=y''_n=0$.
     
For convenience, we define $g_0(\vect{y})$ and $g_{BFT}(\vect{y})$ as follows: $g_0(\vect{y}):=\min_{\vect{d}} f_1(\vect{d},\vect{y}) \ s.t. \eqref{11b}-\eqref{11g}, g_{BFT}(\vect{y}):=\min_d f_1(\vect{d},\vect{y}) \ s.t. \eqref{11b},\eqref{11d}-\eqref{11g}$. \yding{Therefore,  $g_0(\vect{y})$ and $g_{BFT}(\vect{y})$ give the objective function value of the optimization problem that gives 
our bound and BFT bound for a given $\vect{y}$, respectively. Since the optimization formulation for the BFT bound includes an extra constraint \eqref{11b}, we have} \( g_0(\vect{y}) \geq g_{BFT}(\vect{y}) \geq f_1(\vect{d}, \vect{y}) \). Next, we will show that for any \( \vect{y} \) such that \( \sum_{i=1}^n y_i < k \) or \( \sum_{i=1}^n y_i > k + 1 \), we have \( g_0(\vect{y}) > g_0(\vect{y}') \), so that \( \vect{y} \) is not optimal. Thus, at the optimal solution $(\vect{d}^*,\vect{y}^*)$, we must have $\sum_{i=1}^n y^*_i \in [k,k+1]$. 

Recall that $f_1(\vect{d},\vect{y})=(1/n \sum_{i=1}^n (y_i/d_{2i-1}+(1-y_i)/d_{2i}))/\sum_{i=1}^n y_i$. 
By Inequalities \eqref{p41} and \eqref{p42}, we have 
$$f_1(\vect{d}, \vect{y})\geq \frac{1}{n}( \sum_{i=1}^n y_i+\frac{n}{\sum_{i=1}^n y_i}-1),$$
where equality holds when all $d_{2i-1}$ variables are equal for $i \in \{i:y_i \neq 0\}$ and $d_{2i}=1$ for $i \in \{i:y_i \neq 1\}$. For $\vect{y}'$, these conditions are satisfied by $\vect{d}'$. Thus, $\vect{d}' \in \argmin_{\vect{d}} f_1(\vect{d},\vect{y}')$. Furthermore, $(\vect{d}',\vect{y}')$ satisfies all the constraints of Problem \eqref{newop}, and thus we have
$$g_0(\vect{y}')=g_{BFT}(\vect{y}')=f_1(\vect{d}',\vect{y}')=\frac{1}{n}( \sum_{i=1}^n y'_i+\frac{n}{\sum_{i=1}^n y'_i}-1)=\frac{1}{n}(k+\frac{n}{k}-1).$$
Similarly, we have
$$g_0(\vect{y}'')=g_{BFT}(\vect{y}'')=f_1(\vect{d}'',\vect{y}'')=\frac{1}{n}( \sum_{i=1}^n y''_i+\frac{n}{\sum_{i=1}^n y''_i}-1)=\frac{1}{n}(k+\frac{n}{k+1}). $$

(a) For any \( \vect{y} \) such that \( \sum_{i=1}^n y_i < k \), we have
$$g_0(\vect{y}) \geq g_{BFT}(\vect{y}) \geq \frac{1}{n}( \sum_{i=1}^n y_i+\frac{n}{\sum_{i=1}^n y_i}-1) > \frac{1}{n}( k+\frac{n}{k}-1) = g_{BFT}(\vect{y}') =g_0(\vect{y}'). $$

(b) For any \( \vect{y} \) such that \( \sum_{i=1}^n y_i > k + 1 \), we have
$$g_0(\vect{y}) \geq g_{BFT}(\vect{y}) \geq \frac{1}{n}( \sum_{i=1}^n y_i+\frac{n}{\sum_{i=1}^n y_i}-1) > \frac{1}{n}( k+\frac{n}{k+1}) = g_{BFT}(\vect{y}'') =g_0(\vect{y}''). $$

\epr

\begin{repeatclaim}[Claim \ref{psc2}]
    $y^*_1=y^*_2=\dots=y^*_k=1, y^*_{k+2}=\dots=y^*_n=0.$  
\end{repeatclaim}

\bpr
Since \( y_i \) is decreasing with \( i \), we only need to prove that \( y^*_k = 1 \) and \( y^*_{k+2} = 0 \). Suppose \( y^*_k \neq 1 \), then if \( y^*_k = 0 \), \( y^*_{k+1} = \dots = y^*_n = 0 \) because 
\( y_i \) is decreasing with \( i \); or if \( 0 <y^*_k < 1 \), then \( y^*_{k+1} = \dots = y^*_n = 0 \) as Constraint \eqref{11g} ensures that at most one \( y_i \) can be strictly between 0 and 1. Thus, \( \sum_{i=1}^n y^*_i = \sum_{i=1}^k y^*_i < k \), contradicting Claim \ref{psc1}. Therefore, we have \( y^*_k = 1 \).

Next, we claim that \( y^*_{k+2} = 0 \). To show that, if \( y^*_{k+1} = 0 \), the conclusion holds as \( y_i \) is decreasing with \( i \); if \( 0 < y^*_{k+1} < 1 \), it holds due to Constraint \eqref{11g}; if \( y^*_{k+1} = 1 \), it holds because \( \sum_{i=1}^n y^*_i \) must be no larger than \( k+1 \), as required by Claim \ref{psc1}.
\epr

\begin{repeatclaim}[Claim \ref{c1}]
$c^*_i=1$ for $i=k+2,\dots,n$.
\end{repeatclaim}

\bpr
Suppose, on the contrary, that there exists $l \in \{k+2,\dots,n\}$ such that $c^*_l<1$. Then let $c'_i=c^*_i$ for $i \leq l-1$ and $c'_i=1$ for $i \geq l$. By construction, $\vect{c}'$ satisfies \eqref{12a}. Since the first $k+1$ terms of $\vect{c}'$ and $\vect{c}^*$ coincide, $\vect{c}'$ also satisfies \eqref{12c} and \eqref{12d}, and we conclude that $\vect{c}'$ is feasible for Problem \eqref{newop2}.

Since the objective function \eqref{newop2obj} is a decreasing function of the coefficients $c_l$ for all $l \geq k+2$, $\vect{c}'$ attains a strictly lower objective value than $\vect{c}^*$, which contradicts the optimality of $\vect{c}^*$.
\epr

\newblock

\begin{repeatclaim}[Claim \ref{c2}]
$c_1^*=c_2^*=\dots=c_k^*$. 
\end{repeatclaim}

\bpr
Suppose, on the contrary, Claim \ref{c2} does not hold. Let $c''_i=\left(\sum_{i=1}^k c_i^*\right)/k$ for all $i\leq k$, and let $c''_{i}=c_{i}^*$ for all $i \ge k+1$.  Since $\sum_{i=1}^k c_i^*=\sum_{i=1}^k c''_i$, $\vect{c}''$ is feasible to Problem \eqref{newop2}.

Using Claim \ref{c1}, the objective function in Problem \eqref{newop2} can be reformulated as
\begin{equation*}
    \tilde{g_1}(c_1,\dots,c_{k+1}):=\frac{\sum_{i=1}^{k+1} \frac{1}{c_i}+n-k-1}{k+\frac{1-\sum_{i=1}^k c_i}{c_{k+1}}}.
\end{equation*}
Since the harmonic mean is always no larger than the arithmetic mean, we have
\begin{equation*}
\begin{aligned}
    \tilde{g_1}(c_1^*,\dots,c_{k+1}^*)=\frac{\sum_{i=1}^{k} \frac{1}{c_i^*}+\frac{1}{c_{k+1}^*}+n-k-1}{k+\frac{1-\sum_{i=1}^k c_i^*}{c_{k+1}^*}} &\ge \frac{\frac{k^2}{\sum_{i=1}^k c_i^*}+ \frac{1}{c''_{k+1}}+n-k-1}{k+\frac{1-\sum_{i=1}^k c''_i}{c''_{k+1}}} \\& = \frac{\sum_{i=1}^k \frac{1}{c''_i}+ \frac{1}{c''_{k+1}}+n-k-1}{k+\frac{1-\sum_{i=1}^k c''_i}{c''_{k+1}}} =\tilde{f_1}(c''_1,\ldots,c''_{k+1}),
\end{aligned}
\end{equation*}
where equality holds only when $\vect{c}^*=\vect{c}''$. Therefore, if $\vect{c}^* \ne \vect{c}''$, $\vect{c}''$ yields a strictly lower objective function value than $\vect{c}^*$, 
which contradicts the optimality of $\vect{c}^*$.
\epr

\newblock

\begin{repeatclaim}[Claim \ref{c3}]
$c_{k+1}^*$ is either $\max\{x,1-kx\}$ or $1$.
\end{repeatclaim}

\bpr
Recall that, \td{after removing the constant $1/n$, the objective function \eqref{newop2obj} reduces to: }
\begin{equation*}
    h_1(x,c_{k+1})=\frac{ \frac{k}{x}+\frac{1}{c_{k+1}}+n-k-1}{k+\frac{1-kx}{c_{k+1}}}.
\end{equation*}

The partial derivative of $h_1(x,c_{k+1})$ with respect to $c_{k+1}$ is given by
\begin{equation}\label{hpdc}
    \frac{\partial h_1(x,c_{k+1})}{\partial c_{k+1}}=\frac{1}{(kc_{k+1}+1-kx)^2} \left((n-k-1+\frac{k}{x})(1-kx)-k \right),
\end{equation}
and note that its value is independent of $c_{k+1}$. That is, $\partial h_1(x,c_{k+1})/\partial c_{k+1}$ is either always positive, or always negative for all $c_{k+1}\in \left[\max\{x, 1-kx\},1\right]$. Thus, $h_1(x,c_{k+1})$ attains its minimum at one of the two endpoints of the feasible interval for $c_{k+1}$, implying that $c_{k+1}^*$ is equal to either $\max\{x,1-kx\}$ or $1$.

\epr

\newblock

\begin{repeatclaim}[Claim \ref{c4}]
When $n \geq 3$, $h_1(x,c_{k+1})$ reaches its minimum either at $x=1/(k+1),c_{k+1}=1/(k+1)$ or at $x=1/k,c_{k+1}=1$.
\end{repeatclaim}

\bpr
By Claim \ref{c3}, we know that $c_{k+1}^*$ is either $\max\{x,1-kx\}$ or $1$.

\begin{enumerate}
    \item [(1)] If $c_{k+1}^*=\max\{x,1-kx\}$, then we can characterize $x^*$ as follows.

\begin{enumerate}
    \item [(i)]  If $x\in [1/n,1/(k+1)]$, we have $c_{k+1}^*=1-kx$, and 
\begin{equation*}
    \begin{aligned}
        h_1(x,c_{k+1}^*)=h_1(x,1-kx)=\frac{(\frac{k^2}{kx}+\frac{1}{1-kx})\cdot(kx+1-kx)+n-k-1}{k+1}& \geq \frac{(k+1)^2+n-k-1}{k+1}\\&=k+\frac{n}{k+1},
    \end{aligned}
\end{equation*}
where equality holds when $x=1/(k+1)$.
  \item [(ii)]  If $x\in [1/(k+1),1/k]$, we have $c_{k+1}^*=x$, and 
\begin{equation*}
    \begin{aligned}
        h_1(x,c_{k+1}^*)=h_1(x,x)=k+1+(n-k-1)x\geq k+\frac{n}{k+1},
    \end{aligned}
\end{equation*}
where equality holds when $x=1/(k+1)$.
\end{enumerate}

Thus, in this case, $h_1(x,c_{k+1})$ reaches its minimum at $x=1/(k+1), c_{k+1}=1/(k+1)$.

\item [(2)] If $c_{k+1}^*=1$, then we let $\tilde{h}_1(x):=h_1(x,1)=(k/x+n-k)/(k+1-kx)$. The partial derivative of $\tilde{h}_1(x)$ is given by 
\begin{equation*}
    \tilde{h}_1'(x)=\frac{k(n-k)}{(k+1-kx )^2 x^2} \left( x^2+\frac{2k}{n-k}x-\frac{k+1}{n-k} \right).
\end{equation*}

Note that the sign of $\tilde{h}_1'(x)$ depends on the expression in the last bracket, denoted by $\tilde{h}_0'(x)$. Clearly, $\tilde{h}_0'(x)$ increases with  $x$. Now, we show that $\tilde{h}_0'(1/k)\leq 0$ when $n\geq 4$, and the case $n=3$ will be subsequently considered. Indeed, if it is true, then $\tilde{h}_0'(x) \leq 0$ for all $x  \leq 1/k$, and thus $\tilde{h}_1(x)$ reaches its minimum at $x=1/k$.

Suppose, on the contrary, that $\tilde{h}'_0(1/k) > 0$, which implies that
\begin{equation}
    k^2+n-k > k^3.\label{4.5}
\end{equation}
Let $\overline{x}$ denote the positive root of $\tilde{h}_0'(x)=0$. Then $\tilde{h}_1(x)$ reaches its minimum at $\overline{x}$, and we know that $\overline{x}<1/k$. 

By optimality of the solution $(\overline{x},1)$, the value of $h_1(x,c_{k+1})$ at $x=\overline{x},~c_{k+1}=1$ is smaller than that at $x=1/(k+1),~c_{k+1}=1/(k+1)$, i.e.,
$$\frac{\frac{k}{\overline{x}}+n-k}{k+1-k \overline{x}} \leq \frac{(k+1)^2+n-k-1}{k+1},$$
which implies that
\begin{equation}
\begin{aligned}
    (k^2+k+n) \overline{x}^2-(k+2)(k+1) \overline{x} +k+1 \leq 0.
\end{aligned}
\end{equation}

By \eqref{4.5}, we have
\begin{equation}
\begin{aligned}
    0 &\geq (k^2+k+n) \overline{x}^2-(k+2)(k+1) \overline{x} +k+1\\ &> (k^3+2k) \overline{x}^2 -(k+2)(k+1)\overline{x}+k+1
    \\ &=(k \overline{x}-1)((k^2+2)\overline{x}-k-1)+k(k-2).
\end{aligned}
\label{4.7}
\end{equation}

When $n\geq 4$, we have $k=\lfloor \sqrt{n} \rfloor \geq 2$. Furthermore, since $\overline{x}<1/k$ and $k\geq 2$, it follows that $k \overline{x}-1 < 0, (k^2+2)\overline{x}-k-1 <2/k-1\leq 0$, and $k(k-2) \geq 0$, which contradicts \eqref{4.7}. Therefore, $h_0(1/k) \leq 0$ and thus $x^*=1/k$ when $n\geq 4$. 

We will next show that $n=3$ does not fall under this case, but rather under the previous case, i.e., $c_{k+1}^*=\max\{x,1-kx\}=1/(k+1)$. We have $k=\lfloor \sqrt{3} \rfloor=1$. It is sufficient to show that $h_1(x,1)>h_1(1/2,1/2)=5/2$. Indeed,
$$h_1(x,1)-\frac{5}{2}=\frac{\frac{5}{2}x+\frac{1}{x}-3}{2-x}\geq \frac{\sqrt{10}-3}{2-x}>0.$$
\end{enumerate}
\epr

\newblock

\begin{repeatclaim}[Claim \ref{c6}]
When $n=2$, $c_1^*=\sqrt{3}-1, c_2^*=1$.
\label{sec:clm6}
\end{repeatclaim}

\bpr
    Note that for $n=2, k=\lfloor \sqrt{2} \rfloor=1$. By $\eqref{12a}$, we have $x\geq 1/2$ and $c_2\leq 1$. Thus, the partial derivative of $h_1(x,c_2)$ with respect to $c_2$, as shown in \ref{hpdc}, is non-positive, i.e.,
    \begin{equation*}
    \frac{\partial h_1(x,c_2)}{\partial c_2}=\frac{1-2x}{x(c_2+1-x)^2}\leq 0.
\end{equation*}
Thus, we have $c_2^*=1$. Then,  $$\tilde{h}_1(x)=h_1(x,1)=\frac{1+x}{x(2-x)},$$ whose derivative is:
\begin{equation*}
    \tilde{h}'_1(x)=\frac{x^2+2x-2}{2(2x-x^2)^2}\begin{cases}
    \quad\leq 0, &x \in [\frac{1}{2},\sqrt{3}-1];\\
    \quad > 0, &x \in (\sqrt{3}-1,1].
\end{cases}\\
\end{equation*}
Thus, $\tilde{h}_1(x)$ reaches its minimum at $x=\sqrt{3}-1$ and $\tilde{h}_1(x)=(2+\sqrt{3})/4$, i.e., $c_1^*=\sqrt{3}-1$.

\epr




\newblock

\section{Proof of Lemma \ref{lem:Deltan}\label{sec:Deltan}}

\begin{repeatclaim}[Lemma \ref{lem:Deltan}]
For any $a\in \mathbb{N}^+$, $\Delta(n)$ reaches a local maximum when $n=a(a+1)$.
\end{repeatclaim}

\bpr
For any integer $n \geq 2$, there exists $a\in \mathbb{N}^+$ such that $n\in [a^2,(a+1)^2)$. We will show that when $n\in [a^2,a(a+1)]$, $\Delta(n)$ increases with respect to $n$; and when $n\in [a(a+1),(a+1)^2)$, $\Delta(n)$ decreases with respect to $n$. Therefore, $\Delta(n)$ reaches a local maximum when $n=a(a+1)$. Note that $k=\floor{\sqrt{n}}=a, \epsilon=\sqrt{n}-a$.
\begin{enumerate}[(1)]
    \item When $n\in [a^2,a(a+1))$, by \eqref{delta}, we have $\Delta(n)=(\sqrt{n}-a)^2/(a(n-2\sqrt{n}+1))$, whose derivative is:
    $$\Delta'(n)=\frac{(\sqrt{n}-a)(a-1)}{a\sqrt{n}(\sqrt{n}-1)^3}\geq 0.$$
    
    \item When $n\in [a(a+1),(a+1)^2)$, by \eqref{delta}, we have $\Delta(n)=(1-\sqrt{n}+a)^2/((a+1)(\sqrt{n}-1)^2)$, whose derivative is:
    $$\Delta'(n)=-\frac{(1-\sqrt{n}+a)a}{(a+1)\sqrt{n}(\sqrt{n}-1)^3}< 0.$$
\end{enumerate}

\epr

\section{Proof of Propositions \ref{prop:4}, \ref{proppfworst} and \ref{rmk1}} \label{sec:proofprop689}

\begin{repeatproposition}[Proposition \ref{prop:4}]
Suppose $U \subseteq \{\vect{u}\, |\, 0 \leq u_i \leq L_i, i \in N\}$ is a convex and compact utility set, and $\max_{\vect{u} \in U} u_i=L_i$ for $i \in N$. Then there exist $c_i \in [1/nL_i,1/L_i],~i \in N$, such that $U \subseteq U'=\{\vect{u}\, |\, \sum_{i=1}^n c_i u_i \leq 1, \,0 \leq u_i \leq L_i, \,i \in N \}$, and $\max \{ \sum_{i=1}^n \log{u_i} \,| \,\vect{u} \in U \} = \max \{ \sum_{i=1}^n \log{u_i}\, |\, \vect{u} \in U' \}$. Consequently, $POF(U;PF) \le POF(U';PF)$.
\end{repeatproposition}

\bpr
By definition, $U$ is a subset of a polyhedron with constraints $0\leq u_i \leq L_i, i \in N$. Let $K$ denote a transformation of $U$ such that $K=\{(k_1,k_2,\dots,k_n)\,|\,(k_1 L_1,k_2 L_2,\dots,k_n L_n) \in U\}$. Then $K$ is a subset of the unit cubic, i.e., $K \subseteq [0,1]^n$.

By Proposition \ref{prop:1}, there exist $a_i \in [1/n,1],~ i \in N$, such that $K \subseteq K'=\{k:=(k_1,k_2,\dots,$ $k_n)\,|\,\sum_{i=1}^n a_i k_i \leq 1, 0 \leq k_i \leq 1, i \in N\}$, and $\max \{ \sum_{i=1}^n \log{k_i} \,|\, k\in K \} = \max \{ \sum_{i=1}^n \log{k_i} \,|\, k\in K' \}$.

Finally, we derive the expression of $U'$ corresponding to $K'$. Let $c_i = a_i /L_i \in [1/nL_i,1/L_i],~i \in N$. That is, $U'=\{\vect{u} \,|\, (u_1/L_1,u_2/L_2,\dots,u_n/L_n) \in K'\}=\{\vect{u} \,|\, \sum_{i=1}^n c_i u_i \leq 1, 0 \leq u_i \leq L_i, i \in N \}$. Then, $\max \{ \sum_{i=1}^n \log{u_i} \,|\, \vect{u} \in U \} = \max \{ \sum_{i=1}^n \log{k_i} \,|\, k\in K \} + \sum_{i=1}^n \log{L_i} = \max \{ \sum_{i=1}^n \log{k_i} \,|\, k\in K' \}+ \sum_{i=1}^n \log{L_i}= \max \{ \sum_{i=1}^n \log{u_i} \,|\, \vect{u} \in U' \}$.
\epr

\begin{repeatproposition}[Proposition \ref{proppfworst}]
    $$\sup_{\vect{L}:~1 \ge L_1  \geq ... \geq L_n>0} UB(n,\vect{L};PF) = 1- \frac{1}{n},$$
\yding{where the supremum is approached when $\vect{L}=(1-\epsilon, \epsilon/(n-1), \dots, \epsilon/(n-1))$ with $\epsilon \rightarrow 0$.}
\end{repeatproposition}

\bpr

From Theorem \ref{t2}, $UB(n,\vect{L};PF)$ has the following two possible expressions:
\begin{subequations}
\begin{alignat}{2}
    &UB(n,\vect{L};PF) = 1- \min_{l\in \mathbb{N}} \frac{(\sum_{i=1}^l \sqrt{L_i})^2 +\sum_{i=l+1}^n L_i}{n\sum_{i=1}^l L_i};& & \label{25a} \\
&UB(n,\vect{L};PF) = 1- \frac{(\sqrt{L_2^2+2 L_1 L_2 +(L_1+L_2) \sum_{i=3}^n L_i}+\sqrt{L_1 L_2})^2}{n (L_1+L_2)^2} (&&\text{Only possible when } \label{25b} \\
& &&\sum_{i=1}^{n} L_i \cdot L_{2} > L_1^2). \nonumber
\end{alignat}
\end{subequations}

(1) If $UB(n,\vect{L};PF)$ is given by \eqref{25a}, then
\begin{equation}
    \begin{aligned}
        UB(n,\vect{L};PF) &= 1- \min_{l\in \mathbb{N}} \frac{(\sum_{i=1}^l \sqrt{L_i})^2 +\sum_{i=l+1}^n L_i}{n\sum_{i=1}^l L_i} \leq 1-  \min_{l\in \mathbb{N}} \frac{\sum_{i=1}^l L_i +\sum_{i=l+1}^n L_i}{n\sum_{i=1}^l L_i}\\
        & = 1-  \frac{1}{n} - \min_{l\in \mathbb{N}} \frac{\sum_{i=l+1}^n L_i}{n\sum_{i=1}^l L_i} < 1-\frac{1}{n}.
    \end{aligned}
\end{equation}

(2) If $UB(n,\vect{L};PF)$ is given by \eqref{25b}, then
\begin{equation}
    \begin{aligned}
        UB(n,\vect{L};PF) &= 1- \frac{(\sqrt{L_2^2+2 L_1 L_2 +(L_1+L_2) \sum_{i=3}^n L_i}+\sqrt{L_1 L_2})^2}{n (L_1+L_2)^2}  \\
        &= 1-  \frac{(\sqrt{ (L_1+L_2) \sum_{i=1}^n L_i -L_1^2 }+\sqrt{L_1 L_2})^2}{n (L_1+L_2)^2}  \\
        &< 1-  \frac{(\sqrt{ (L_1+L_2) \frac{L_1^2}{L_2}  - L_1^2 }+\sqrt{L_1 L_2})^2}{n (L_1+L_2)^2} \qquad (\text{By } \sum_{i=1}^{n} L_i \cdot L_{2} > L_1^2) \\
        & = 1-   \frac{ ( L_1 \sqrt{  \frac{L_1}{L_2} }+\sqrt{L_1 L_2})^2}{n (L_1+L_2)^2}\\ & < 1-   \frac{ ( L_1 +L_2)^2}{n (L_1+L_2)^2} =1-\frac{1}{n}.
    \end{aligned}
\end{equation}

Thus, for any vector $\vect{L}$ of  maximum achievable utilities, $UB(n,\vect{L};PF) < 1-1/n$.  We next prove that $\sup_{\vect{L}} UB(n,\vect{L};PF) = 1 - 1/n$.

Indeed, we will show that for any positive $\delta$ and a number smaller than $1 - 1/n - \delta$, there exists a vector of maximum achievable utilities $\vect{L}^*$, for which $UB(n,\vect{L}^*;PF)>1 -1/n - \delta$. Specifically, let $L^*_1 :=1-\epsilon$, and $L^*_i := \epsilon/(n-1),  i = 2,3 ..., n$, where $\epsilon$ is sufficiently small (e.g., $\epsilon \leq (2n-1-\sqrt{4n-3})/(2(n-1))$) so that $\sum_{i=1}^{n} L^*_i \cdot L^*_{2} \leq L_1^{*2}$. Then, for any given $\delta>0$, by choosing $\epsilon = \min\{n \delta/(2+n \delta),(2n-1-\sqrt{4n-3})/(2(n-1))\}$, we have

\begin{equation*}
    \begin{aligned}
        UB(n,\vect{L}^*;PF)&=1- \min_{l\in \mathbb{N}} \frac{(\sqrt{1-\epsilon}+(l-1)\sqrt{\frac{\epsilon}{n-1}})^2 +(n-l) \frac{\epsilon}{n-1}}{n(1-\epsilon+(l-1) \epsilon)}\\ &\geq 1-  \frac{1}{n(1-\epsilon) }\geq 1 - \frac{1}{n} - \frac{\delta}{2} > 1-\frac{1}{n}-\delta.
    \end{aligned}
\end{equation*}

\epr

 \begin{repeatproposition}[Proposition \ref{rmk1}]

    $$\sup_{\vect{L}:~1 \ge L_1  \geq ... \geq L_n>0} BFT(n,\vect{L};PF) = 1-\frac{1}{n}+\frac{\left(1-\sqrt{\frac{2\sqrt{n}-1}{n}}\right)^2}{n-1}=1-O(\frac{1}{n^{1.25}}),$$
where that supremum is reached when 
\begin{equation*}
\vect{L}=\left(\sqrt{\delta}, \frac{1-\sqrt{\delta}}{n-1},\dots, \frac{1-\sqrt{\delta}}{n-1}\right) \text{ and } \delta=\frac{2\sqrt{n}-1}{n}.
\end{equation*}
\end{repeatproposition}

\bpr

From BFT (2011), we have
\begin{equation*}\label{pfbft}
    BFT(n,\vect{L};PF)=1-\frac{2\sqrt{n}-1}{n}\frac{L_n}{L_1}-\frac{1}{n}+\frac{L_n}{\sum_{i=1}^n L_i}.
\end{equation*}

Note that we can assume, without loss of generality that $\sum_{i=1}^n L_i=1$, since for a normalized vector $\vect{L}'$ with $\sum_{i=1}^n L'_i=1, BFT(n,\vect{L};PF)=BFT(n,\vect{L}';PF)$. For convenience, define $\delta:=(2\sqrt{n}-1)/n$. Then, we have

$$BFT(n,\vect{L};PF)=1-\frac{1}{n}+(1-\frac{\delta}{L_1})L_n.$$

By definition, since $L_n$ is the smallest component of $\vect{L}$, the average of $L_2, L_3,\dots, L_{n-1}$ is greater than or equal to $L_n$, i.e.,
$$\frac{1-L_1-L_n}{n-2}\geq L_n \Rightarrow L_1\leq 1-(n-1)L_n,$$

which implies that
\begin{equation}\label{sl1}
    BFT(n,\vect{L};PF)=1-\frac{1}{n}+(1-\frac{\delta}{L_1})L_n \leq 1-\frac{1}{n}+(1-\frac{\delta}{1-(n-1)L_n})L_n.
\end{equation}

To find the maximum of the right hand side of \eqref{sl1}, we use the first order condition, which yields, at optimality, that
$$L_1=1-(n-1)L_n=\sqrt{\delta}.$$

In addition, observe that the sum of middle $L_i$'s, i.e., $\sum_{i=2}^{n-1} L_i$ is exactly equal to $(n-2)L_n$, implying, since $L_n$ is the smallest element, that $L_2=L_3=\dots=L_n$.

We next verify that $L_1 \geq L_2=L_3=\dots=L_n$. Indeed,
$$L_1=\sqrt{\delta}\geq \frac{1-\sqrt\delta}{n-1} \Leftrightarrow n(2\sqrt{n}-1) \geq 1,$$
where the second inequality is valid for any $n\geq 1$.

Thus, the vector $\vect{L}$ derived above is feasible and for which
$$BFT(n,\vect{L};PF)=1-\frac{1}{n}+\frac{\left(1-\sqrt{\frac{2\sqrt{n}-1}{n}}\right)^2}{n-1}. $$

Finally, we have
\begin{equation*}
    \begin{aligned}
        \frac{1}{n}-\frac{\left(1-\sqrt{\frac{2\sqrt{n}-1}{n}}\right)^2}{n-1}&=\frac{-2\sqrt{n}+2n\sqrt{\frac{2\sqrt{n}-1}{n}}}{n(n-1)}=\frac{-2n^{-0.25}+2\sqrt{2-n^{-0.5}}}{n^{0.25}(n-1)}\\
        &\leq \frac{2\sqrt{2}}{n^{0.25}(n-1)}\leq \frac{4\sqrt{2}}{n^{1.25}}=O(\frac{1}{n^{1.25}}).
    \end{aligned}
\end{equation*}
The last inequality holds when $n \geq 2$. Thus,
$$\sup_{\vect{L}:L_i \in (0,1], L_1  \geq \cdots \geq L_n} BFT(n,\vect{L};PF) = 1-\frac{1}{n}+\frac{\left(1-\sqrt{\frac{2\sqrt{n}-1}{n}}\right)^2}{n-1}=1-O(\frac{1}{n^{1.25}}).$$
\epr

\section{Proof of Claims \ref{claim:cstaruneq} - \ref{clmfeasible}}\label{sec:proofclaims712}

\begin{repeatclaim}[Claim \ref{claim:cstaruneq}]
$\overline{c}_i=1/L_i$, for $i=l+2,\dots,n.$
\end{repeatclaim}

\bpr

Recall that the objective function \eqref{pofpfueobj} is
$$\frac{\sum_{i=1}^n \frac{1}{n c_i}}{\sum_{i=1}^{l} L_i + \frac{1-\sum_{i=1}^{l} c_i L_i}{c_{l+1}}}.$$

Note that $\overline{c}_i=1/L_i,i=l+2,\dots,n$, is feasible for Problem \eqref{pofpfue}, and since  $c_i,i=l+2,\dots,n$, only appear in the  nominator of the objective function, \eqref{pofpfueobj}, which is to be minimized, at optimality, they should attain their upper bounds. 


\epr

\begin{repeatclaim}[Claim \ref{clm:cstartol}]
    $\overline{c}_1 \sqrt{L_1}=\overline{c}_2 \sqrt{L_2}=\dots=\overline{c}_l \sqrt{L_{l}}$.
\end{repeatclaim}
\bpr

By Cauchy–Schwarz inequality, we have
\begin{equation}
    \sum_{i=1}^{l} \frac{1}{c_i} \geq \frac{(\sum_{i=1}^{l} \sqrt{L_i})^2}{\sum_{i=1}^{l} c_i L_i}, 
    \label{cauchy}
\end{equation}
where equality holds when $c_1^2 L_1=c_2^2 L_2=\dots=c_{l}^2 L_{l}$. When adjusting the values of $c_i$ for $i=1,2,\dots,l$, while keeping $\sum_{i=1}^l c_i L_i$ constant, the denominator of the objective function \eqref{pofpfueobj} remains invariant, and the numerator of the objective function \eqref{pofpfueobj} reaches its minimum when the equality in \eqref{cauchy} is satisfied. So, at optimality, $\overline{c}_1 \sqrt{L_1}=\overline{c}_2 \sqrt{L_2}=\dots=\overline{c}_l \sqrt{L_{l}}$.

\epr

\begin{repeatclaim}[Claim \ref{clmx}]
 $x^*=1/L_{l^*+1}$.
\end{repeatclaim}

\bpr
Recall that 
\begin{equation*}
    f_2(x,y,l;\vect{L})=\frac{\frac{1}{y} A(l)+ \frac{1}{x}+B(l)}{M(l) + \frac{1-y A(l)}{x}}, \quad A(l)=\sum_{i=1}^l \sqrt{L_i}, \quad B(l)= \sum_{i=l+2}^n L_i, \quad M(l)=\sum_{i=1}^l L_i.
\end{equation*}

Then, the partial derivative of $f_2(x,y,l;\vect{L})$ with respect to $x$, is:

\begin{equation*}
\begin{aligned}
    \frac{\partial f_2(x,y,l;\vect{L})}{\partial x}=\frac{\frac{1}{y}A(l) -y A(l) B(l)-M(l)+B(l)-(A(l))^2}{(xM(l) + 1-y A(l))^2}.
\end{aligned}
\end{equation*}


Note that since, by definition, $A(l)$ and $B(l)$ are positive, the numerator of $\partial f_2(x,y,l;\vect{L}) / \partial x$, which is independent of $x$, strictly decreases with $y$. Moreover, $\partial f_2(x,y,l;\vect{L}) / \partial x>0$ when $y \rightarrow 0^+$ and  $\partial f_2(x,y,l;\vect{L}) / \partial x<0$ when $y=1/A(l)$. As a result, there exists $\overline{y}(l) \in (0, 1/A(l))$ at which $\partial f_2(x,y,l;\vect{L}) / \partial x=0$. Thus, $\partial f_2(x,y,l;\vect{L}) / \partial x > 0$ when $y \in (0, \overline{y}(l))$, and $\partial f_2(x,y,l;\vect{L}) / \partial x \leq 0$ when $y \in [\overline{y}(l), 1/A(l)]$. Next we prove that at optimality, $y^* \in [\overline{y}(l^*), 1/A(l^*)]$. Therefore, the objective function is minimized when $x$ reaches its upper limit, $1/L_{l^*+1}$, which proves the claim.

We prove that $y^* \in [\overline y(l^*), 1/A(l^*)]$ by contradiction. Suppose, on the contrary, that $y^* \in (0,\overline{y}(l^*))$, and let us focus on $y\in (0,\overline{y}(l^*))$, for which we have that $\partial f_2(x,y,l^*;\vect{L}) / \partial x > 0$. Since, by \eqref{yx1}, $x>(1-y A(l^*))/L_{l^*+1}$, we have
$$
f_2(x,y,l^*;\vect{L})>f_2(\frac{1-yA(l^*)}{L_{l^*+1}},y,l^*;\vect{L})=\frac{B(l^*)+\frac{A(l^*)}{y}+\frac{L_{l^*+1}}{1-y A(l^*)}}{M(l^*+1)}.
$$
By Cauchy–Schwarz inequality, we have
\begin{equation*}
    \begin{aligned}
        \frac{A(l^*)}{y}+\frac{L_{l^*+1}}{1-y A(l^*)}=\left(\frac{A(l^*)^2}{yA(l)}+\frac{L_{l^*+1}}{1-y A(l^*)}\right) \cdot (yA(l^*) + 1-y A(l^*)) &\geq (A(l^*)+\sqrt{L_{l^*+1}})^2\\
        &=A(l^*+1)^2.
    \end{aligned}
\end{equation*}

Thus, 
\begin{equation*}
        \begin{aligned}
        f_2( \frac{1-yA(l^*)}{L_{l^*+1}},y,l^*;\vect{L}) &\geq  \frac{B(l^*)+(A(l^*+1))^2}{M(l^*+1)} \\ & = \frac{(A(l^*+1))^2+L_{l^*+2}+B(l^*+1)}{M(l^*+1)}= f_2(\frac{1}{L_{l^*+2}},\frac{1}{A(l^*+1)}, l^*+1;\vect{L}).
        \end{aligned}
\end{equation*}

Thus, as long as $y^* \in (0, \overline{y}(l^*))$, for all feasible $x$, we have $f_2(x,y,l^*;\vect{L})>f_2((1-yA(l^*)) /L_{l^*+1},$ $ y, l^*;\vect{L}) \geq f_2(1/L_{l^*+2}, 1/A(l^*+1), l^*+1;\vect{L})$. Therefore, any $l^* \le n-2$ is sub-optimal as a lower objective function value can be attained by replacing $l^*$ with $l^*+1$, contradicting the optimality of $l^*$. We conclude that the only possible case for $y^*\in (0,\overline{y}(l^*))$ is when $l^*=n-1$, since, in this case, replacing $l^*$ with $l^*+1$, would lead to $l^*=n$, which is infeasible. 

Next, we prove that when $l^*=n-1$, $y^*\in (0,\overline{y}(l^*))$ is also impossible.

Since $B(l)=\sum_{l+2}^n L_i$, we have $B(n-1)=0$, and $\partial f_2(x,y,n-1;\vect{L})/\partial x= (A(n-1)/y-M(n-1)-(A(n-1))^2 )/(xM(n-1)+1-yA(n-1))^2$. Thus, $\overline{y}(n-1)$, the solution to $\partial f_2(x,y,n-1;\vect{L})/\partial x=0$, has the following expression,
    $$\overline{y}(n-1)=\frac{A(n-1)}{(A(n-1))^2+M(n-1)}.$$

Thus, for $y \in [0,\overline{y}(n-1)]$, we have $f_2(x,y,n-1;\vect{L})>f_2((1-yA(n-1))/L_n,y,n-1;\vect{L})$. The partial derivative of $f_2((1-yA(n-1))/L_n,y,n-1;\vect{L})$ with respect to $y$ has the following form,
$$\frac{\partial f_2(\frac{1-yA(n-1)}{L_n},y,n-1;\vect{L})}{\partial y}=\frac{A(n-1) \cdot(A(n)y-1)\cdot((\sqrt{L_n}-A(n-1))y+1)}{M(n)\cdot y^2\cdot (A(n-1)\cdot y-1)^2},$$
and it is negative when $y \in (0,1/A(n))$, and it is positive when $y \in (1/A(n),1/A(n-1)]$. Thus, $f_2((1-yA(l^*))/L_{l^*+1},y,l^*;\vect{L})$ decreases in $y$ when $y \in (0,1/A(n)]$. We observe that
    $$\overline{y}(n-1)=\frac{A(n-1)}{(A(n-1))^2+M(n-1)}\leq \frac{A(n-1)}{(A(n-1))^2+A(n-1) \sqrt{L_n}}=\frac{1}{A(n-1)+\sqrt{L_n}}=\frac{1}{A(n)}.$$

Thus, $f_2((1-yA(n-1))/L_{n},y,n-1;\vect{L})$ decreases over $[0,\overline{y}(n-1)]$. Consequently,
   \begin{equation*}
        \begin{aligned}
            f_2(x,y,n-1;\vect{L})&>f_2((1-y A(n-1))/L_{n},y,n-1;\vect{L}) \\ 
            &\ge f_2((1-\overline{y}(n-1) \cdot A(n-1))/L_{n},\overline{y}(n-1),n-1;\vect{L})\\
            &=\frac{(A(n-1))^2+M(n-1)}{M(n-1)} \geq\frac{(A(n-1))^2+L_n}{M(n-1)}=f_2(1/L_n,1/A(n-1),n-1;\vect{L}),
        \end{aligned}
    \end{equation*}
and we conclude that, given that $l^*=n-1$, the objective function value for any feasible pair $(x,y)$ such that $y\in (0,\overline{y}(n-1))$ is greater than that at $x=1/L_n, y=1/A(n-1)$, which implies that $y^*\in (0,\overline{y}(l^*))$ is impossible. 
Therefore, $y^* \ge \overline{y}(l^*)$ and $x^*=1/L_{l^*+1}$.
\epr

\begin{repeatclaim}[Claim \ref{clmy}]
(1) $y^*=1/A(l^*)$ when $l^*= 1$ with $\sum_{i=1}^n L_i \cdot L_{2} \leq L_1^2$, or when $l^* \geq 2$. \\ (2) $y^*=\tilde{y}:=(-\sqrt{L_1}+\sqrt{B(0) + L_1+ B(0)L_1/L_2})/B(0)$ when $l^*=1$ with $ \sum_{i=1}^n L_i \cdot L_{2} > L_1^2$.
\end{repeatclaim}

\bpr

According to Claim \ref{clmx}, $x^*=1/L_{l^*+1}$, and thus, the objective function of Problem \eqref{minxy} can be written as:
\begin{equation*}
g_2(y,l):=f_2(\frac{1}{L_{l+1}},y,l; \vect{L})=
    \frac{\frac{A(l)}{y}+B(l-1)}{M(l+1)-y A(l) L_{l+1}}.
\end{equation*}

The partial derivative of $g_2(y,l)$ with respect to $y$ has the following form:
$$\frac{\partial g_2(y,l)}{\partial y}=\frac{A(l)}{y^2 (M(l+1)-A(l) L_{l+1} y)^2}\left((B(l)L_{l+1}+L_{l+1}^2)y^2+2 A(l) L_{l+1} y-M(l+1)\right).$$
Let $\tilde{y}$ denote the positive root of $\partial g_2(y,l)/\partial y = 0$. Then by the quadratic formula, 
$$\tilde{y}=\frac{-A(l) L_{l+1}+\sqrt{ (A(l))^2 L_{l+1}^2 + L_{l+1} M(l+1) B(l-1)}}{B(l-1)L_{l+1}},$$
and the above partial derivative $\partial g_2(y,l)/\partial y$ is negative when $y < \tilde{y}$, and it is positive when $y > \tilde{y}$. Thus, if $\tilde{y}\geq 1/A(l)$, we have, by \eqref{15c}, that for all feasible $y$, $y \leq 1/A(l)$, $\partial g_2(y,l)/\partial y \leq 0$, and thus $g_2(y,l)$ attains its minimum at $y=1/A(l)$. If, on the other hand, $\tilde{y} < 1/A(l)$, then $\partial g_2(y,l)/\partial y \leq 0$ for $y\in (0,\tilde{y}]$, and $\partial g_2(y,l)/\partial y > 0$ for $y\in (\tilde{y},1/A(l)]$, implying that the objective function attains its minimum at $y=\tilde{y}$. Thus, we have $y^* = \min\{\tilde y, 1/A(l^*)\}$.

Recall that $A(l)=\sum_{i=1}^l \sqrt{L_i}, M(l)=\sum_{i=1}^l L_i, B(l)=\sum_{i=l+2}^n L_i$. 

If $l^*=1$ with $\tilde{y} < 1/A(1) \Leftrightarrow \sum_{i=1}^n L_i \cdot L_{2} > L_1^2$, which proves Case (2) of the claim.

If $l^*=1$ with $\tilde{y} \geq 1/A(1) \Leftrightarrow  \sum_{i=1}^n L_i \cdot L_{2} \leq L_1^2$, which proves Case (1) of the claim for $l^*=1$.

When $l^* \ge 2$, $\tilde{y}\geq 1/A(l^*) \Leftrightarrow (A(l^*))^2 \cdot M(l^*) \geq [(A(l^*))^2+B(l^*-1)] \cdot L_{l^*+1}$. We next prove this inequality by contradiction, which would imply that $y^*=1/A(l^*)$ when $l^* \ge 2$.

So, suppose, on the contrary, that
\begin{equation}
    (A(l^*))^2 \cdot M(l^*) < [(A(l^*))^2+B(l^*-1)] \cdot L_{l^*+1}, \label{3.20}
\end{equation}
then $\tilde{y}<1/A(l^*)$, and we have $y^*=\tilde{y}$.

Since, by definition, the objective function, $g_2(y,l)$, attains its minimum at $y=y^*,l=l^*$, its value at this point is smaller than or equal to its value at $y=1/A(l^*+1),l=l^*+1$, i.e.,
\begin{equation*}
    \frac{\frac{A(l^*)}{y^*}+B(l^*-1)}{M(l^*+1)-y^* A(l^*)L_{l^*+1}} \leq \frac{(A(l^*+1))^2+B(l^*)}{M(l^*+1)}
\end{equation*}
\begin{equation*}
\begin{aligned}
    \Rightarrow [(A(l^*))^2+2 A(l^*) \sqrt{L_{l^*+1}}+B(l^*-1)]L_{l^*+1} &y^{*2}  \\-(A(l^*)+2\sqrt{L_{l^*+1}})\cdot M(l^*+1) \cdot &y^*+M(l^*+1)\leq 0.
\end{aligned}
\end{equation*}

By \eqref{3.20}, we have
\begin{equation}
\begin{aligned}
    0 \geq &[(A(l^*))^2+2 A(l^*) \sqrt{L_{l^*+1}}+B(l^*-1)]L_{l^*+1} y^{*2} -(A(l^*)+2\sqrt{L_{l^*+1}})\cdot M(l^*+1) \cdot y^*\\& +M(l^*+1)\\ > & [(A(l^*))^2 M(l^*) + 2 A(l^*) L_{l^*+1}\sqrt{L_{l^*+1}}] y^{*2}-(A(l^*)+2\sqrt{L_{l^*+1}})M(l^*+1)\cdot y^*+M(l^*+1) \\
    =& (A(l^*) y^*-1)[(A(l^*)M(l^*)+2L_{l^*+1} \sqrt{L_{l^*+1}})y^*-M(l^*+1)]+(A(l^*)-2\sqrt{L_{l^*+1}}) M(l^*).
\end{aligned}
\label{exception}
\end{equation}

We will next show that the bottom expression in \eqref{exception}, denoted by $Q$, is positive, which reveals that \eqref{exception} does not hold. Consider the first term in $Q$, which consists of the product of two terms, say $P_1$ and $P_2$. By supposition, we know $\tilde{y}=y^* < 1/A(l^*)$. Thus, $P_1=A(l^*) y^*-1 < 0$. Next, consider the term $P_2$. We have $P_2=(A(l^*)M(l^*)+2L_{l^*+1} \sqrt{L_{l^*+1}})y^*-M(l^*+1) <(A(l^*) y^*-1)M(l^*)+2 L_{l^*+1} \sqrt{L_{l^*+1}} y^*-L_{l^*+1} <2 L_{l^*+1} \sqrt{L_{l^*+1}} /A(l^*)-L_{l^*+1} = L_{l^*+1}(2 \sqrt{L_{l^*+1}}-A(l^*))/A(l^*)$.

When $l^*\geq 2$, $2 \sqrt{L_{l^*+1}}-A(l^*) \leq 0.$ Thus, the first product in the right-hand side of Equation \eqref{exception} is positive. Also the second product is non-negative, which means \eqref{exception} does not hold. We conclude that, \eqref{3.20} does not hold, i.e., $(A(l^*))^2 \cdot M(l^*) \geq [(A(l^*))^2+B(l^*-1)] \cdot L_{l^*+1}$.

\epr

Recall that $$h_2(l):=f_2(\frac{1}{L_{l+1}},\frac{1}{A(l)},l;\vect{L})=\frac{(A(l))^2+L_{l+1}+B(l)}{M(l)}=\frac{(A(l))^2+B(l-1)}{M(l)}.$$

\begin{repeatclaim}[Claim \ref{clml}]
$h_2(l)$ is unimodal, i.e., there exists $\ell\in [1,n-1]$ such that $h_2(l)$ is decreasing for $l \in [1,\ell]$, increasing for $l \in [\ell,n-1]$, and 
\begin{equation*}
\ell =\begin{cases}
1,  & \mbox{ if } S^P_1 =\varnothing, \\
\max(S^P_1),   & \mbox{ otherwise },
    \end{cases}
    \end{equation*}
where $S^P_1:=\{l\in \mathbb{N}^+\cap [2,n-1]\,|\, \sqrt{L_{l}}[(\sum_{i=1}^l \sqrt{L_i})^2+\sum_{i=l+1}^n L_i] \geq 2 \sum_{i=1}^l L_i \cdot \sum_{i=1}^{l-1} \sqrt{L_i}\}$. 
\end{repeatclaim}

\bpr
Since $l \in \{1, 2, \dots, n-1\}$, there exists an $\ell$ minimizing $h(l)$. If there are multiple values of $l$ that minimize $h(l)$, we define $\ell$ to be the largest such value. Thus, for that $\ell$, we have
\begin{numcases}{}
    h_2(\ell) < h_2(\ell+1),\label{in} \\
    h_2(\ell)\leq h_2(\ell-1).\label{de}
\end{numcases}

From \eqref{in}, we have
\begin{equation*}
    \begin{aligned}
         \frac{(A(\ell))^2+B(\ell-1)}{M(\ell)} &< \frac{(A(\ell+1))^2+B(\ell)}{M(\ell+1)},\\
        \Rightarrow (M(\ell)+L_{\ell+1})[(A(\ell))^2+B(\ell-1)] &< M(\ell)[(A(\ell)+\sqrt{L_{\ell+1}})^2+B(\ell-1)-L_{\ell+1}],\\
        \Rightarrow \qquad \quad \, \sqrt{L_{\ell+1}}[(A(\ell))^2+B(\ell-1)] &< 2 A(\ell) M(\ell).
    \end{aligned}
\end{equation*}

Then we have
\begin{equation*}
\begin{aligned}
     h_2(\ell+2)-h_2(\ell+1)&=\frac{\sqrt{L_{\ell+2}}}{M(\ell+1) \cdot M(\ell+2)} \cdot   \left\{2A(\ell) M(\ell) +  2A(\ell) L_{\ell+1}+2 M(\ell+1) \sqrt{L_{\ell+1}} \right. \\& \qquad \qquad \qquad \quad \qquad \qquad \left. -[(A(\ell))^2+2 A(\ell) \sqrt{L_{\ell+1}}+B(\ell-1)]\sqrt{L_{\ell+2}}\right\}\\
    &  >  \frac{\sqrt{L_{\ell+2}}}{M(\ell+1) \cdot M(\ell+2)} \cdot \left \{[(A(\ell))^2+B(\ell-1)+2A(\ell) \sqrt{L_{\ell+1}}]\cdot \right. \\&\qquad \qquad \qquad \qquad \qquad \quad \left. (\sqrt{L_{\ell+1}}-\sqrt{L_{\ell+2}})+ 2 M(\ell+1) \sqrt{L_{\ell+1}}\right\}\\
    &\geq  0.
\end{aligned}
\end{equation*}

Similarly, from \eqref{de}, we have
\begin{equation*}
    \begin{aligned}
        \frac{(A(\ell))^2+B(\ell-1)}{M(\ell)} &\leq \frac{(A(\ell-1))^2+B(\ell-2)}{M(\ell-1)}, \\
        \Rightarrow  (M(\ell)-L_{\ell})[(A(\ell))^2+B(\ell-1)] &\leq M(\ell)[(A(\ell)-\sqrt{L_{\ell}})^2+B(\ell-1)+L_{\ell}],
    \end{aligned}
\end{equation*}
\begin{equation}
         \  \Rightarrow \qquad \quad \   \sqrt{L_{\ell}}[(A(\ell))^2+B(\ell-1)] \geq 2 M(\ell) A(\ell-1). \qquad \qquad \qquad \  \label{ec9}
\end{equation}

Then we have
\begin{equation*}
\begin{aligned}
     h_2(\ell-2)-h_2(\ell-1)&=\frac{\sqrt{L_{\ell-1}}}{M(\ell-1) \cdot M(\ell-2)}\cdot \left\{\sqrt{L_{\ell-1}}[(A(\ell))^2-2 A(\ell) \sqrt{L_{\ell}}+2 L_{\ell}+B(\ell-1)]\right.\\&\left. \qquad \qquad\qquad \qquad\qquad \quad+ (2\sqrt{L_{\ell-1}}  - 2A(\ell-1))M(\ell-1) \right\}  \\
      & \geq \frac{\sqrt{L_{\ell-1}}}{M(\ell-1) \cdot M(\ell-2)} \cdot \left[2A(\ell-1)\cdot (\sqrt{L_{\ell-1}}-\sqrt{L_{\ell}})\cdot \frac{M(\ell-1)}{\sqrt{L_{\ell}}}\right.\\  & \left. \qquad \qquad \qquad \qquad\qquad \quad + 2\sqrt{L_{\ell-1}}\cdot M(\ell-1) \right]\\
      &\geq  0.
\end{aligned}
\end{equation*}

Thus, $h_2(l)$ is unimodal. To characterize the minimizer $\ell$, we define $S_1^P$ as the set of $l$’s such that $h_2(l)$ is no larger than $h_2(l-1)$, i.e., $S^P_1:=\{l\in \mathbb{N}^+\cap [2,n-1]\,|\, h_2(l)\leq h_2(l-1)\}=\{l\in \mathbb{N}^+\cap [2,n-1]\,|\, \sqrt{L_{l}}[(\sum_{i=1}^l \sqrt{L_i})^2+\sum_{i=l+1}^n L_i] \geq 2 \sum_{i=1}^l L_i \cdot \sum_{i=1}^{l-1} \sqrt{L_i}\}$, where the last expression follows from \eqref{ec9}. Then, we have
\begin{equation*}
\ell =\begin{cases}
1,  & \mbox{ if } S^P_1 =\varnothing, \\
\max(S^P_1),   & \mbox{ otherwise.}
    \end{cases}
    \end{equation*}

\epr

\begin{repeatclaim}[Claim \ref{clmfeasible}]
$(x^*,y^*,l^*)$, as characterized in Claims \ref{claim:cstaruneq}-\ref{clml}, is feasible and thus optimal to Problem \eqref{pb2}.
\end{repeatclaim}

\bpr
It suffices to show that the optimal solution $(x^*,y^*,l^*)$ to Problem \eqref{minxy} satisfies the two constraints in Problem \eqref{pb2} that were removed from Problem \eqref{minxy}, that is,
\begin{subequations}
\begin{alignat}{2}
&\frac{y^*}{\sqrt{L_{l^*}}} \leq x^*, \label{18a}\\
        &\frac{1}{n \sqrt{L_{l^*}}} \leq y^*. \label{17b}
\end{alignat}
\end{subequations}

When $l^*= 1$ with $(B(0)+L_1)L_{2} \leq L_1^2$, or when $l^* \geq 2$, by Claims \ref{clmx} and \ref{clmy}, the optimal solution for Problem \eqref{minxy} is $x^*=1/L_{l^*+1}, y^*=1/A(l^*)$. Since $A(l^*)\sqrt{L_{l^*}} \geq L_{l^*} \geq L_{l^*+1}$, \eqref{18a} holds. 

Let us consider \eqref{17b}. Then, we have $A^2(l^*)=M(l^*)+2 \sum_{1\leq i < j \leq l^*} \sqrt{L_i L_j} \leq l^* M(l^*)$, and by \eqref{ec9},
\begin{equation*}
    \begin{aligned}
        \frac{A(l^*-1)}{\sqrt{L_{l^*}}} &\leq \frac{A^2(l^*)+B(l^*-1)}{2 M(l^*)} \leq \frac{l^* M(l^*)+\frac{n-l^*}{l^*}M(l^*)}{2M(l^*)}\leq \frac{1}{2}(l^*+\frac{n}{l^*}-1)\leq n-1,\\
        &\Rightarrow A(l^*) \leq n\sqrt{L_{l^*}},\qquad \Rightarrow \frac{1}{n \sqrt{L_{l^*}}} \leq y^*.
    \end{aligned}
\end{equation*}

Thus, \eqref{17b} holds.

When $l^*= 1$ with $ \sum_{i=1}^n L_i \cdot L_{2} > L_1^2$, by Claims \ref{clmx} and \ref{clmy}, the optimal solution for Problem \eqref{minxy} is $x^*=1/L_2, y^*=(-\sqrt{L_1}+\sqrt{B(0) + L_1+ B(0)L_1/L_2})/B(0)$. Since $y^* \leq 1/\sqrt{L_1} \leq \sqrt{L_1}/L_2$, \eqref{18a} holds. For \eqref{17b}, since $B(0)\leq (n-1)L_2$, we have
\begin{equation*}
    \begin{aligned}
        y^*&=\frac{-\sqrt{L_1}+\sqrt{B(0)+L_1+B(0)\frac{L_1}{L_2}}}{B(0)}=\frac{1+\frac{L_1}{L_2}}{\sqrt{B(0)+L_1+B(0)\frac{L_1}{L_2}}+\sqrt{L_1}}\\
        &\geq \frac{1+\frac{L_1}{L_2}}{\sqrt{(n-1)L_2+n L_1}+\sqrt{L_1}}=\frac{1}{\sqrt{L_1}}\cdot \frac{1+\frac{L_1}{L_2}}{\sqrt{(n-1)\frac{L_2}{L_1}+n}+1}\geq \frac{1}{n \sqrt{L_1}},
    \end{aligned}
\end{equation*}
so \eqref{17b} holds.
\epr


\section{Proof of Corollary \ref{crl1}}\label{sec:prooflemma2}

\begin{repeatcorollary}[Corollary \ref{crl1}]
When the players have equal maximum achievable utilities, the bound in Theorem \ref{t2} reduces to the bound in Theorem \ref{t1}. 
\end{repeatcorollary}

\bpr
\discuss{To see that, let $L_i \equiv 1$, for $i \in N$. In this case, $\sum_{i=1}^{n} L_i \cdot L_{2} - L_1^2  =n-1 >0$. Then, by Theorem \ref{t2}, we have
\begin{equation}
    \begin{aligned}
        UB(n,\vect{1};PF)&= 1- \min  \bigg\{  \frac{(\sum_{i=1}^{\ell} \sqrt{L_i})^2 +\sum_{i=\ell+1}^n L_i}{n \sum_{i=1}^{\ell} L_i}, \\ & \qquad \qquad \quad \ \frac{(\sqrt{L_2^2+2 L_1 L_2 +(L_1+L_2) \sum_{i=3}^n L_i}+\sqrt{L_1 L_2})^2}{n (L_1+L_2)^2}  \bigg\}\\
        &=1-\frac{1}{n} \min  \bigg\{   \frac{\ell^2 +n-\ell}{\ell},  \frac{\sqrt{2n-1}+n}{2} \bigg\},
        \label{e2}
    \end{aligned}
\end{equation}}
    where
\begin{equation*}
\ell =\begin{cases}
1,  & \mbox{ if } S^P_1 =\varnothing, \\
\max(S^P_1),   & \mbox{ otherwise },
    \end{cases}
    \end{equation*}
and $S^P_1=\{l\in \mathbb{N}^+\cap [2,n-1]\,|\, l(l-1)\leq n\}$.

\discuss{For $n=2$, we have $S^P_1=\varnothing$, $\ell=1$, and
\begin{equation*}
    \begin{aligned}
        UB(n,\vect{1};PF) =1-\frac{1}{2}\min \bigg\{2, \frac{\sqrt{3}+2}{2}\bigg\}=1-\frac{1}{2}\cdot\frac{\sqrt{3}+2}{2}=\frac{2-\sqrt{3}}{4},
    \end{aligned}
\end{equation*}
which corresponds to Case (a) in Theorem \ref{t1}.}

For $n \geq 3$, we have $S^P_1=\{l\in \mathbb{N}^+\cap [2,n-1]\,|\, l(l-1)\leq n\}$. Recall that we define $k=\lfloor\sqrt{n}\rfloor,\epsilon=\sqrt{n}-k$, then we have
\begin{equation*}
   \ell =
    \begin{cases}
        k, & \text{ if }n < k(k+1);\\
        k+1, & \text{ if }n \geq k(k+1),
    \end{cases}
\end{equation*}
and

\begin{equation*}
    \begin{aligned}
        UB(n,\vect{1};PF) &=  1-\frac{1}{n} \min  \bigg\{  \frac{{\ell}^2 +n-\ell}{\ell},  \frac{\sqrt{2n-1}+n}{2} \bigg\} = 1-\frac{1}{n} (\ell+\frac{n}{\ell}-1)\\
        &=
    \begin{cases}
        1-\tfrac{1}{n}(k+\tfrac{n}{k}-1)=1-\tfrac{2\sqrt{n}-1+\frac{\epsilon^2}{k}}{n}, & \text{ if }n\leq k(k+1);\\
        \qquad \\
        1-\tfrac{1}{n}(k+\tfrac{n}{k+1})=1-\tfrac{2\sqrt{n}-1+\frac{(1-\epsilon)^2}{k+1}}{n}, & \text{ if }n> k(k+1),
    \end{cases}
    \end{aligned}
\end{equation*}
which corresponds to Case (b) in Theorem \ref{t1}. The second equality follows from Lemma \ref{lem:tight}, whose proof is provided below.

\discuss{Therefore, we conclude that the upper bound for the proportional POF when the maximum achievable utilities by the players are not necessarily equal, given by Theorem \ref{t2}, reduces to the bound given by Theorem \ref{t1} when the maximum achievable utilities of all players are equal.}

\begin{lemma}\label{lem:tight}
For $n \geq 3$,
\begin{equation}
     \frac{{\ell}^2+n-\ell}{\ell} \leq \frac{\sqrt{2n-1}+n}{2},
    \label{e1}
\end{equation}
where \begin{equation*}
   \ell =
    \begin{cases}
        k, & \text{ if }n < k(k+1);\\
        k+1, & \text{ if }n \geq k(k+1).
    \end{cases}
\end{equation*}
\end{lemma}

\bpr
First consider the case where $n\geq 5$, and let $\tilde{h}_2(l):=(l^2 +n-l)/l$. Then, $\tilde{h}_2(l)$ decreases monotonically when $l \in [1,\sqrt{n}]$ and increases monotonically when $l \in [\sqrt{n},n-1]$. Note that $\ell$ is an integer that minimizes $\tilde{h}_2(l)$, and it is either $k$ or $k+1$, where $k=\lfloor \sqrt{n} \rfloor$. Also, note that there must be an integer between $\sqrt{n}$ and $\sqrt{n}+1$. So we have
$$\frac{{\ell}^2+n-\ell}{\ell} \leq \tilde{h}_2 (\sqrt{n}+1)=\sqrt{n}+\frac{n}{\sqrt{n}+1}.$$
Let
$$D(n):=\frac{\sqrt{2n-1}+n}{2}-\sqrt{n}-\frac{n}{\sqrt{n}+1}.$$

We first show that for $n\geq 5$, $D(n)\geq 0$, which implies that \eqref{e1} holds for this case. Now, the derivative of $D(n)$ is:
$$D'(n)=\frac{(n-3)\sqrt{n}-1}{\sqrt{n}(1+\sqrt{n})^2}+\frac{1}{\sqrt{2n-1}},$$
which is positive for $n\geq 5$. Thus, $D(n)\geq D(5)=(21-9\sqrt{5})/4>0$ for $n\geq 5$.

It remains to consider the cases when $n=3,4$.

When $n=3, \ell =2, {\ell}^2+n-\ell/\ell=2.5, (\sqrt{2n-1}+n)/2=(3+\sqrt{5})/2 \approx 2.62$, thus \eqref{e1} holds.

When $n=4, \ell =2, {\ell}^2+n-\ell/\ell=3, (\sqrt{2n-1}+n)/2=(4+\sqrt{7})/2 \approx 3.32$, thus \eqref{e1} holds.

\epr

\section{Proof of Propositions \ref{prop6}-\ref{mmfbftwc}}\label{sec:proofprop1012}

\begin{repeatproposition}[Proposition \ref{prop6}]
Suppose $U \subseteq \{\vect{u}\, |\, 0 \leq u_i \leq L_i, i\in N\}$ is a convex and compact utility set, and $\max\{u_i\,|\, \vect{u} \in U\}=L_i$ for $i \in N$. Then there exist $c_i \in \mathbb{R}$ for $i \in N$, such that $U \subseteq U':=\{\vect{u}\, |\, \sum_{i=1}^n c_i u_i \leq 1, 0 \leq u_i \leq L_i,~ i \in N\}$, and $\max \{ \min_{i\in N} (u_i/L_i) \,|\, \vect{u} \in U \} = \max \{ \min_{i\in N} (u_i/L_i) \,|\, \vect{u} \in U' \}$. Consequently,  
\begin{equation*}
    POF(U;MMF) \leq 1-\frac{\sum_{i=1}^n L_i}{\sum_{i=1}^n c_i L_i \sum_{i=1}^n u^*_i(U')}.
\end{equation*}
\end{repeatproposition}

\bpr
The MMF solution first maximizes the minimum among the ratios of the players' utilities to their respective maximum achievable utilities, and so on. We define $k_i$ as the ratio of player $i$'s utility to their maximum achievable utility, i.e., $k_i:=u^{MMF}_i(U)/L_i$, then $\vect{u}^{MMF}(U)=(k_1L_1,\dots,k_nL_n)$.

Let $\phi$ denote the maximum ratio of a player's utility to their respective maximum achievable utility that all players can derive simultaneously. Then $k_i \geq \phi$ for all $i \in N$. Since every player $i$ can achieve in $U$ their maximum achievable utility, $(0,\dots, L_i, \dots, 0) \in U$, for all $i\in N$. Then, by convexity, $(L_1/n,\dots,L_n/n) \in U$, so $\phi$ is at least $1/n$. Further, since players cannot achieve a utility that exceeds their respective maximum achievable utility, $\phi \in [1/n,1]$.

Then, $\vect{u}^{MMF}(U)\geq (\phi L_1,\dots,\phi L_n)$, and we have
$$POF(U;MMF)=1-\frac{\sum_{i=1}^n u^{MMF}_i(U)}{\sum_{i=1}^n u^*_i(U)} \leq 1- \frac{\phi \sum_{i=1}^n L_i}{\sum_{i=1}^n u^*_i(U)}. $$

Since $\overline{\vect{u}}:=(\phi L_1,\dots,\phi L_n)$ is a boundary point in $U$, by the supporting hyperplane theorem, there exists a vector $\vect{v} \neq 0$, such that 
\begin{equation}\label{supp}
    \vect{v}^T \overline{\vect{u}} \leq \vect{v}^T \vect{u}, \forall \vect{u} \in U.
\end{equation}
Since $0 \in U$, we have $\vect{v}^T \overline{\vect{u}}\leq 0$. We claim that $\vect{v}^T \overline{\vect{u}}<0$. Indeed, suppose, on the contrary, that $\vect{v}^T \overline{\vect{u}} = 0$, i.e., $\sum_{i=1}^n v_i L_i = 0$. Then, since $(0,\dots,L_i,\dots,0) \in U$ for all $i$, by \eqref{supp}, we have $v_i L_i \geq 0$ for all $i$, which implies that $v_i L_i=0$ for all $i$. However, since $L_i >0$, we must have that $v_i = 0$ for all $i$, contradicting the fact that $\vect{v} \neq 0$. 

Let $c_i:=v_i/ \sum_{i=1}^n v_i \overline{u}_i=v_i/ \sum_{i=1}^n v_i L_i \phi$. Then, the supporting hyperplane can be written as $\sum_{i=1}^n c_i u_i \leq 1$ for all $\vect{u} \in U$, and $\sum_{i=1}^n c_i L_i=1 / \phi$, implies that $\phi = 1 / \sum_{i=1}^n c_i L_i$.

Let $U'=\{(u_1,u_2,\dots,u_n)\,|\, \sum_{i=1}^n c_i u_i \leq 1, \ 0 \leq u_i \leq L_i,\ i = 1,\dots,n\}$. Then, $U \subseteq U', \sum_{i=1}^n u^*_i(U)\leq \sum_{i=1}^n u^*_i(U')$, and
\begin{equation}
POF(U;MMF)\leq 1- \frac{\phi \sum_{i=1}^n L_i}{\sum_{i=1}^n u^*_i(U)} \leq 1- \frac{\phi \sum_{i=1}^n L_i}{\sum_{i=1}^n u^*_i(U')} =  1- \frac{\sum_{i=1}^n L_i}{\sum_{i=1}^n c_i L_i \sum_{i=1}^n u^*_i(U')}.
\label{eq1}
\end{equation}

\epr

\begin{repeatproposition}[Proposition \ref{mmfwc}]
    $$\sup_{\vect{L}:L_i \in (0,1], L_1 \geq \cdots \geq L_n} UB(n,\vect{L};MMF) = 1- \frac{1}{n}.$$
where the supremum is approached when $\vect{L}=(1-\epsilon, \epsilon/(n-1), \dots, \epsilon/(n-1))$ with $\epsilon \rightarrow 0$.
\end{repeatproposition}

\bpr

From Theorem \ref{thme}, $UB(n,\vect{L};MMF)$ has the following two possible expressions:
\begin{subequations}
\begin{alignat}{3}
    &(1)\ UB(n,\vect{L};MMF) = 1- \frac{4L_{l^*+1}\sum_{i=1}^n L_i}{(\sum_{i=1}^{l^*} L_i+(n-l^*+1)L_{l^*+1})^2}, \quad & &l^*=\max S^M_1,  \qquad  & &\text{if } S^M_1\neq \varnothing; \label{37a}    \\
&(2)\ UB(n,\vect{L};MMF) = 1 - \frac{\sum_{i=1}^n L_i}{\sum_{i=1}^{l^*} L_i(n-l^*+1)}, \quad & & l^*=\min S^M_2, & & \text{if }S^M_1=\varnothing. \label{37b}
\end{alignat}
\end{subequations}
where $S^M_1:=\{l\in \mathbb{N}^+\,|\,(n-l-1)L_{l+1}< \sum_{i=1}^l L_i \leq (n-l+1)L_{l+1}\},$ $S^M_2:=\{l\in \mathbb{N}^+\,|\,\sum_{i=1}^l L_i>(n-l+1)L_{l+1}\}.$

\begin{enumerate}
    \item [(1)] If $S^M_1\neq \varnothing$, we have $(n-l^*-1)L_{l^*+1}< \sum_{i=1}^{l^*} L_i \leq (n-l^*+1)L_{l^*+1}$, and it follows from \eqref{37a} that 
\begin{equation}
    \begin{aligned}
        UB(n,\vect{L};MMF) &= 1- \frac{4L_{l^*+1}\sum_{i=1}^n L_i}{(\sum_{i=1}^{l^*} L_i+(n-l^*+1)L_{l^*+1})^2}
        \leq 1-  \frac{4L_{l^*+1}(\sum_{i=1}^{l^*} L_i+L_{l^*+1})}{(\sum_{i=1}^{l^*} L_i+(n-l^*+1)L_{l^*+1})^2}. \label{38}
    \end{aligned}
\end{equation}
For convenience, let $$q(x)=\frac{4L_{l^*+1}(x+L_{l^*+1})}{(x+(n-l^*+1)L_{l^*+1})^2}.$$
The derivative of $q(x)$ is: 
$$q'(x)=\frac{4L_{l^*+1}((n-l^*-1)L_{l^*+1}-x)}{(x+(n-l^*+1)L_{l^*+1})^3}<0. \qquad \text{(By the condition of \eqref{37a})}$$
Thus, 
$$q(x)\geq \frac{4L^2_{l^*+1}(n-l^*+2)}{4(n-l^*+1)^2 L_{l^*+1}^2}=\frac{(n-l^*+2)}{(n-l^*+1)^2}\geq \frac{n+1}{n^2}> \frac{1}{n},$$
where the second-to-last inequality follows since $(n-l+2)/(n-l+1)$ is increasing with $l\in [1,n-1]$.
Thus,
$$UB(n,\vect{L};MMF) \leq 1-q(x) <1-\frac{1}{n}.$$

\item[(2)]
If $S^M_1 = \varnothing$, we have $\sum_{i=1}^{l^*} L_i>(n-l^*+1)L_{l^*+1}$, and it follows from \eqref{37b} that 
\begin{equation}
    \begin{aligned}
        UB(n,\vect{L};MMF) &= 1- \frac{\sum_{i=1}^n L_i}{\sum_{i=1}^{l^*} L_i(n-l^*+1)} < 1-  \frac{\sum_{i=1}^{l^*} L_i}{\sum_{i=1}^{l^*} L_i(n-l^*+1)}  \\
        &= 1-  \frac{1}{n-l^*+1} \leq 1-   \frac{1}{n}.
    \end{aligned}
\end{equation}
Thus, we have $UB(n,\vect{L};MMF) < 1 - 1/n$. 

\end{enumerate}

We next prove that $\sup_{\vect{L}} UB(n,\vect{L};MMF) = 1 - 1/n$. Indeed, if there exists a strictly smaller upper bound $1 - 1/n - \delta$, where $\delta > 0$, let $L^*_1 = L > 0, L^*_i = \epsilon$ for $i = 2, 3, \dots, n$. Then, there exists a small $\epsilon$, say $\epsilon=n L \delta/(2(n-1))$, such that the upper bound of the price of MMF with the given $\vect{L}^*$ is greater than $1 - 1/n - \delta$. The condition $\sum_{i=1}^{l} L^*_i>(n-l^*+1)L^*_{l+1}$ is satisfied for all possible $l$ when $\epsilon$ is close to 0, thus, $S^M_1=\varnothing$, $l^*=1$, and
$$UB(n,\vect{L}^*;MMF)=1- \frac{L+ (n-1)\epsilon }{\sum_{i=1}^{l^*} L_i(n-l^*+1)} \geq 1-  \frac{L +(n-1) \epsilon}{n L} =1 - \frac{1}{n} - \frac{\delta}{2} >1-\frac{1}{n}-\delta.$$

\epr

\begin{repeatproposition}[Proposition \ref{mmfbftwc}]
 $$\sup_{\vect{L}:~1 \ge L_1 \geq \cdots \geq L_n>0} BFT(n,\vect{L};MMF) = 1- \frac{4}{(n+1)^2}=1-O(\frac{1}{n^2}).$$
where the supremum is approached when $\vect{L}=(1-\epsilon, \epsilon/(n-1), \dots, \epsilon/(n-1))$ with $\epsilon \rightarrow 0$.
\end{repeatproposition}

\bpr
From BFT (2011), we have
$$BFT(n,\vect{L};MMF)=1-\frac{4n}{(n+1)^2}\frac{1/n \sum_{i=1}^n L_i}{L_1}.$$

\yding{To achieve the supremum, we assign the term $\sum_{i=1}^n L_i/L_1$ its lowest possible value, which is achieved by letting $\sum_{i=2}^n L_i \rightarrow 0$, resulting in $\sum_{i=1}^n L_i/L_1 \rightarrow 1$. Then, we obtain the supremum as $1-4/(n+1)^2$.  Moreover, since
$$\frac{4}{(n+1)^2}\leq \frac{4}{n^2} \Rightarrow \frac{4}{(n+1)^2}=O(\frac{1}{n^2}),$$
we have
$$\sup_{\vect{L}:~1 \ge L_1 \geq \cdots \geq L_n>0} BFT(n,\vect{L};MMF) = 1- \frac{4}{(n+1)^2}=1-O(\frac{1}{n^2}).$$
}
\epr

\section{Proof of Claims \ref{clm1}-\ref{clm10}}\label{sec:proofclaims1317}

\begin{repeatclaim}[Claim \ref{clm1}]
$\overline{c}_i=1/L_i$ for $i=l+2, \dots, n$.
\end{repeatclaim}

\bpr
Note first that $c_i,i=l+2, \dots, n$, solely appear in the first term of the denominator of the objective function \eqref{66a}. Therefore, to minimize \eqref{66a}, these variables should be assigned their maximum values within the feasible domain.

Next, we show that we can assume, without loss of generality, that $L_{l+2} \geq L_{l+3} \geq \cdots \geq L_n$. Indeed, suppose, on the contrary, that this assumption is not satisfied, and that $L_j<L_{j+1}$, for some $j\in [l+2,n-1]$. Then Constraints \eqref{66b} and \eqref{66c} imply that $c_j\leq c_{j+1}\leq 1/L_{j+1} < 1/L_j$. Thus, $\sum_{i=l+2}^n c_i L_i < n-l-1$. On the other hand, let $L'_i=L_i, i =1,\dots,l+1$, and let $L'_i, i=l+2, \dots, n$, denote the partial vector derived from the partial vector $(L_i), i=l+2, \dots,n$, after rearranging the components of the latter in a descending order. That is, $L'_{l+2} \geq L'_{l+3} \geq \cdots \geq L'_n$. Similarly, let $c'_i=c_i, i=1,\dots,l+1$, and $c'_i=1/L'_i, i=l+2,\dots,n,$ denote a feasible solution to the optimization Problem \eqref{originalpb} corresponding to the parameters $L'_i, i=1,\dots,n$. Thus, $\sum_{i=l+2}^n c'_i L'_i = n-l-1 >\sum_{i=l+2}^n c_i L_i$, while the other terms in the objective function remain invariant for these two sets of parameters and variables. Consequently, we have $f_3(\vect{c}',l;\vect{L}')\leq f_3(\vect{c},l;\vect{L})$. Thus, we can assume $L_{l+2} \geq L_{l+3} \geq \cdots \geq L_n$.

This assumption ensures that $c_i=1/L_i, i=l+2, \dots, n$ satisfies Constraint \eqref{66b} and hence is feasible. Incorporating the above analysis that $c_i, i=l+2, \dots, n$ should be assigned their maximum values within the feasible domain, we have $c_i^*=1/L_i, i=l+2, \dots, n$.

\epr

\begin{repeatclaim}[Claim \ref{clm2}]
  If $A \leq (n-l-1)L_{l+1} $, 
    $$x^* \in [\frac{1}{A+L_{l+1}},\frac{1}{L_{l+1}}],\quad y^*=1-x^*L_{l+1},\quad g_3(x^*,y^*;l)=\frac{1}{(n-l)(A+L_{l+1})};$$

    If $(n-l-1)L_{l+1} < A \leq (n-l+1)L_{l+1}$, 
    $$x^*=\frac{1}{L_{l+1}},\quad y^*=\frac{1}{2}\bigg(\frac{A}{L_{l+1}}-n+l+1 \bigg),\quad  g_3(x^*,y^*;l)=\frac{4L_{l+1}}{(A+(n-l+1)L_{l+1})^2};$$

    If $A > (n-l+1)L_{l+1} $, 
    $$x^*=\frac{1}{L_{l+1}},\quad y^*=1,\quad g_3(x^*,y^*;l)=\frac{1}{A(n-l+1)}.$$
\end{repeatclaim}

\bpr
    If $A \leq (n-l-1)L_{l+1}$, let $x^*$ be any value in $[1/(A+L_{l+1}),1/L_{l+1}]$, $y^*=1-x^*L_{l+1}$, then 
    $$g_3(x^*,y^*;l)=\frac{x^*}{(y^*+n-l-1+x^* L_{l+1})(Ax^*+1-y^*)}=\frac{1}{(n-l)(A+L_{l+1})}.$$
    
    For any feasible $x,y$, we have
    \begin{equation*}
    \begin{aligned}
        g_3(x,y;l)-&g_3(x^*,y^*;l)\\&=\frac{(1-y-xL_{l+1})x\cdot A+(n-l-1+y)(y+xL_{l+1}-1)}{(n-l)(A+L_{l+1})(y+n-l-1+x L_{l+1})(Ax+1-y)}\\
        &\geq\frac{(1-y-xL_{l+1})(n-l-1)x L_{l+1}+(n-l-1+y)(y+xL_{l+1}-1)}{(n-l)(A+L_{l+1})(y+n-l-1+x L_{l+1})(Ax+1-y)}\\ &\qquad \qquad \qquad \qquad \qquad \qquad \qquad \qquad \qquad \qquad  \text{(since, by \eqref{66e}, $1-y-xL_{l+1}<0$)}\\
        &=\frac{(y+xL_{l+1}-1)((n-l-1)(1-xL_{l+1})+y)}{(n-l)(A+L_{l+1})(y+n-l-1+x L_{l+1})(Ax+1-y)}\\
        &\geq 0.
    \end{aligned}
    \end{equation*}

    If $(n-l-1)L_{l+1} < A \leq (n-l+1)L_{l+1}$, let $x^*=1/L_{l+1}$, $y^*=(A/L_{l+1}-n+l+1)/2$, then 
    $$g_3(x^*,y^*;l)=\frac{1/L_{l+1}}{(A/L_{l+1}+n-l+1)^2/4}=\frac{4L_{l+1}}{(A+(n-l+1)L_{l+1})^2}.$$
    
    For any feasible $x,y$, we have
    \begin{equation*}
    \begin{aligned}
        g_3(x,y;l)-&g_3(x^*,y^*;l)\\&=\frac{x(A-L(-3-l+n+2xL_{l+1}+2y))^2+4L_{l+1}(1-x L_{l+1})(xL_{l+1}+y-1)}{(A+(n-l+1)L_{l+1})^2(y+n-l-1+x L_{l+1})(Ax+1-y)}\\
        &\geq \frac{4L_{l+1}(1-x L_{l+1})(xL_{l+1}+y-1)}{(A+(n-l+1)L_{l+1})^2(y+n-l-1+x L_{l+1})(Ax+1-y)}\\
        &\geq 0.
    \end{aligned}
    \end{equation*}

    If $A> (n-l+1)L_{l+1}$, let $x^*=1/L_{l+1}$, $y^*=1$, then 
    $$g_3(x^*,y^*;l)=\frac{x^*}{(y^*+n-l-1+x^* L_{l+1})(Ax^*+1-y^*)}=\frac{1}{A(n-l+1)}.$$    

    For any feasible $x,y$, we have
    \begin{equation*}
    \begin{aligned}
        g_3(x,y;l)-&g_3(x^*,y^*;l)\\&=\frac{(2-y-L_{l+1}x)Ax-(y+x L_{l+1}+n-l-1) (1 - y)}{A(n-l+1)(y+n-l-1+x L_{l+1})(Ax+1-y)}\\
        &> \frac{(2-y-L_{l+1}x)(n-l+1)xL_{l+1}-(y+x L_{l+1}+n-l-1) (1 - y)}{A(n-l+1)(y+n-l-1+x L_{l+1})(Ax+1-y)}\\
        &=\frac{(1-y)^2+(1-xL_{l+1})(xL_{l+1}+(n-l)(xL_{l+1}+y-1))}{A(n-l+1)(y+n-l-1+x L_{l+1})(Ax+1-y)}\\
        &\geq 0.
    \end{aligned}
    \end{equation*}

The last inequality holds due to $1-xL_{l+1}\geq 0$ by \eqref{yxcc}, $xL_{l+1}+y-1\geq 0$ by \eqref{yxce}, $n\geq l+1$ by definition and $1-y\geq 0$ by \eqref{yxcd}.

\epr

\begin{repeatclaim}[Claim \ref{clm3}]
    $p(A,L_{l+1};l)$ is decreasing both in $A$ and $L_{l+1}$.
\end{repeatclaim}

\bpr
    We first show that $p(A,L_{l+1};l)$ decreases in $A$. Recall that $p(A,L_{l+1};l)$ is defined in \eqref{funcp}, and $A=\sum_{i=1}^l L_i$. Now, note that $A$ only appears in the denominator of $p(A,L_{l+1};l)$, and in all three ranges for which $p$ is defined, an increase in $A$ increases the denominator of $p(A,L_{l+1};l)$. Thus, in each of these three ranges, $p$ is decreasing in $A$.

    Further, note that if $A=(n-l-1)L_{l+1}$, then $1/((n-l)(A+L_{l+1}))=4L_{l+1}/(A+(n-l+1)L_{l+1})^2$; and if $A=(n-l+1)L_{l+1}$, then $4L_{l+1}/(A+(n-l+1)L_{l+1})^2=1/(A(n-l+1))$. Thus, $p(A,L_{l+1};l)$ is continuous in $A$ over the entire region, and we conclude that $p(A,L_{l+1};l)$ is decreasing in $A$.

    To prove that $p(A,L_{l+1};l)$ decreases in $L_{l+1}$, we rewrite $p(A,L_{l+1};l)$ as follows:
    $$p(A,L_{l+1};l):=
    \begin{cases}
        \frac{1}{A(n-l+1)}, & L_{l+1} < \frac{A}{n-l+1};\\
        \frac{4L_{l+1}}{(A+(n-l+1)L_{l+1})^2}, & \frac{A}{n-l+1} \leq L_{l+1} < \frac{A}{n-l-1};\\
        \frac{1}{(n-l)(A+L_{l+1})}, &L_{l+1} \geq \frac{A}{n-l-1}.
    \end{cases} $$

    Clearly, if $L_{l+1}< A/(n-l+1)$ or $L_{l+1} \geq A/(n-l-1)$, then $p(A,L_{l+1};l)$ is decreasing in $L_{l+1}$. For the region where $A/(n-l+1) \leq L_{l+1} < A/(n-l-1)$, we calculate the partial derivative of $p(A,L_{l+1};l)$ with respect to $L_{l+1}$, 

    $$\frac{\partial p}{\partial L_{l+1}}= \frac{4 (A-(n-l+1)L_{l+1})}{(A+(n-l+1)L_{l+1})^3},$$
    which is obviously negative in this region. Thus, $p(A,L_{l+1};l)$ is decreasing in $L_{l+1}$ in this region as well. Finally, it is easy to verify that $p(A,L_{l+1};l)$ is continuous in $L_{l+1}$, and thus $p(A,L_{l+1};l)$ is decreasing in $L_{l+1}$.
\epr

\begin{repeatclaim}[Claim \ref{clm9}]
The value of $p(A,L_{l+1};l)$ is minimized with respect to the $L_i, i=1,\dots,n$, if they are arranged in a descending order.
\end{repeatclaim}

\bpr
We have previously shown that the permutation of the $L_i's$ that would minimize $p(A,L_{l+1};l)$ should satisfy:
$L_1 \geq L_2 \geq \cdots \geq L_l \geq L_{l+2}$,
$L_{l+1} \geq L_{l+2}$ and $L_{l+2} \geq L_{l+3} \geq \cdots \geq L_n$. Thus, what remains to show is $L_{l+1} \leq L_l$. Suppose it is not true, then we can construct $L'$ which is in descending order and for which $p$ attains a smaller value.

    Let $\Delta=L_{l+1}-L_l>0, L'_{l+1}=L_{l+1}-\Delta, A'=A+L_{l+1}-L'_{l+1}=A+\Delta$. The relationship among $A',(n-l-1)L'_{l+1}$ and $(n-l+1)L'_{l+1}$, as well as the relationship among $A,(n-l-1)L_{l+1}$ and $(n-l+1)L_{l+1}$, affect which segments of the piecewise functions $p(A',L'_{l+1};l)$ and $p(A,L_{l+1};l)$ are valid. 

    The following two diagrams present the valid expressions for $p(A',L'_{l+1};l)$ and $p(A,L_{l+1};l)$ in seven possible different cases.

\begin{enumerate}
    \item [(1)] If $(n-l-1)L_{l+1}\leq (n-l+1)L_{l+1}-(n-l-2)\Delta$, then the following five different cases, depending on the range of $A$, are possible:
    \begin{center}
    \begin{tikzpicture}[scale=3.5]
  \draw[->] (-1.2, 0) -- (2, 0);
  \draw[dashed,color=red] (-1.2, -.65) -- (-1.2, -.85);
  \draw[dashed,color=red] (-.8, -.65) -- (-.8, -.85);
  \draw[dashed,color=red] (-.1, -.65) -- (-.1, -.85);
  \draw[dashed,color=red] (.6, -.65) -- (.6, -.85);
  \draw[dashed,color=red] (1.3, -.65) -- (1.3, -.85);
  \draw[dashed,color=red] (2, -.65) -- (2, -.85);

    \foreach \x/\xtext in {-.8/\small$ (n-l-1)L_{l+1}-(n-l)\Delta $, .6/\small$(n-l+1)L_{l+1}-(n-l-2)\Delta$}
        \draw[shift={(\x,0)}](0pt,1pt)--(0pt,-1pt)node[below]{\xtext};
    \foreach \x/\xtext in {-.1/\small$(n-l-1)L_{l+1}$, 1.3/\small$(n-l+1)L_{l+1}$}
        {\draw[shift={(\x,0)}](0pt,1pt)--(0pt,-1pt);
        \node[above] at (\x,.05) {\xtext};}

        \draw[thick,decorate,decoration={calligraphic brace,amplitude=10pt}](-1.2,.25)--(-.11,.25)node[pos=0.5,above=15pt]{$\frac{1}{(n-l)(A+L_{l+1})}$};
        \draw[thick,decorate,decoration={calligraphic brace,amplitude=10pt}](-.09,.25)--(1.29,.25)node[pos=0.5,above=15pt]{$\frac{4L_{l+1}}{(A+(n-l+1)L_{l+1})^2}$};
        \draw[thick,decorate,decoration={calligraphic brace,amplitude=10pt}](1.31,.25)--(2,.25)node[pos=0.5,above=15pt]{$\frac{1}{A(n-l+1)}$};

        \draw[thick,decorate,decoration={calligraphic brace,amplitude=10pt}](-.81,-.25)--(-1.2,-.25)node[pos=0.5,below=13pt]{$\frac{1}{(n-l)(A'+L'_{l+1})}$};
        \draw[thick,decorate,decoration={calligraphic brace,amplitude=10pt}](.59,-.25)--(-.79,-.25)node[pos=0.5,below=10pt]{$\frac{4L'_{l+1}}{(A'+(n-l+1)L'_{l+1})^2}$};
        \draw[thick,decorate,decoration={calligraphic brace,amplitude=10pt}](2,-.25)--(.61,-.25)node[pos=0.5,below=13pt]{$\frac{1}{A'(n-l+1)}$};

  \node[anchor = west] at (2, 0) {$ A $};
  \node[anchor =south] at (-1.8, .4) {$ p(A,L_{l+1}\text{;\ }l)= $};
  \node[anchor =south] at (-1.8, -.6) {$ p(A',L'_{l+1}\text{;\ }l)= $};
  \node[anchor =south] at (-1.8, -.8) {Index};
  \node[anchor =south] at (-1, -.83) {\textcircled{\scriptsize{1}}};
  \node[anchor =south] at (-.45, -.83) {\textcircled{\scriptsize{2}}};
  \node[anchor =south] at (.25, -.83) {\textcircled{\scriptsize{3}}};
  \node[anchor =south] at (.95, -.83) {\textcircled{\scriptsize{4}}};
  \node[anchor =south] at (1.65, -.83) {\textcircled{\scriptsize{5}}};


\end{tikzpicture}
\end{center}

 \item [(2)] If $(n-l-1)L_{l+1}> (n-l+1)L_{l+1}-(n-l-2)\Delta$, then the following five different cases, depending on the range of $A$, are possible:
     \begin{center}
    \begin{tikzpicture}[scale=3.5]
  \draw[->] (-1.2, 0) -- (2, 0);
  \draw[dashed,color=red] (-1.2, -.65) -- (-1.2, -.85);
  \draw[dashed,color=red] (-.8, -.65) -- (-.8, -.85);
  \draw[dashed,color=red] (-.1, -.65) -- (-.1, -.85);
  \draw[dashed,color=red] (.6, -.65) -- (.6, -.85);
  \draw[dashed,color=red] (1.3, -.65) -- (1.3, -.85);
  \draw[dashed,color=red] (2, -.65) -- (2, -.85);

    \foreach \x/\xtext in {-.8/\small$ (n-l-1)L_{l+1}-(n-l)\Delta $, .6/\small$(n-l-1)L_{l+1}$}
        \draw[shift={(\x,0)}](0pt,1pt)--(0pt,-1pt)node[below]{\xtext};
    \foreach \x/\xtext in {-.1/\small$(n-l+1)L_{l+1}-(n-l-2)\Delta$, 1.3/\small$(n-l+1)L_{l+1}$}
        {\draw[shift={(\x,0)}](0pt,1pt)--(0pt,-1pt);
        \node[above] at (\x,.05) {\xtext};}

        \draw[thick,decorate,decoration={calligraphic brace,amplitude=10pt}](-1.2,.25)--(.59,.25)node[pos=0.5,above=15pt]{$\frac{1}{(n-l)(A+L_{l+1})}$};
        \draw[thick,decorate,decoration={calligraphic brace,amplitude=10pt}](.61,.25)--(1.29,.25)node[pos=0.5,above=15pt]{$\frac{4L_{l+1}}{(A+(n-l+1)L_{l+1})^2}$};
        \draw[thick,decorate,decoration={calligraphic brace,amplitude=10pt}](1.31,.25)--(2,.25)node[pos=0.5,above=15pt]{$\frac{1}{A(n-l+1)}$};

        \draw[thick,decorate,decoration={calligraphic brace,amplitude=10pt}](-.81,-.25)--(-1.2,-.25)node[pos=0.55,below=13pt]{$\frac{1}{(n-l)(A'+L'_{l+1})}$};
        \draw[thick,decorate,decoration={calligraphic brace,amplitude=10pt}](-.11,-.25)--(-.79,-.25)node[pos=0.3,below=10pt]{$\frac{4L'_{l+1}}{(A'+(n-l+1)L'_{l+1})^2}$};
        \draw[thick,decorate,decoration={calligraphic brace,amplitude=10pt}](2,-.25)--(-.09,-.25)node[pos=0.5,below=13pt]{$\frac{1}{A'(n-l+1)}$};

  \node[anchor = west] at (2, 0) {$ A $};
  \node[anchor =south] at (-1.8, .4) {$ p(A,L_{l+1}\text{;\ }l)= $};
  \node[anchor =south] at (-1.8, -.6) {$ p(A',L'_{l+1}\text{;\ }l)= $};
  \node[anchor =south] at (-1.8, -.8) {Index};
  \node[anchor =south] at (-1, -.83) {\textcircled{\scriptsize{1}}};
  \node[anchor =south] at (-.45, -.83) {\textcircled{\scriptsize{2}}};
  \node[anchor =south] at (.25, -.83) {\textcircled{\scriptsize{6}}};
  \node[anchor =south] at (.95, -.83) {\textcircled{\scriptsize{7}}};
  \node[anchor =south] at (1.65, -.83) {\textcircled{\scriptsize{5}}};

\end{tikzpicture}
\end{center}
\end{enumerate}

Next, we proceed to analyze the seven cases identified above:

\begin{enumerate}
    \item [\textcircled{\scriptsize{1}}] $A\leq (n-l-1)L_{l+1}-(n-l)\Delta$.
\begin{equation*}
    \begin{aligned}
        p(A',L'_{l+1};l)&=\frac{1}{(n-l)(A'+L'_{l+1})}=\frac{1}{(n-l)(A+\Delta+L_{l+1}-\Delta)}\\&=\frac{1}{(n-l)(A+L_{l+1})}=p(A,L_{l+1};l).
    \end{aligned}
\end{equation*}

\item [\textcircled{\scriptsize{2}}]$(n-l-1)L_{l+1}-(n-l)\Delta < A \leq \min\{(n-l-1)L_{l+1},(n-l+1)L_{l+1}-(n-l-2)\Delta \}$.
\begin{equation*}
    \begin{aligned}
        p(A',L'_{l+1};l)&=\frac{4L'_{l+1}}{(A'+(n-l+1)L'_{l+1})^2}=\frac{4(L_{l+1}-\Delta)}{(A+L_{l+1}+(n-l)(L_{l+1}-\Delta))^2}\\&\leq \frac{4(L_{l+1}-\Delta)}{4(A+L_{l+1})(n-l)(L_{l+1}-\Delta)}=\frac{1}{(A+L_{l+1})(n-l)}=p(A,L_{l+1};l).
    \end{aligned}
\end{equation*}

\item [\textcircled{\scriptsize{3}}]$(n-l-1)L_{l+1} < A \leq (n-l+1)L_{l+1}-(n-l-2)\Delta$.
$$p(A',L'_{l+1};l)=\frac{4L'_{l+1}}{(A'+(n-l+1)L'_{l+1})^2}=\frac{4(L_{l+1}-\Delta)}{(A+(n-l+1)L_{l+1}-(n-l)\Delta)^2}.$$
\begin{equation*}
    \begin{aligned}
        &(A+(n-l+1)L_{l+1}-(n-l)\Delta)^2 L_{l+1}\\=&(A+(n-l+1)L_{l+1})^2 L_{l+1}+(n-l)^2 \Delta^2L_{l+1}-2\Delta(A+(n-l+1)L_{l+1})(n-l)L_{l+1}\\
        &+\Delta(A+(n-l+1)L_{l+1})^2+(n-l)^2L_{l+1}^2 \Delta-\Delta(A+(n-l+1)L_{l+1})^2-(n-l)^2L_{l+1}^2 \Delta\\
        =&\Delta(A+L_{l+1})^2-(n-l)^2\Delta L_{l+1}(L_{l+1}-\Delta)+(A+(n-l+1)L_{l+1})^2 (L_{l+1}-\Delta)\\
        \geq& \Delta^2(n-l)^2 L_{l+1}+(A+(n-l+1)L_{l+1})^2 (L_{l+1}-\Delta) \\
        \geq& (A+(n-l+1)L_{l+1})^2 (L_{l+1}-\Delta).
    \end{aligned}
\end{equation*}
Then
\begin{equation*}
    \begin{aligned}
       p(A',L'_{l+1};l) \leq \frac{4(L_{l+1}-\Delta)L_{l+1}}{(A+(n-l+1)L_{l+1})^2 (L_{l+1}-\Delta)}
        = \frac{4L_{l+1}}{(A+(n-l+1)L_{l+1})^2}=p(A,L_{l+1};l).
    \end{aligned}
\end{equation*}

\item [\textcircled{\scriptsize{4}}]$(n-l+1)L_{l+1}-(n-l-2)\Delta < A \leq (n-l+1)L_{l+1}$. From the corresponding diagram of this case, we can observe $(n-l-1)L_{l+1}\leq (n-l+1)L_{l+1}-(n-l-2)\Delta \Rightarrow (n-l-2)\Delta \leq 2L_{l+1}$.
\begin{equation*}
    \begin{aligned}
       p(A',L'_{l+1};l) &=\frac{1}{A'(n-l+1)}=\frac{1}{(A+\Delta)(n-l+1)}\\ &=\frac{1}{\Delta(n-l+1)+\frac{(A+(n-l+1)L_{l+1})^2}{4L_{l+1}}-\frac{(A-(n-l+1)L_{l+1})^2}{4L_{l+1}}}.
    \end{aligned}
\end{equation*}
According to the range of $A$ in this case, we have
$$\frac{(A-(n-l+1)L_{l+1})^2}{4L_{l+1}}<\frac{(n-l-2)^2\Delta^2}{4L_{l+1}}\leq \frac{(n-l-2)\Delta}{2},$$
$$\Rightarrow \Delta(n-l+1)-\frac{(A-(n-l+1)L_{l+1})^2}{4L_{l+1}}> \Delta(n-l+1) - \frac{(n-l-2)\Delta}{2}=\frac{\Delta}{2}(n-l+4)\geq 0.$$
\begin{equation*}
    \begin{aligned}
       p(A',L'_{l+1};l) &=\frac{1}{\Delta(n-l+1)+\frac{(A+(n-l+1)L_{l+1})^2}{4L_{l+1}}-\frac{(A-(n-l+1)L_{l+1})^2}{4L_{l+1}}}\\& <\frac{1}{\frac{(A+(n-l+1)L_{l+1})^2}{4L_{l+1}}}=\frac{4L_{l+1}}{(A+(n-l+1)L_{l+1})^2}=p(A,L_{l+1};l).
    \end{aligned}
\end{equation*}

\item [\textcircled{\scriptsize{5}}]$A >(n-l+1)L_{l+1} $.
$$p(A',L'_{l+1};l)=\frac{1}{A'(n-l+1)}< \frac{1}{A(n-l+1)} =p(A,L_{l+1};l).$$

\item [\textcircled{\scriptsize{6}}]$(n-l+1)L_{l+1}-(n-l-2)\Delta < A \leq (n-l-1)L_{l+1} $.
\begin{equation*}
    \begin{aligned}
    p(A',L'_{l+1};l)&=\frac{1}{A'(n-l+1)}= \frac{1}{(A+\Delta)(n-l+1)}\\&=\frac{1}{(A+L_{l+1})(n-l)+A+L_{l+1}-(n-l+1)(L_{l+1}-\Delta)}\\& < \frac{1}{(A+L_{l+1})(n-l)+(n-l+2)L_{l+1}-(n-l-2)\Delta-(n-l+1)(L_{l+1}-\Delta)}\\&=\frac{1}{(A+L_{l+1})(n-l)+L_{l+1}+3\Delta}\\&<\frac{1}{(A+L_{l+1})(n-l)}=p(A,L_{l+1};l).
    \end{aligned}
\end{equation*}

\item [\textcircled{\scriptsize{7}}]$(n-l-1)L_{l+1}<A\leq (n-l+1)L_{l+1} $.
From the corresponding diagram of this case, we can observe $(n-l+1)L_{l+1}-(n-l-2)\Delta < (n-l-1)L_{l+1} \Rightarrow 2L_{l+1} < (n-l-2)\Delta$.
\begin{equation*}
    \begin{aligned}
       p(A',L'_{l+1};l) &=\frac{1}{A'(n-l+1)}=\frac{1}{(A+\Delta)(n-l+1)}\\ &=\frac{1}{\Delta(n-l+1)+\frac{(A+(n-l+1)L_{l+1})^2}{4L_{l+1}}-\frac{(A-(n-l+1)L_{l+1})^2}{4L_{l+1}}}.
    \end{aligned}
\end{equation*}
According to the range of $A$ in this case, we have
$$\frac{(A-(n-l+1)L_{l+1})^2}{4L_{l+1}}<\frac{(2L_{l+1})^2}{4L_{l+1}}=L_{l+1} < \frac{(n-l-2)\Delta}{2},$$
$$\Rightarrow \Delta(n-l+1)-\frac{(A-(n-l+1)L_{l+1})^2}{4L_{l+1}}> \Delta(n-l+1) - \frac{(n-l-2)\Delta}{2}=\frac{\Delta}{2}(n-l+4)\geq 0.$$
\begin{equation*}
    \begin{aligned}
       p(A',L'_{l+1};l) &=\frac{1}{\Delta(n-l+1)+\frac{(A+(n-l+1)L_{l+1})^2}{4L_{l+1}}-\frac{(A-(n-l+1)L_{l+1})^2}{4L_{l+1}}}\\& <\frac{1}{\frac{(A+(n-l+1)L_{l+1})^2}{4L_{l+1}}}=\frac{4L_{l+1}}{(A+(n-l+1)L_{l+1})^2}=p(A,L_{l+1};l).
    \end{aligned}
\end{equation*}
\end{enumerate}

Thus, in each of the cases, we have $p(A',L'_{l+1};l)\leq p(A,L_{l+1};l)$, and, actually, in four of the cases, $p(A',L'_{l+1};l)<p(A,L_{l+1};l)$, and the proof of Claim \ref{clm9} is complete.
\epr

\begin{repeatclaim}[Claim \ref{clm10}]
Let $l^*$ denote an optimal value of $l$. If $S^M_1\neq \varnothing$, $l^*=\max S^M_1$, otherwise $l^*=\min S^M_2$.
\end{repeatclaim}

\bpr
    By Claim \ref{clm9}, $L_1 \geq L_2 \geq \cdots \geq L_n$, and recall that $S^M_0=\{l\in \mathbb{N}^+:\sum_{i=1}^l L_i \leq (n-l-1)L_{l+1}\}, S^M_1=\{l\in \mathbb{N}^+:(n-l-1)L_{l+1}< \sum_{i=1}^l L_i \leq (n-l+1)L_{l+1}\}, S^M_2=\{l\in \mathbb{N}^+:\sum_{i=1}^l L_i>(n-l+1)L_{l+1}\}$. Thus, the values of the boundary points defining $S^M_0,S^M_1,S^M_2$, i.e., $(n-l-1)L_{l+1},(n-l+1)L_{l+1}$, decrease as $l$ increases, while $\sum_{i=1}^l L_i$ increases as $l$ increases. We conclude that if $l\in S^M_0$, then $l+1$ could belong to $S^M_0, S^M_1$ or $S^M_2$; if $l\in S^M_1$, then $l+1$ could belong to $S^M_1$ or $S^M_2$, and if $l\in S^M_2$, then $l+1$ must belong to $S^M_2$. Furthermore, for any $l_0 \in S^M_0, l_1 \in S^M_1, l_2 \in S^M_2$, we have $l_0<l_1<l_2$.

    To prove Claim \ref{clm10}, we consider the following cases:

    \textbf{Case 1}: If $l \in S^M_0$, then we show that $p(\sum_{i=1}^l L_i,L_{l+1};l) \geq p(\sum_{i=1}^{l+1} L_i,L_{l+2};l+1)$, regardless of which set $l+1$ belongs to.

    \textbf{Case 2}: If $l \in S^M_1$, then we show that $p(\sum_{i=1}^l L_i,L_{l+1};l) > p(\sum_{i=1}^{l+1} L_i,L_{l+2};l+1)$ when $l+1 \in S^M_1$, and $p(\sum_{i=1}^l L_i,L_{l+1};l) \leq p(\sum_{i=1}^{l+1} L_i,L_{l+2};l+1)$ when $l+1 \in S^M_2$.

     \textbf{Case 3}: If $l \in S^M_2$, then $l+1\in S^M_2$, and we show that $p(\sum_{i=1}^l L_i,L_{l+1};l) < p(\sum_{i=1}^{l+1} L_i,L_{l+2};l+1)$.

    Clearly, if we prove the assertions in the above three cases, then $l^*=\max S^M_1$, if $S^M_1$ exists, and $l^*=\min S^M_2$, otherwise.
    


    \textbf{Case 1}: $l\in S^M_0$.

    \begin{enumerate}

    \item [] If $l+1 \in S^M_0$, we have $A+L_{l+1}\leq (n-l-2)L_{l+2}$,
    \begin{equation*}
        \begin{aligned}
            p(A+ L_{l+1},L_{l+2};l+1)&=\frac{1}{(n-l-1)(A+L_{l+1}+L_{l+2})}\\&=\frac{1}{(n-l)(A+L_{l+1})+(n-l-1)L_{l+2}-A-L_{l+1}}\\&<\frac{1}{(n-l)(A+L_{l+1})}=p(A,L_{l+1};l);
        \end{aligned}
    \end{equation*}

    \item[] if $l+1 \in S^M_1$,
 \begin{equation*}
        \begin{aligned}
            p(A+ L_{l+1},L_{l+2};l+1)&=\frac{4L_{l+2}}{(A+L_{l+1}+(n-l)L_{l+2})^2}\\ &\leq \frac{4L_{l+2}}{4(A+L_{l+1})(n-l)L_{l+2}}=\frac{1}{(n-l)(A+L_{l+1})}=p(A,L_{l+1};l);
        \end{aligned}
    \end{equation*}

    \item[] if $l+1 \in S^M_2$,
 \begin{equation*}
        \begin{aligned}
            p(A+ L_{l+1},L_{l+2};l+1)=\frac{1}{(n-l)(A+L_{l+1})}=p(A,L_{l+1};l).
        \end{aligned}
    \end{equation*}
\end{enumerate}

\textbf{Case 2}: $l \in S^M_1$. Then, we have $(n-l)L_{l+1} <A+L_{l+1} \leq (n-l+2)L_{l+1}$.

 \begin{enumerate}
     \item [] If $l+1 \in S^M_1$,
    $$p(A+ L_{l+1},L_{l+2};l+1)=\frac{4L_{l+2}}{(A+L_{l+1}+(n-l)L_{l+2})^2}=\frac{4L_{l+2}L_{l+1}}{L_{l+1}(A+L_{l+1}+(n-l)L_{l+2})^2},$$
    \begin{equation*}
        \begin{aligned}
            &L_{l+1}(A+L_{l+1}+(n-l)L_{l+2})^2=L_{l+1}((A+L_{l+1})^2+2(n-l)L_{l+2}(A+L_{l+1})+(n-l)^2L_{l+2}^2)\\
            =& L_{l+1}((A+L_{l+1})^2+(n-l)^2L_{l+1}^2)+2(n-l)L_{l+2}L_{l+1}(A+L_{l+1})\\&+L_{l+2}((A+L_{l+1})^2+(n-l)^2L_{l+1}^2)-L_{l+2}((A+L_{l+1})^2+(n-l)^2L_{l+1}^2)\\
            =& L_{l+2}((A+L_{l+1})^2+(n-l)^2L_{l+1}^2)+2(n-l)L_{l+2}L_{l+1}(A+L_{l+1})\\&+(L_{l+1}-L_{l+2})((A+L_{l+1})^2-(n-l)^2L_{l+1}L_{l+2})\\ > & L_{l+2} (A+(n-l+1)L_{l+1})^2+(L_{l+1}-L_{l+2})((n-l)^2L_{l+1}^2-(n-l)^2L_{l+1}L_{l+2})\\=& L_{l+2} (A+(n-l+1)L_{l+1})^2+(L_{l+1}-L_{l+2})^2(n-l)^2L_{l+1} \\ \geq & L_{l+2} (A+(n-l+1)L_{l+1})^2,
        \end{aligned}
    \end{equation*}
    \begin{equation*}
        \begin{aligned}
            p(A+ L_{l+1},L_{l+2};l+1)&=\frac{4L_{l+2}L_{l+1}}{L_{l+1}(A+L_{l+1}+(n-l)L_{l+2})^2}\\ &< \frac{4L_{l+1}}{(A+(n-l+1)L_{l+1})^2}=p(A,L_{l+1};l);
        \end{aligned}
    \end{equation*}

\item [] if $l+1 \in S^M_2$,
\begin{equation*}
    \begin{aligned}
        p(A+ L_{l+1},L_{l+2};l+1)&=\frac{1}{(n-l)(A+L_{l+1})}=\frac{4L_{l+1}}{4L_{l+1}(A+L_{l+1})(n-l)} \\&\geq \frac{4L_{l+1}}{(A+L_{l+1}+(n-l)L_{l+1})^2}=p(A,L_{l+1};l).
    \end{aligned}
\end{equation*}

\end{enumerate}

\textbf{Case 3}: $l\in S^M_2$. Then, $l+1 \in S^M_2, A>(n-l+1)L_{l+1}$, and we have,

    \begin{equation*}
        \begin{aligned}
        p(A+ L_{l+1},L_{l+2};l+1)&=\frac{1}{(A+L_{l+1})(n-l)} =\frac{1}{A(n-l+1)+L_{l+1}(n-l)-A}\\
        &>\frac{1}{A(n-l+1)}=p(A,L_{l+1};l).
        \end{aligned}
    \end{equation*}

\epr

\section{Proof of Corollary \ref{crl2}}\label{sec:crl2}

\begin{repeatcorollary}[Corollary \ref{crl2}]
     \begin{enumerate}
        \item [(a)] For $L_1=L_2=\dots=L_n$, our upper bound reduces to the BFT bound for the case of equal maximum achievable utilities, i.e.,
        $$UB(n,\vect{L};MMF)=1-\frac{4n}{(n+1)^2}.$$
    \item [(b)] For $L_1=L_2=\dots=L_{\lfloor n/2 \rfloor +1}\geq L_{\lfloor n/2 \rfloor +2} \geq \cdots \geq L_n$, our upper bound reduces to the BFT bound for the case of unequal maximum achievable utilities, i.e.,
        $$UB(n,\vect{L};MMF)=1-\frac{4\sum_{i=1}^n L_i}{(n+1)^2 L_1}.$$
    \end{enumerate}
\end{repeatcorollary}

\bpr
\begin{enumerate}
    \item [(a)] Denote $\overline L:=L_1=L_2=\cdots=L_n$. We have $S^M_1=\{l \in \mathbb{N}^+\,|\, (n-l-1)L_{l+1} < \sum_{i=1}^l L_i \leq (n-l+1)L_{l+1} \}=\{l \in \mathbb{N}^+: (n-1)/2 < l \leq (n+1)/2 \}\neq \varnothing$. Then by Theorem \ref{thme}, the upper bound is
   
$$UB(n,\vect{L};MMF)=1- \frac{4L_{l^*+1}\sum_{i=1}^n L_i}{(\sum_{i=1}^{l^*} L_i+(n-l^*+1)L_{l^*+1})^2}=1- \frac{4 n \overline{L}^2}{(n+1)^2 \overline{L}^2} =1-\frac{4n}{(n+1)^2},$$ 
which coincides with the BFT bound for the price of max-min fairness when all maximum achievable utilities are equal (BFT 2011).

    \item [(b)] We separately consider two cases: $n$ odd and $n$ even. 
    \begin{enumerate}
        \item [(1)] If $n$ is odd, let $L_1=L_2=\cdots=L_{(n+1)/2}$. Then, by Theorem \ref{thme}, if $L_{(n+3)/2}=L_1$, we have $l^*=\max S^M_1=(n+1)/2$ and the upper bound is 
        $$UB(n,\vect{L};MMF)=1- \frac{4L_{l^*+1}\sum_{i=1}^n L_i}{(\sum_{i=1}^{l^*} L_i+(n-l^*+1)L_{l^*+1})^2}=1-\frac{4 \sum_{i=1}^n L_i}{(n+1)^2 L_1};$$
        if $L_{(n+3)/2}<L_1$, we have $S_1^M=\varnothing$,  $l^*=\min S^M_2=(n+1)/2$, and the upper bound is
$$UB(n,\vect{L};MMF)=1- \frac{\sum_{i=1}^n L_i}{\sum_{i=1}^{l^*}L_i (n-l^*+1)}=1-\frac{4\sum_{i=1}^n L_i}{(n+1)^2 L_1},$$
which coincides with the BFT bound. \footnote{Our notation differs from that in BFT (2011): namely, $L_1=\max_{j\in \{1,\dots,n\}} u_j^*$, and $\sum_{i=1}^n L_i =\sum_{i=1}^n u_j^*$.} \td{The BFT bound is verified in an example where $n$ is odd and $L_1=L_2=\dots=L_{(n+1)/2}\geq 1$ and $L_{(n+3)/2}=L_{(n+5)/2}=\dots=L_n=1$, which is a stronger condition than our setting.}

\item [(2)]If $n$ is even, we let $L_1=L_2=\cdots=L_{n/2+1}$. Then by Theorem \ref{thme}, we have $l^*=\max S^M_1=n/2$, and the upper bound is
$$UB(n,\vect{L};MMF)=1- \frac{4L_{l^*+1}\sum_{i=1}^n L_i}{(\sum_{i=1}^{l^*} L_i+(n-l^*+1)L_{l^*+1})^2}=1-\frac{4 \sum_{i=1}^n L_i}{(n+1)^2 L_1} ,$$
which coincides with the BFT bound. 

\end{enumerate}
    
\end{enumerate}

\epr

\theendnotes